# Observation-Driven Configuration of Complex Software Systems

**Aled Sage**

Thesis submitted for the Ph.D. degree
St Andrews

18th July 2003

School of Computer Science, University of St Andrews
St Andrews, Fife, KY16 9SS, Scotland

# Abstract

The ever-increasing complexity of software systems makes them hard to comprehend, predict and tune due to emergent properties and non-deterministic behaviour. Complexity arises from the size of software systems and the wide variety of possible operating environments: the increasing choice of platforms and communication policies leads to ever more complex performance characteristics. In addition, software systems exhibit different behaviour under different workloads.

Many software systems are designed to be configurable so that policies (e.g. communication, concurrency and recovery strategies) can be chosen to meet the needs of various stakeholders. For complex software systems it can be difficult to accurately predict the effects of a change and to know which configuration is most appropriate.

This thesis demonstrates that it is useful to run automated experiments that measure a selection of system configurations. Experiments can find configurations that meet the stakeholders' needs, find interesting behavioural characteristics, and help produce predictive models of the system's behaviour. The design and use of ACT (*Automated Configuration Tool*) for running such experiments is described, in combination a number of *search strategies* for deciding on the configurations to measure.

Design Of Experiments (DOE) is discussed, with emphasis on *Taguchi Methods*. These statistical methods have been used extensively in manufacturing, but have not previously been used for configuring software systems. The novel contribution here is an industrial case study, applying the combination of ACT and Taguchi Methods to DC-Directory, a product from Data Connection Ltd (DCL). The case study investigated the applicability of Taguchi Methods for configuring complex software systems. Taguchi Methods were found to be useful for modelling and configuring DC-Directory, making them a valuable addition to the techniques available to system administrators and developers.

## Declarations

I, Aled Sage, hereby certify that this thesis, which is approximately 42,000 words in length, has been written by me, that it is the record of work carried out by me and that it has not been submitted in any previous application for a higher degree.

date 18th July 2003 signature of candidate ………………………….

I was admitted as a research student in October 1999 and as a candidate for the degree of PhD in July 2003; the higher study for which this is a record was carried out in the University of St Andrews between 1999 and 2003.

date 18th July 2003 signature of candidate ………………………….

I hereby certify that the candidate has fulfilled the conditions of the Resolution and Regulations appropriate for the degree of PhD in the University of St Andrews and that the candidate is qualified to submit this thesis in application for that degree.

date 18th July 2003 signature of supervisor ………………………….

In submitting this thesis to the University of St Andrews I understand that I am giving permission for it to be made available for use in accordance with the regulations of the University Library for the time being in force, subject to any copyright vested in the work not being affected thereby. I also understand that the title and abstract will be published, and that a copy of the work may be made and supplied to any bona fide library or research worker.

date 18th July 2003 signature of candidate ………………………….

## Acknowledgements

I would like to thank my supervisors, Graham Kirby and Ron Morrison, for their frequent guidance and advice. Their insightful comments and unending patience were invaluable. Thanks also to Al Dearle for useful discussions, and for helping me find employment!

Cooperation with Data Connection Ltd (DCL) was vital to this research. Thanks especially to Richard Stamp, David Court and Edward Hibbert for their suggestions and their help in the industrial case studies.

Harry Staines rescued me during my initial sorties into the field of statistics, and has contributed greatly to the ideas and work in this thesis. He introduced me to Taguchi Methods and provided expert help in applying the techniques and analysing the results.

I'd like to thank Alice for her unending encouragement and for proof reading sections of this thesis, but I'm sure I'll have to reciprocate in a year or two… Thanks to my parents for much valuable support over the years, and to my flatmates and officemates for their distractions and support. Finally thanks to the St Andrews University Mountaineering Club for providing the frequent escapes from St Andrews, both physical and spiritual, that maintained my sanity and sometimes my hangover.

# Contents

# 1 Introduction

Software systems play an increasingly important role in organisations and in everyday life, the UK software market being worth £8.5 billion in 2001 [18]. Software systems are also growing in complexity: they are hard to comprehend, predict and tune due non-deterministic behaviour and *emergent properties* (i.e. behaviour of the whole system cannot be inferred from its parts) [32]. Complexity arises from the size of software systems, the amount of data and the scale of distributed systems. The choice of platforms, network configurations and communication policies has also increased markedly over the years, leading to a wide variety of operating environments with ever more complex performance characteristics. This is compounded by different versions of a software system having different characteristics, with behaviour being dependent on the *workload* (i.e. usage pattern).

Software systems attempt to meet the potentially conflicting needs of various stakeholders. Some users demand reliability and low response times, while others may desire consistently high throughput with minimal resource requirements. The appropriate trade-off between these needs depends on the particular set of users.

Many software systems are designed to be configurable so that policies (e.g. communication, concurrency and recovery strategies) can be chosen to give desired behaviour. Such systems expose a group of *configurable aspects*, which are implementation details that can be controlled explicitly. The system's *configuration* refers to a group of values that specifies a setting for each configurable aspect.

Comprehending the behaviour of a software system is related to the problem of configuring it. It can be difficult to accurately predict a configurable aspect's effect on behaviour, and to know which configuration is most appropriate for a given *condition* (i.e. environment and workload). This is especially true for complex software systems that expose many configurable aspects: the number of configurations increases exponentially with the number of configurable aspects, referred to as *the curse of dimensionality* [27].

## 1.1 Hypothesis

This thesis proposes the following hypothesis: automating the empirical measurement of a selection of configurations can:

- find a robust configuration (if any) that exhibits desired behaviour under a particular condition – useful for configuring the software system when it is first deployed or when adapting its configuration to cope with changes in the conditions of use;
- find characteristics of interest, such as conditions that cause substantial deterioration in performance – useful for guiding system usage and for focusing development efforts;
- help construct a predictive model of the software system that can estimate behaviour for untested configurations under given conditions.

The aim of the research described in this thesis is to develop software support, and an associated methodology, to test the above hypothesis. The approach involves empirically measuring the behaviour of a selection of configurations under various conditions, without assuming *a priori* knowledge of the system's implementation. This is done before the software system goes into use.

The hypothesis was tested in an industrial case study involving DC-Directory [10], an LDAP and X.500 product from Data Connection Ltd (DCL). The case study supported the hypothesis, demonstrating that automated measurement of a selection of configurations can achieve the goals listed above. Discussions with system developers revealed that measurements also improved comprehension of the system's behaviour. The software and methodology described in this thesis therefore makes a valuable contribution towards solving the problem of configuring software systems.

## 1.2 Overview of approach

The following chapters describe the design of ACT (*Automated Configuration Tool*) [93], and its use to explore the behaviour of software systems. ACT provides an infrastructure to run a sequence of *trials*: it *tests* (i.e. empirically measures the behaviour of) a configuration during each trial. The software system's configuration and conditions of use are adapted between trials. ACT is generic in that it can explore the behaviour of a wide variety of software systems, and can use a variety of *search strategies* to decide on the configurations to test.

Figure 1.1 illustrates the semi-automated process of running *experiments* to measure a sequence of configurations. The software system under test is called the *target system*. The *experimenter* (i.e. the user of ACT) decides on the configurable aspects to vary and the values to test for each, and chooses an appropriate search strategy. The experimenter provides information about the configurable aspects, including the locations of functions to set each configurable aspect's value, and the locations of functions to run a trial and recover from target system failure. For automation, these functions must run without human intervention. ACT dynamically loads the functions and uses them to run the trials, and to configure both the target system and the conditions under which it operates.

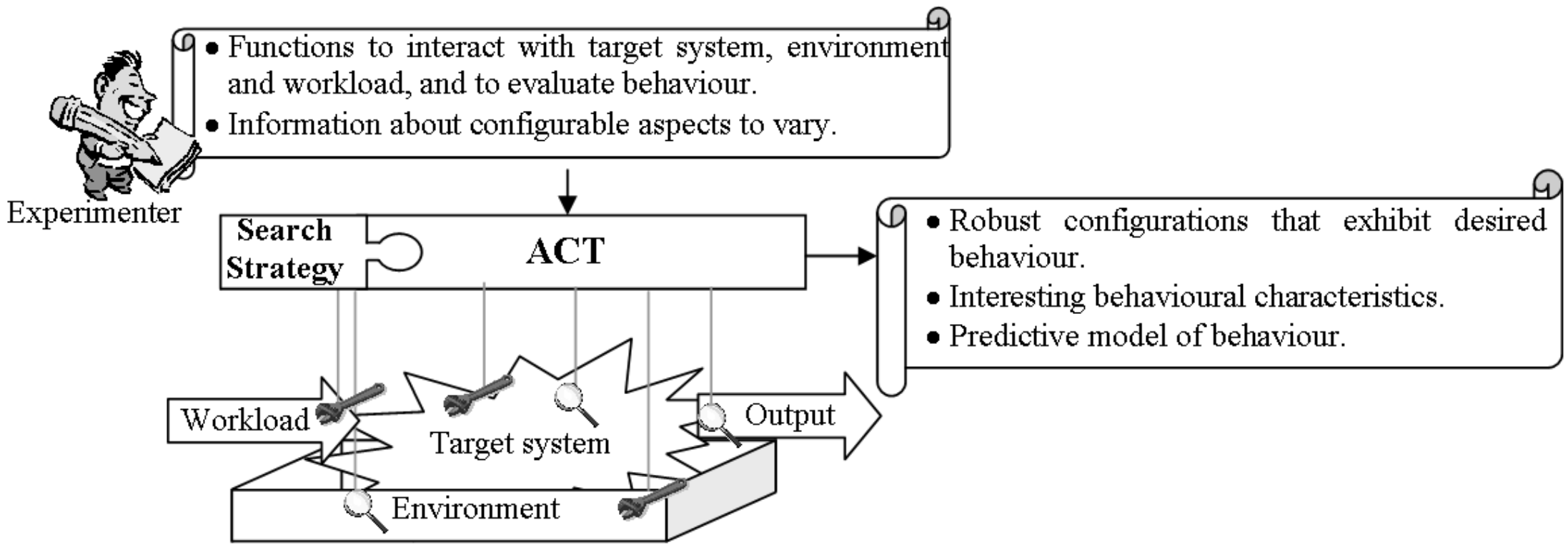

**Figure 1.1: System overview**

A search strategy generates a sequence of target system configurations and conditions to test. There are two categories of search strategy for ACT. The first is a feedback-based search, which uses observations of previous configurations to decide on the next configuration to test. The second category makes use of Design Of Experiments (DOE) [107], where the configurations to test and the number of times to test each are decided in advance.

*Taguchi Methods* [101] is a standardised statistical approach for DOE, to select a small subset of the possible configurations and conditions. It assumes that experts provide a list of (at most dozens of) configurable aspects and suggest (up to four) values for each. Research into the use of Taguchi Methods for configuring software systems is in collaboration with Professor Harry Staines of Abertay University, who is expert in their use for other fields such as biology and manufacturing.

Experiments designed using Taguchi Methods are conducted in two phases. The first phase involves testing a selection of configurations to produce a predictive model of the target system's behaviour. This is used to predict which configuration will perform optimally. The second phase involves testing the predicted optimal configuration and other configurations that have similar settings for the configurable aspects. Results from the second phase are used to more accurately predict a configuration that will consistently deliver high performance.

Taguchi Methods were used to explore the behaviour of DC-Directory. In-built observation and adaptation mechanisms were used to measure its performance and adapt its configuration. DC-Directory has a broad customer base, leading to a wide variety of usage patterns, operating environments and performance requirements. It exposes hundreds of configurable aspects, including caching policies, concurrency policies and queue management policies. Their effects on DC-Directory's behaviour depend on the conditions of use, and vary between system versions. DCL currently undertakes performance analysis and tuning by hand, which relies heavily upon costly expertise and only permits testing of a few configurations due to time constraints. The following example illustrates the problem: given only ten configurable aspects and four possible values for each, it would take almost sixty years to measure every configuration (given the requirement of thirty minutes per measurement).

## 1.3 Contribution

Comprehending the behaviour of complex software systems is difficult: some target systems may never be fully understood. Configuring complex software systems is therefore also difficult since it is hard to predict which configuration will be most appropriate for a particular condition. Use of ACT can reveal information about the behaviour of complex software systems. The approach is based on empirically measuring a selection of target system configurations, without assuming *a priori* knowledge of the system's implementation or architecture.

ACT provides a generic infrastructure to run experiments for target systems, which contributes to the state of the art in two ways:

- Running experiments assists human administrators to comprehend and configure target systems.
- ACT can use various search strategies, and investigate the strengths and weaknesses of each for exploring target system behaviour.

An important example is the use of ACT to investigate the suitability of Taguchi Methods for software configuration. The novelty of this approach is in the application of Taguchi Methods to help comprehend and configure complex software systems. Taguchi Methods have been used extensively in manufacturing for almost five decades, but the techniques have not previously been used before for configuring software systems. This thesis develops the techniques and applies them to an industrial case study: the combination of Taguchi Methods and ACT, yielding the semi-automated *TACT process*, is used to model and configure DC-Directory. Benefits of Taguchi Methods over other search strategies include:

- Only a small number of configurations need be tested to infer the effects on target system behaviour of many configurable aspects, and of selected interactions[1] between them. This gives a predictive model of the target system's behaviour and associated confidence levels in the results (see section 6.2 for a discussion of the assumptions of Taguchi Methods).
- The standardised approach for the design of experiments is simple and accessible to non-statisticians.

Some target systems exhibit variability in behaviour. The importance of *robustness* (i.e. consistently high performance with low variability) is emphasised in this thesis: sometimes improving the worst cases is more useful than improving the average case. Taguchi proposes a metric for estimating the robustness of a configuration, given replicated measurements of its performance. Use of this metric is novel to the field of software systems.

## 1.4 Thesis structure

Chapter 2 presents a literature review of the broad range of work related to configuring software systems. A glossary of the terminology used can be found in Appendix A.

Chapter 3 presents the Automated Configuration Tool in detail. Chapter 4 describes some possible search strategies, with particular emphasis on Taguchi Methods and on the statistical analysis of results.

Chapter 5 describes two industrial case studies. Results for DC-MailServer, a back-end mail server from Data Connection Ltd (DCL), illustrate the importance of replicating observations to measure robustness. The second case study illustrates the use of the TACT process to configure and model DC-Directory. Chapter 6 uses results from the case studies to evaluate the usefulness of running experiments, and discusses the benefits and limitations of Taguchi Methods.

---

[1] An *interaction effect* differs from interaction between components of a system (e.g. message passing). See section 4.3.3.2 for a description of interaction effects.

Chapter 7 gives details of future work, proposing further experiments and uses of ACT. There is a discussion of future versions of ACT to support on-the-fly evolution of target systems. Chapter 8 presents conclusions.

# 2 Literature review

Configuring target systems is important for tuning their performance, repairing faults and modifying or adding functionality. This is true both before the target system goes into use and *on-the-fly* (i.e. at run-time, while in use). The *configuration process* refers to the set of activities involved in tuning or evolving a target system. *Tuning* is adapting a target system at a given time, to find a configuration that behaves in a desired way. *Evolution* is the strategic adaptation of a target system's configuration over time, to progress to new and improved versions of the target system.

There are two main stages to configuring a target system: first deciding which changes are required and when, and second making the changes. Many projects aim to do this (e.g. [4, 9, 28]), often sharing with ACT the following common features:

- Observation mechanisms are used to measure the target system's behaviour and conditions of use.
- Adaptation mechanisms are used to configure the target system.
- *Controllers* (search strategies in ACT) are used to decide on an appropriate configuration of the target system.

This chapter first discusses techniques for performance evaluation, introduces the notion of *quality* and describes target system models. It then characterises complex software systems and justifies the need for adaptation. The three bullet points above are then addressed to describe observation mechanisms, adaptation mechanisms and systems for coordinating the configuration process. A glossary of the terminology used can be found in Appendix A.

## 2.1 Performance evaluation

When configuring a target system, it is necessary to determine how possible configurations will perform. Techniques for performance evaluation include *analytical modelling*, *simulation analysis* and *empirical measurement* [64].

*Analytical modelling* involves analysis of a target system's design and algorithms to develop a mathematical model of a target system's behaviour. It can be used early in the development life cycle to make predictions before implementation is complete. However, analytical modelling is difficult for complex software systems: complexity necessitates (unrealistic) simplifying assumptions, leading to a low level of accuracy compared to other performance evaluation techniques. Additionally, models require calibration using empirical measurements of the running target system.

*Simulation analysis* involves producing an executable model (i.e. a simulation) of a target system, which is run to predict the real target system's behaviour. Source code analysis and profiling information can be used to produce a simulation of a target system's control flow, resource requirements and communication patterns [22, 66]. The advantage of running a simulation over running the target system is that the execution time is usually less, and that adapting the simulation is easier than adapting the target system's configuration. Additionally, algorithms can be simulated early in the development life cycle before implementation of the target system is complete. However, an executable model produced using source code analysis is unlikely to exhibit the emergent properties and non-determinism of a complex software system.

*Empirical measurement* involves observing the running target system. There is a high level of confidence in results from empirical measurement, compared to predictions from a model or simulation, because observations are of the running target system. This approach can be used for software systems that are too complex to analyse or simulate accurately, assuming there is the capability to observe the running system (see section 2.6). This thesis focuses on the use of empirical measurement.

When measuring behaviour before the target system goes into use, it is important that the workloads be representative of the customer's likely usage patterns. Logging a customer's usage, where possible, allows an identical set of inputs to be used during experiments, or a synthetic workload to be developed based on characteristics of the logged input. An alternative is to use domain-specific benchmarks, such as *DirectoryMark* [17], that describe common usage patterns.

## 2.2 Quality of target system configurations

A target system configuration and a condition under which it operates is called a *combination*. This specifies a value for each configurable aspect of the target system and aspect of the conditions of use that can be configured. In statistics terminology, these aspects are called *factors* and their values are called *levels* [107].

The combinations form a multi-dimensional *input space* whose dimensions correspond to the factors. A point in the space represents a combination, giving a level for every factor.

A multi-dimensional input space is not the only way to view the set of possible combinations. Constraint programmers, in contrast, would describe the input space as a *search tree* [21]. A branch in the tree corresponds to a choice of level for a factor, and a path from the root to a leaf node gives a single level for every factor. Each leaf node therefore represents a combination. The following discussion uses the metaphor of a multi-dimensional input space.

The process of configuring a software system is driven by a *configuration goal*, which specifies the desired behaviour in terms of a potentially conflicting set of *fitness metrics*. An example configuration goal for a database system is to maintain a latency of less than 500ms for 99% of requests, while maximising throughput. Each time a combination is tested, the fitness metrics' values are recorded.

Values for each fitness metric could be qualitative or quantitative; values could have a scale type of *nominal*, *ordinal*, *interval* or *ratio* [51]. These mean respectively that the values refer to categories, that the values are ordered, that the values increase in regular step sizes, and that there is a fixed zero point so that relations such as "twice the value" are meaningful.

Identifying the best configurations, in terms of the fitness metrics, is a multi-objective optimisation problem. Some researchers search for *Pareto optimal solutions*, where no fitness metric can be improved without causing at least one other fitness metric to deteriorate [36]. An alternative approach is to combine the fitness metrics using an *aggregating function* to produce a single *response*, the higher the better[2]. This makes comparing configurations simpler. It is possible if there is a well-defined configuration goal: choice of aggregating function requires knowledge of the relative importance of each fitness metric, their expected range of values and the desired values for each. The following discussion assumes that a single response is calculated using such an aggregating function.

The *quality* of a combination is a measure of how well it meets the configuration goal. Taguchi defines a high quality combination as one that imparts little loss to society from the time the target system is shipped [101]. He suggests measuring quality in terms of *robustness*: consistently high performance with low variability. This view of quality is often used in manufacturing industries, but is just one possible definition (e.g. it ignores time to market). According to Taguchi's definition, quality is determined by the costs incurred whenever the target system fails to meet the configuration goal. This implies that high-quality configurations deliver consistent performance, but some complex software systems produce a different response each time a given combination is run. Replicating trials is therefore important for estimating the quality of a combination – the more replications, the greater the confidence in the estimate. The set of responses from replicated trials combine to give a single value of the *response variable* for each combination.

[2] Maximising the response involves finding the global maximum on a response surface, and is discussed later in the section. Maximising the response is an arbitrary choice: for some problems, such as measuring latency, it may be more appropriate to talk of minimising the response.

The acceptable level of variability in a target system's behaviour depends on the consistency demanded by the customer. If the desired consistency is not achieved, the target system may be classed as *non-deterministic*. The distribution of responses is also important.

It is arguably more useful in some situations to improve the worst cases than the average case of a software system's performance: bad worst cases give a negative impression of the system and can lead to higher costs due to reduced sales and more calls to the helpdesk [99]. Taguchi's robustness metric (called *signal to noise ratio*[3], described in section 4.3.2) takes into account both how high responses are and their consistency. Ranking combinations in terms of robustness can give a different order than if the *mean* of the responses is used. For example, Figure 2.1 shows the results of a hypothetical experiment: it shows the observed responses (represented by triangles) for combinations *A*, *B* and *C*, and the mean response for each (denoted **M**). Combination *C* usually gives a higher response than combination *A* and therefore might be considered preferable. However, one response for *A* was 100, which gives *A* the highest mean. Combination *B* has two observed responses that are very low – bad worst cases – making this configuration unfavourable. Ranking the combinations according to robustness gives *C*, *B*, *A* – the reverse of the ordering produced using the mean.

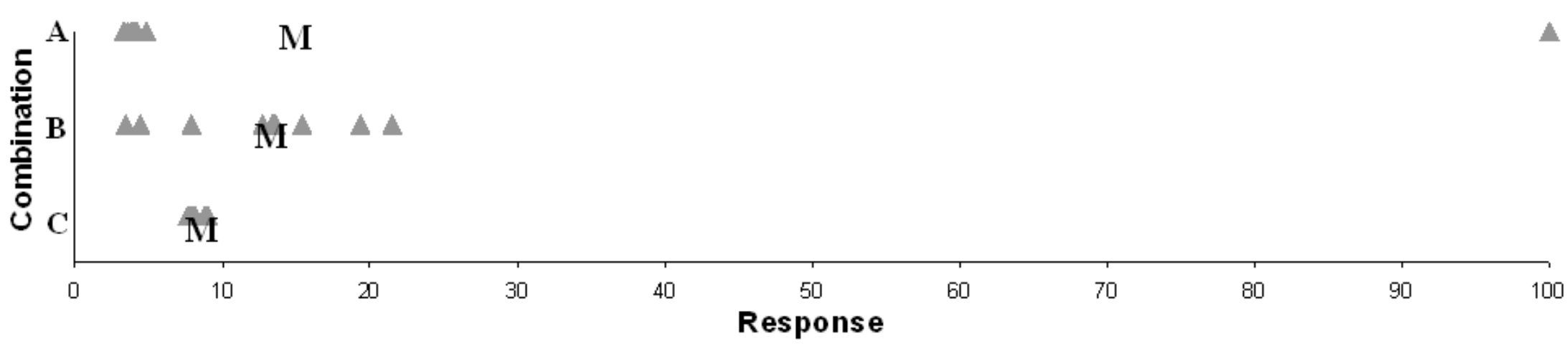

**Figure 2.1: Quality of responses**

[3] *Signal to noise ratio* used here is very different to that used in communications, where it is the ratio of a desired signal's amplitude to the noise's amplitude at a given time on a communication channel.

A *response surface* can be used to show the relationship between the levels of factors and the value of the response variable. The response surface lies over the input space, using the dimensions of the input space and an additional "response variable dimension". For the case of measuring quality (i.e. robustness, according to Taguchi's definition), each point on the surface shows how well a combination meets the configuration goal. Figure 2.2 shows a hypothetical response surface for a two-dimensional input space consisting of factors *A* and *B*. The surface has a *phase change*, where there is a sharp change in the shape of the surface. The surface also includes: a *global maximum*, which corresponds to the configuration delivering optimal quality; a *local maximum*, i.e. a combination better than its neighbours but not the best in the space; and a *local minimum*, i.e. a combination worse than its neighbours.

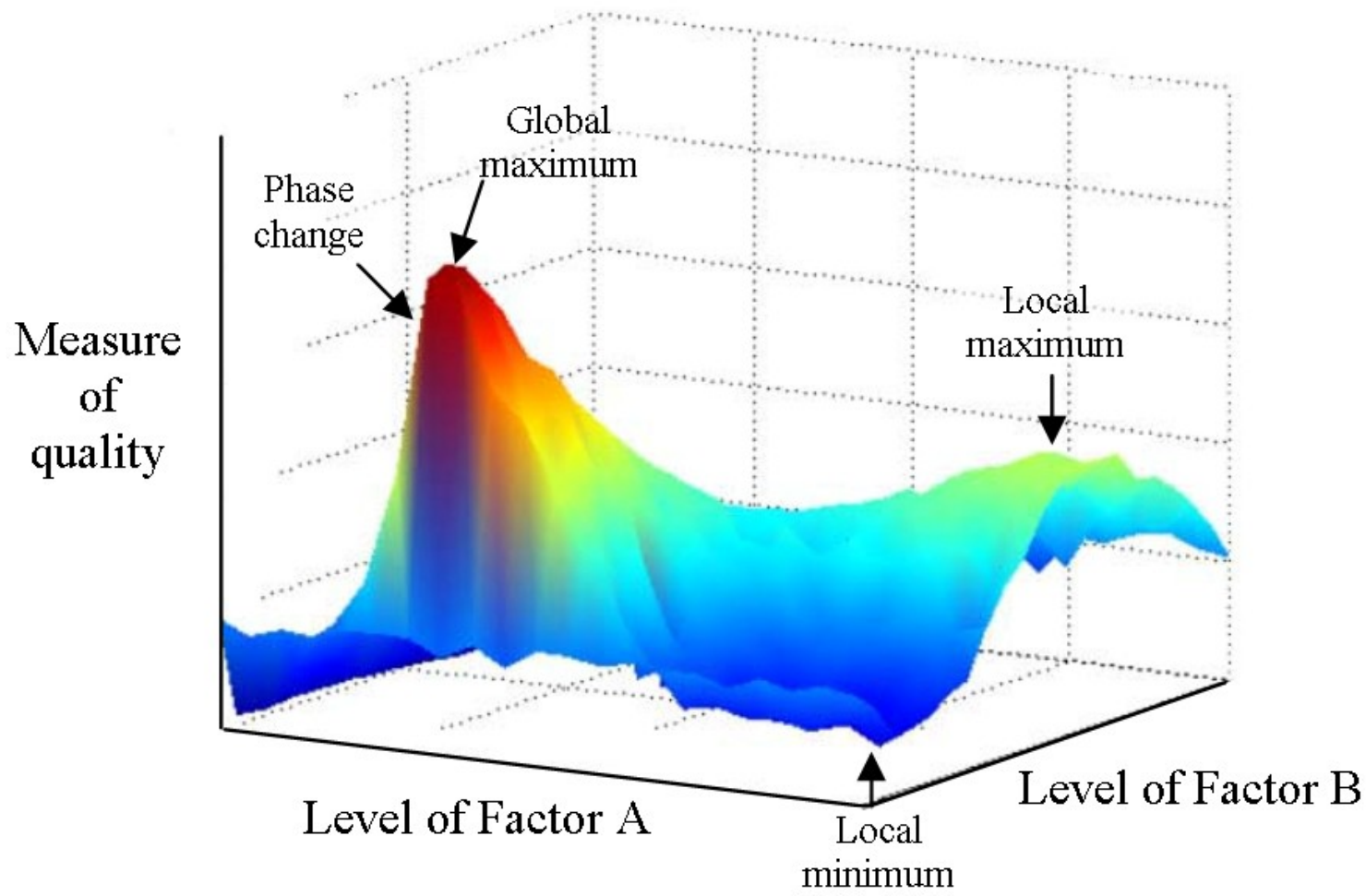

**Figure 2.2: Example response surface**

A search of the response surface is required to identify good configurations, interesting behaviour and the effects of each factor. Depending on the search strategy employed, this can be a non-trivial task due to phase changes, local maxima and interactions among factors. The problem is compounded if there are a large number of possible combinations and a limit to the number of combinations that can be tested. Running experiments that assist human administrators in configuring target systems is therefore a difficult problem.

## 2.3 Target system models

A *model* is a representation that exhibits some property of the target system. Models are useful when configuring target systems because they make explicit the modeller's understanding and allow the target system to be considered at some level of abstraction. A model may form the basis for planning and coordinating adaptation of a target system's configuration: it may predict the target system's behaviour and provide a context to describe and decide upon adaptations.

Models can be categorised by:

- **Properties the model represents**. For example, quality of service (QoS), resource usage, architecture, or a combination of these.
- **The temporal scope of the model**. For example, time the requirements were specified, time the target system's execution was last observed, and/or future time when predicting behaviour.

A model may be *complete* (though not necessarily accurate) or *incomplete*. A complete model represents a target system property over the full input space. An incomplete model provides partial information, describing the target system property in only some situations.

A model may represent the target system in a simplified form. By ignoring some details, complexity is reduced but at the cost of reducing the accuracy of information. Conversely, extra information may be contained in a model (e.g. descriptions of other implementations, version history, etc).

Techniques such as analytical modelling and empirical measurement can contribute towards comprehension of a target system's behaviour and help in the production of a predictive model. As the model is refined, it converges to the most accurate model possible: running the real target system (see Figure 2.3).

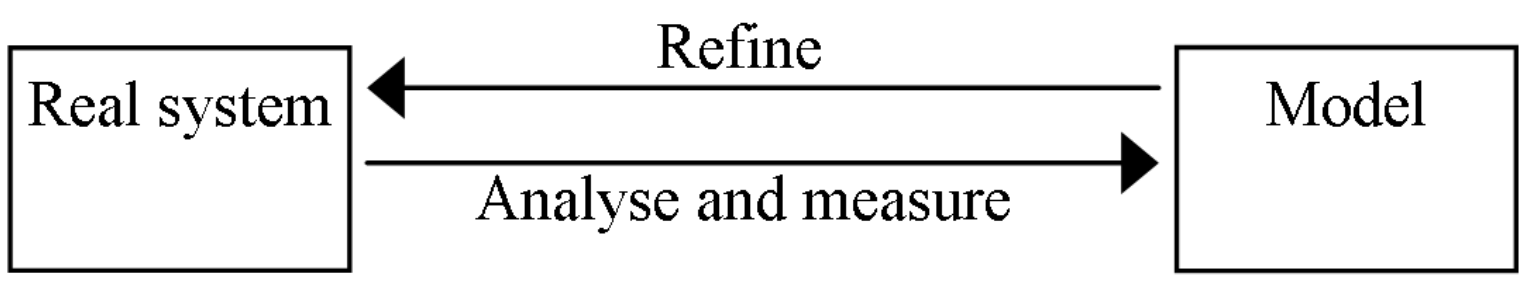

**Figure 2.3: Refining a model**

For a model of an evolving target system to be continually useful (i.e. not become out-of-date), it should be maintained throughout the lifetime of the target system. Bindings between the executable, source code, documentation, requirements and models help in maintaining a target system: changes made to the target system are mapped to changes in the model, and *vice versa*. When there is a causality relationship between the model and the subject being modelled, it is called an *active model*. A *passive model*, in contrast, is independent of the subject [104].

## 2.4 Complex systems

Of particular interest here are complex software systems lacking previously-known predictive behaviour models. These are inherently difficult to configure for the following reasons [32]:

- Emergent properties are only apparent at run-time, which necessitates observation of the running system.
- Non-deterministic behaviour makes drawing conclusions from observations difficult and modelling problematic.
- Non-linearities and phase changes make behaviour hard to predict.
- Trends toward larger software systems, more data and more devices make the system harder to manage.
- Choice of platforms, network configurations and communication policies has increased markedly over the years, leading to a wide variety of operating environments with ever more complex behaviours.
- Different versions of a software system can have different characteristics, which necessitates configuring the system every time it is upgraded.
- Conditions of use can change, which necessitates repeated adaptation of the target system's configuration.
- Many software systems are poorly documented and are maintained in an *ad hoc* manner, which makes comprehending and configuring the system difficult.

## 2.5 The need for adaptability

### 2.5.1 Compliant systems

Morrison *et al* argue that software systems should accommodate, and thus be *compliant* to, the needs of particular applications and customers [83]. This requires that the needs of the application be known and that the software system be configured to meet those needs.

Kiczales *et al* [69] observe that some policy decisions are "crucial strategy issues whose resolution will invariably bias the performance of the resulting implementation." These policy decisions are called *mapping dilemmas* (in an OS, they relate to how an abstraction is mapped onto the underlying hardware). The choice of policy is called a *mapping decision*, and a *mapping conflict* occurs when a policy decision is inappropriate for a given application.

The traditional view of designing systems is to use static abstract components or layers, using encapsulation to encourage software reuse. Policy decisions that are believed to suit the requirements of "typical" applications are hidden from the application, even though the application may have vital information about which policies are best suited to its needs.

An *open implementation* approach aims to expose policy decisions, to avoid the danger of encapsulation outlined above [68]. ACT can run experiments to measure the effects of various policies, to infer which policies are best suited to a given application.

The *Compliant Systems Architecture* (CSA) project describes an architectural approach to the construction of configurable systems: components are designed, top down, with the philosophy of fitting the architecture to the needs of the particular application [83]. The key technique in CSA is to separate *mechanism* and *policy*, allowing the architecture to be tailored to the policy needs of the application. A component's functionality is delivered by a set of mechanisms, and the policy for using these mechanisms can be supplied by other components. Additionally, *downcalls* provide a way to exploit knowledge of a component's internal structure, to extend the interface exposed by a component.

### 2.5.2 Evolution of software systems

Greenwood *et al* discuss the characteristics of evolution in [60], defining it as a transformation; it is an adaptive change with a time dimension. Evolution can be *directed* or *spontaneous* – it can be externally imposed or internally driven. Evolution can be *focused* or *diffuse* – it can be the result of a purposeful strategy decision or consist of many small logically separate changes.

*Co-evolution* refers to the situation where the evolving target system has a dependency on another system [60]. Changes can cause a ripple effect: a change in one system necessitates evolution of dependent systems to maintain consistency between them. With *quasi-independent evolution*, a target system can evolve independently, but only to the extent that its neighbours can accommodate such change.

Lehman's first law states that an *E-type* software system (i.e. a target system used and embedded in a real-world domain) must continually change or become increasingly less useful [74]. Change is driven by the need to repair software faults, cope with new operating environments, and add or modify functionality. It is estimated that these activities comprise 17%, 18% and 65% of software maintenance respectively [97].

Change in expectations of a target system's behaviour drives adaptation. Expectations change due to changing demands of the market, an organisation's desire to stay ahead of the competition and changes in the structure of the organisation itself. The last is driven by feedback loops within the organisation in which the target system operates. Development and use of a software system changes the organisation, and causes a mismatch between the target system and its operational domain. If the target system does not evolve to meet the changing goals of the organisation, the target system's functionality increasingly diverges from meeting the organisation's needs.

Evolution can be divided into *planned* and *unplanned* change. Planned changes are catered for in the target system's design, but unplanned changes are unanticipated. Planned change is generally accepted to be easier than unplanned change, but is still difficult unless one can somehow predict or find an appropriate configuration. Use of ACT can help to find a suitable configuration by empirically measuring a selection of configurations.

When evolving a target system, it is important to strive to reduce complexity and enhance the structure of the target system. Unstructured change can make subsequent evolution harder; it can decrease cohesion and increase coupling both amongst components and amongst non-functional aspects of the target system; it can make the target system harder to comprehend. As Brooks says: "All repairs tend to destroy the structure, to increase the entropy and disorder of the system. Less and less effort is spent on fixing original design flaws; more and more is spent on fixing flaws introduced by earlier fixes. As time passes, the system becomes less and less well-ordered" [31].

The "structure" of a system is its *architecture*: "the fundamental organisation of a system embodied in its components, their relationships to each other, and to the environment, and the principles guiding its design and evolution" [1]. Maintaining an explicit first-class representation of the architecture at run-time, e.g. using an *Architecture Description Language* (ADL) [55], can help preserve a target system's structure. Such a description typically formally identifies the *components* of the target system and the inter-component communication, defined by *connectors*. A connector is a link between two or more components, across which they can interact. A connector could itself be a component, or it could be a binding between two or more components.

## 2.6 Observing behaviour

The approach presented here for configuring software systems is based on empirical measurement of the running target system. This requires observation and interpretation of the target system's behaviour, done by *probes* and *gauges* respectively [79]. Probes collect data, possibly at run-time (e.g. count of operations performed), by interacting with the target system and its environment. Gauges gather and interpret this data in a context meaningful for evaluating behaviour (e.g. in terms of the configuration goal's fitness metrics). Much of the work on probes and gauges described here is part of the DASADA programme, discussed in section 2.9.5.

### 2.6.1 Probes

Probes observe the target system to monitor and measure its behaviour and state, and generate *events* to describe this information. Probing is a form of *reification*: probes provide a mapping from an entity (i.e. the target system) to a concrete representation (i.e. the events).

#### 2.6.1.1 Desirable qualities

There are many qualities desirable in probes, including:

- **Correctness and dependability**. Observation of the target system should closely mirror the behaviour or state of the target system. Probes should consistently observe, and not miss, behaviour of interest.
- **Little or no *probe effects***. Deploying and activating the probe should not cause perturbation in the target system's behaviour; the target system's behaviour and state should be the same both when the probe is and is not present. This implies safety: the probe will not cause the target system to malfunction.
- **Separation of concerns**. The probe technology should be separate from the target system implementation, to promote reusability of probes and ease of extensibility.
- **Probe control and adaptation**. It may be desirable to deploy, activate and/or deactivate probes at run-time. Facilities to tailor probes for particular tasks (i.e. adapt their configuration at run-time) may also be desirable.
- **Security**. Control of probes and dissemination of information should be restricted to trusted parties.

#### 2.6.1.2 Categories of probes

There are many techniques for observing the behaviour of a running target system. Probes can use third party software, such as tools to monitor resource usage, profilers to collect execution time information, and target system specific tools for extracting diagnostic information. Other probe technologies include instrumenting a target system's code or executable, intercepting calls to shared libraries [26], reflection [23], and measuring a server's behaviour at the client side by monitoring responses to requests.

Below is a description of characteristics that categorise types of probes. This summarises and augments the discussions in [57, 90].

<u>Aspect of the target system</u>

Observation (and adaptation) mechanisms can be categorised according to the aspect of the system on which they act. Categorisation promotes reuse of probes by identifying the set of systems that each probe can be used to observe. Mechanisms specific to bespoke components of the target system are useful for only that target system, while mechanisms that observe or adapt the environment are reusable for any system that operates in that environment.

Wells *et al* use the terms *AppliProbe* and *EnviroProbe* to describe probes that observe the application and environment respectively [105]. Aspects of the system can be categorised further, as listed below in order of increasing generality and reusability:

- target system *components* and *architecture* (i.e. parts of the target system and the way they are connected) that are unique to the target system;
- the workload (i.e. facets of how the target system is used);
- shared infrastructure, such as shared libraries or middleware;
- the environment in which the target system operates.

<u>Location in the target system architecture</u>

Probes can be classified according to the location in the target system architecture in which the probe is inserted. These include the categories below (the first three are described in [57]):

- *Component boundary intrusive* probes observe the target system from inside a component. Any changes to the target system required for probing are contained within the component being observed.

- *Connector intrusive* probes only require changes to a connector in the target system.
- *Architecturally intrusive* probes require changes to the target system that are visible at the architectural level, e.g. adding a new monitoring component or changing the interface of a component.
- *Inbuilt* probes are already part of the target system, requiring no target system alterations for their use.
- *External* probes lie outside of the target system: they observe side effects of the running target system. A special case is *user-centred* probes, which observe the behaviour of the target system from the perspective of the user (e.g. response time).

<u>Mechanism for probe insertion</u>

Probes can be categorised according to the mechanism used for insertion. Some possible mechanisms include:

- source code modification;
- binary or byte-code modification during or before load-time;
- connector indirection, to intercept communications between components;
- redirection during compile-time or run-time linking, for example to use an alternative component;
- inbuilt facilities for adaptation in the target system, such as behavioural or structural reflection [72];
- changes to the execution environment, such as modifying or replacing the virtual machine.

<u>Time of insertion</u>

Probes can be inserted (or removed) before the target system goes into use or on-the-fly. The former can occur any time prior to the execution of the probed target system (e.g. composition time, compilation time, or during run-time linking). On-the-fly insertion is performed at run-time, while the target system is in use.

<u>Knowledge of the internals of a target system</u>

Probes can be categorised as *black box* probes or *white box* probes. *Black box* probes require no knowledge of the internals of the target system, including its architecture and implementation. *White* *box* probes require a (limited) understanding of the internals of the target system. Gill's definition of black box probes includes those that use, but have no understanding of, the target system's source code [57]. In this thesis, such probes are classified as white box because availability of source code implies delving behind the interface exposed by the target system (see section 2.8.3).

Probe dependencies

Probes have *implementation dependencies* and *conceptual dependencies* [57]. Implementation dependencies of a probe technology are requirements that can be overcome by developing new tools. Conceptual dependencies are absolute requirements of the probe technology. For example, a tool for instrumenting Java source code has an implementation dependency of working only with Java, but there is a conceptual dependency of requiring write access to the source code prior to (or during) compilation.

Triggering mechanism

Probes can be categorised by the *triggering mechanism* (i.e. the type of activity that causes a probe to generate an event):

- *Passive probes* are reactive, generating events entirely in response to activities in the target system. For example, a passive probe could generate an event whenever a particular function in an API is invoked. Such probes may be *autonomous* (i.e. generate an event every time they are triggered) or *controllable* (e.g. capable of being activated and deactivated). Below are two possible execution mechanisms for such probes:
    - A probe may be executed as part of the natural flow of control of the target system.
    - A probe may have its own thread of control, and observe the external behaviour of the target system's processes.
- *Active probes* are proactive, generating events in response to activities external to the target system. Triggers include:
    - a query, which is a *pull* mechanism that allows an external agent to trigger event generation (e.g. a target system administrator requests the current CPU usage statistics);
    - a timing mechanism, which uses some schedule to control the probe's activation (e.g. measure network usage every five minutes).

- *Hybrid probes* combine characteristics of both passive and active probes: they can passively observe activities in the target system, and generate events when a constraint is satisfied or an external activity triggers event generation. For example, a hybrid probe may observe the processing of database search operations, and generate an event every minute reporting the throughput.

Awareness of scope

Finally, types of probes can be categorised according to their awareness of what triggered them. That is, the ability to identify who queried the probe, in what component an observed event occurred and what caused the event.

### 2.6.2 Gauges

Gauges gather and interpret observations; they can aggregate, compute, analyse and then disseminate high-level events that describe the target system and its conditions of use [5, 56]. Gauges differ from probes in two ways:

- Gauges can consume events produced by other probes and gauges, while probes take no such input.
- Gauges interpret observations in the context of a model of the target system (e.g. in terms of fitness metrics of the configuration goal), while events generated by probes need not be directly meaningful in the context of any high-level model.

### 2.6.3 Probe and gauge run-time infrastructures

The purpose of a *probe run-time infrastructure* is to standardise the run-time deployment and control of (potentially distributed) probes, and the delivery of events relating to probes. Such an infrastructure has been developed as part of the DASADA programme by the *probe run-time infrastructure working group* [25, 57]. Communication in the probe run-time infrastructure is through events, which are disseminated over a *probe reporting bus* (e.g. using the *Siena* publish/subscribe event notification service [106]). A set of event types has been defined [57], partitioned into *infrastructure events* (i.e. control instructions for probes) and events that probes can generate. A *probe adapter* is installed *a priori* on each participating host to receive and send events, and to interact with probes on that host.

As part of the DASADA programme, the *gauge infrastructure working group* have developed a conceptual architecture for the use and management of gauges. It "defines (1) a common framework for describing, developing and integrating gauges, which can be used as a standard that is shared between gauge developers and gauge consumers/integrators; and (2) a common set of services that support run-time communication between gauges and the consumers of their outputs" [5]. The *gauge reporting bus* provides a publish/subscribe service for dissemination of information: gauges publish information (in the form of events), while consumers subscribe to such information.

### 2.6.4 Confidence in observations

The *confidence level* in an observation is the probability that behaviour is as suggested by the observation, versus the probability of a *false positive* result due to an alternative explanation. For example, an observation of a component's failure may be due to the component having failed or due to packet loss on the network between the observer and the component. The appropriate reaction to an observation depends on the associated confidence level and on the cost of the action. For example, it is often not beneficial to dynamically change network routing tables when a single observation suggests that a route is down: the risk of having to restore the routing tables outweighs the benefits [78].

Corroborative observations from independent sources decrease the probability of a false positive result, which is the product of the probability for each individual observation being a false positive. Thus confidence greatly increases when there is evidence from multiple sources.

Bayesian statistics can help to determine the confidence level. Bayes' rule states that the probability of a hypothesis, *h*, being true, given evidence, *e*, is:

$$P(h \mid e) = \frac{P(e \mid h)P(h)}{P(e \mid h)P(h) + P(e \mid \sim h)P(\sim h)}$$

Consider the hypothesis that a component has failed, given an observation of failure. The probability of the component having failed is proportional to the probability of making the observation when the component has failed and to the probability of the component failing. The two terms on the denominator are: the probability of observing failure when the component has failed, multiplied by the probability of failure; and the probability of making the observation when the component has not failed, multiplied by the probability of no component failure. If these terms can be estimated, the confidence level in an observation can be calculated.

## 2.7 Adapting target systems

Adaptation mechanisms provide the capability to change a target system's configuration or its conditions of use.

### 2.7.1 Dimensions of adaptation

Many issues need to be addressed when designing adaptation mechanisms, particularly for on-the-fly adaptation. These issues are categorised under eight dimensions [46]:

- **The interface for triggering adaptations**. Adaptations can either be *directed* or *spontaneous* (i.e. externally imposed or internally driven). Mechanisms for triggering directed adaptation can be *declarative* (e.g. based on specifying the required behaviour) or *procedural* (e.g. exposing hooks to which new code can be bound). By definition, there is no interface to trigger spontaneous adaptation.
- **Authorisation of adaptation requests**. Who can trigger adaptations, and when?
- **Feasibility of adaptation**. Is the suggested adaptation possible in the current situation?
- **Dependency management**. Interdependencies among components mean that a change in one component can necessitate change in other components. Making changes requires a mechanism to determine these dependencies (e.g. based on a formal description of the target system's architecture). Where necessary, a mechanism is required to perform a set of adaptations as an atomic transaction that potentially spans multiple sites [102].
- **State transfer**. If an existing component is replaced with a new version, how is state information transferred from the old to the new component?
- **Source of new code**. If introducing new code into the target system or replacing components, where does the new code come from?
- **Binding**. What is the mechanism for binding new components into the target system and activating them?
- **Security issues**. Are new components safe and what execution privileges do they have?

### 2.7.2 Complexity and flexibility of adaptation mechanisms

Adaptation mechanisms can be categorised according to complexity, illustrated in Figure 2.4. Compared to simple adaptation mechanisms, complex adaptation mechanisms generally provide more flexibility in terms of changes that can be made. However, they cause greater perturbation to the target system and are harder to automate. The vertical arrow in Figure 2.4 represents the progress of the target system. The loops show different adaptation mechanisms, the size of the loop indicating the complexity. The simplest mechanism, labelled *tuning knobs*, uses pre-defined configurable aspects exposed by the target system. A more complex adaptation mechanism binds new code into the target system to change or augment behaviour, for example using structural reflection [72]. A third mechanism seeks external help to adapt the target system, for instance requesting additional hardware resources or a non-trivial change to the target system's implementation.

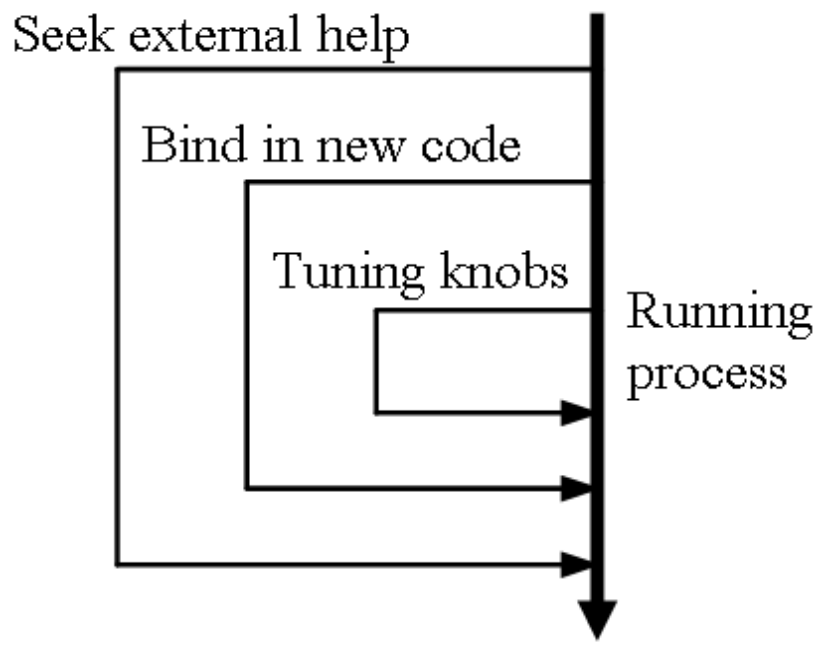

**Figure 2.4: Mechanisms for adaptation**

The adaptation mechanisms used are orthogonal to deciding how to adapt the target system's configuration. Each of these mechanisms can be treated in the same way, provided the set of legal adaptations is known and there is a uniform method of invocation (e.g. sending events or calling wrapper functions). Appendix B includes a list of issues relating to use of adaptation mechanisms.

### 2.7.3 Adaptation mechanisms

There are many mechanisms for adapting a target system, including those discussed below.

<u>In-built facilities for adaptation</u>

Many target systems have in-built configurable aspects. For example, use of the *strategy pattern* in object-oriented design allows a program to be structured such that the choice of algorithm can be made at run-time [46, 54]. The set of supported strategies is always predetermined if there is no mechanism to bind in new code.

Resource allocation and process migration

A target system's behaviour can be modified by reallocating the available resources. This includes process migration, where the execution of a process is moved from one node to another [80]. Below are some advantages in supporting on-the-fly resource allocation:

- Redundant resources can be exploited to balance the load across multiple nodes and to match a task to the resource most suited to its execution (e.g. CPU intensive tasks on the fastest node).
- Communication can be improved by co-locating processes that interact intensively and by locating a process close to the source of its data.
- Fault resilience can be improved by replicating processes and data (i.e. redundancy) and by migrating processes from nodes that are suspected to have experienced partial failure.
- Mobile users can utilise resources in the local proximity.

Late binding and reflection

The capability to modify a target system's implementation, or the meaning of that implementation, allows the target system's functionality and behaviour to be changed. Binding new code into the target system at run-time (i.e. *late binding*) allows policy decisions to be made at run-time, and the set of supported implementations to be extended dynamically. Several techniques are discussed below.

Dynamically Linked Libraries (DLLs) are bound to a target system at load-time, rather than at compilation time. A target system's behaviour depends on the functions used in the shared library: loading a new DLL can change the target system's behaviour. This technique is comparable to the Java mechanism of using a *ClassLoader* [85]. Balzer and Goldman's *mediating connectors* provide a mechanism to intercept calls to shared libraries [26]. Their *mediators* are wrappers that change the behaviour of the wrapped libraries on-the-fly by changing or augmenting calls to the libraries' APIs. These changes are transparent to the target system.

The behaviour of programs may be changed using a combination of first-class procedures, L-value binding and assignment [42]. Languages such as Napier88 [84] and ProcessBase [82] allow one to assign a procedure to a variable and call the procedure using the variable's name. Subsequent assignment to this variable replaces the procedure and thus adapts the behaviour of the procedure calls. Using a persistent store as a repository for the running program allows its behaviour to be modified by changing the contents of locations containing its procedures.

*Reification* and *reflection* provide the capability for a target system to observe and adapt its own behaviour in the course of its evaluation; they provide a means of achieving openness and flexibility. Reification involves making explicit a representation of an aspect of the run-time system, and making this representation accessible to the program itself. Two categories of reflection are *structural reflection* and *behavioural reflection*.

Run-time structural reflection, also called *linguistic reflection* [71], is the ability of a running program to generate new source code, compile it using a dynamically callable compiler, and link it into the program's own execution. Behavioural reflection allows a program to adapt its own meaning by manipulating its *evaluator*. The *meta-object protocol* [67] provides access to the evaluator by defining *meta-objects* that control aspects of a program's behaviour, such as method invocation and object creation. Changes to the meta-objects cause changes in related aspects of the program's behaviour.

<u>Dynamic software architectures</u>

*Component-based Software Engineering* (CBSE) is concerned with developing components, building systems from these components, and evolving the system by replacing and customising components [100]. Building systems from components is not new: Parnas introduced the ideas of modules and abstract interfaces decades ago [89]. However, there is a drive to formalise the process and methodology for building and evolving component-based systems.

The architecture of some target systems permits the addition, replication, removal and replacement of components and connectors. Some Architecture Description Languages, such as π-SPACE [33], can describe the mobility of links between processes, and can thus describe possible adaptations to the target system's architecture. An architecture description maintained at run-time can form the basis for specifying desired adaptations, which map to operations on the target system itself [86].

## 2.8 Design Of Experiments (DOE)

### 2.8.1 Theory

> "To call in the statistician after the experiment is done may be no more than asking him to perform a postmortem examination: he may be able to say what the experiment died of."
>
> *Sir Ronald Fisher, Indian Statistical Congress, Sankhya, ca 1938*

An experiment design is a description of the set of combinations to test and the number of replications for each. Some search strategies, such as Taguchi Methods (see section 4.3), use a structured statistical approach for the Design Of Experiments (DOE) [107] and analysis of results. The aim of DOE is to determine the maximum amount of information about a target system with the minimum of effort. The approach studies simultaneously the effects of multiple factors on the response variable (i.e. on the target system's quality), to infer the effect of each factor and of selected interactions, and to estimate the confidence in the results. An effect is said to be significant at the 5% level if there is 95% confidence that the effect on the response variable is non-zero[4].

A *main effect* is defined as the effect on a target system's behaviour caused by a single factor being varied. An *interaction effect* between factors is defined as the degree to which the factors' effects depend on one another's levels. An interaction effect between two factors is called a *two-factor interaction effect*. See sections 4.3.3.1 and 4.3.3.2 for a more detailed discussion of main and interaction effects.

A *full factorial design* experiment tests all possible combinations, allowing the effects of every factor and every interaction to be inferred. Such experiments are often prohibitively expensive, in terms of time and cost, due to the large number of combinations. Approaches to reduce the size of the experiment include:

- decreasing the number of factors to vary;
- decreasing the number of levels to test for each factor;

[4] The 5% level is commonly used in statistics as an acceptable probability of incorrectly concluding that an effect is significant; other levels can equally be chosen.

- using a *fractional factorial design*, where only some of the possible configurations are tested (given a list of factors and a list of levels for each, without ignoring entire factors or levels in the list).

The first approach assumes that not all factors are important in all situations – often a necessary assumption when there are hundreds of possible factors. Deducing which are important requires either time to determine this experimentally or expert knowledge. The second and third approaches assume that the experimenter can infer from results the effects of factors and interactions, allowing predictions for untested configurations.

One technique, described in [64], is to run the experiment in two phases. The first phase involves identifying which factors have a significant effect on the response by using a full factorial design with just two levels per factor (called a *$2^k$ design* when there are *k* factors). The second phase involves running an experiment for only the significant factors, to test more levels. However, the number of combinations in a $2^k$ design increases exponentially with the number of factors. Some fractional factorial designs use fewer combinations and are therefore often preferable.

The assumption with fractional factorial designs is that the experimenter can interpolate the response of untested combinations from the effects of factors and a selection of interactions. Such designs obtain less information than full factorial designs as some effects cannot be estimated independently: varying multiple factors simultaneously can prevent the experimenter from inferring which change affected the target system's behaviour. This is called *aliasing*, and the effects whose influence cannot be separated are said to be *aliased* [64].

Some effects are assumed to be of no interest to the experimenter [107]. The *hierarchical ordering principle* states that the more factors involved in an interaction effect, the less likely it is to be significant: interactions involving three or more factors are seldom of interest. Similarly, the *effect heredity principle* states that an interaction effect can only be significant if the effect of at least one of the factors involved in the interaction is significant. The experimenter can reduce the number of combinations to test by deliberately aliasing effects that are assumed to be insignificant[5]. Taguchi Methods provide an efficient way of designing experiments, based on these assumptions.

---

[5] These assumptions have been shown to hold in other fields, such as manufacturing. Section 6.2 discusses their applicability to software systems.

This thesis focuses on complex software systems with many possible combinations (i.e. many factors and levels for each). Fractional factorial designs, in particular when using Taguchi Methods, make the testing of such systems tractable (subject to the assumptions listed in section 4.3).

### 2.8.2 Related work

Fractional factorial designs, in the context of software systems, have been described in several texts [38, 64]. Taguchi Methods have been used extensively by engineers for almost five decades to produce such designs, but their use in comprehending and configuring software systems has been limited: Sankar and Thampy suggested their use for performance tuning in [94] but the techniques have not been applied.

Taguchi Methods provide an accessible approach for non-statisticians and have two key benefits over previous work:

- They provide a standardised approach to DOE, which can be automated (e.g. by Minitab ™ [81]). This makes it simple to select combinations to test, such that the effects of each factor and of selected interactions can be inferred from the results.
- The *signal to noise ratio* metric provides a mechanism to calculate the robustness of a combination, given a set of responses from replicated trials.

Other projects that involve testing a sequence of combinations include:

- an automated tool, developed by Vetland and Woodside, for running full factorial experiments [103];
- Courtois and Woodside's use of *multivariate adaptive regression splines* to model a target system's behaviour [39], which requires a much larger set of combinations to produce a model of a response surface than does Taguchi Methods (see section 4.3);
- Diao *et al*'s use of control theory, which requires that the choice of combinations give "dense and uniform coverage" of the input space [43];
- IBM's *AutoTune*, which models a target system's response surface using a neural network – the training set is produced by testing a large set of combinations [28].

### 2.8.3 Controlling the target system

There are several usage options when running experiments, which relate to where the experiments are run, when they are run, and the nature of the interaction with the target system.

Experiments can be run either in a laboratory or at the customer's site. The former involves simulating the expected conditions of use. The latter involves testing the target system in the environment in which it will be used.

The target system can be configured either before it goes into use or on-the-fly. The former is either part of the development and testing phase (i.e. in a laboratory setting) or part of the deployment phase (i.e. at a customer's site). Configuring the target system on-the-fly allows it to adapt dynamically to changes in the conditions of use. This is particularly popular for multimedia applications (e.g. maintaining audio or video streams), and for mobile and grid computing where the platform's characteristics and available resources vary dynamically.

Interaction with the target system depends on the configurable aspects exposed, the availability and control over source code, and the wishes of the experimenter. The infrastructure that controls the experiment can treat the target system as a *black box*, a *white box* or some shade of grey in between:

- The only interaction with a black box system is through the interfaces it exposes: no knowledge is available of its internal workings. Diagnostic output of the target system can be used and policy decisions exposed by the target system can be set. Observation and adaptation mechanisms external to the target system can also be used. For example, one could observe the resource usage and response time of a target system, under various conditions (e.g. various workloads). Possible adaptations include changes to the environment, such as re-allocating resources or adjusting NFS settings that control mounting of remote file systems [96].
- Interaction with a white box system can involve delving behind the interfaces it exposes: there is access to internal information (e.g. source code). This increases the range of possible adaptations and measurements that can be taken, through techniques such as source code instrumentation.
- Use of expert advice to guide the design of experiments falls into a grey area between these two extremes.

The case studies in chapter 5 involve running experiments in a laboratory setting before the target system goes into use. ACT treats the target system as a grey box: in-built configurable aspects are used, and expert advice guides the choice of factors and levels to test.

## 2.9 Systems for performance tuning and evolution

This section first classifies adaptations by confidence in their effects. It then discusses techniques for coordinating the configuration process, to decide how and when to adapt the target system's configuration. A recurring theme is that making beneficial adaptations requires either *a priori* knowledge of the target system's behaviour, or an ability to make experimental adaptations and empirically measure their effects (e.g. using ACT).

### 2.9.1 Version granularity

Two categories of adaptation are *experimental adaptations* and *target adaptations*. Experimental adaptations involve speculative adaptation of the target system's configuration, where the effects of adaptations are not known in advance. Repeatedly making experimental adaptations can find a suitable combination even before the behaviour of the target system is understood. A target adaptation is an adaptation known to produce a desirable combination; it involves adapting the target system *T* to progress to a new target system *T'* believed to meet the configuration goal. Both experimental and target adaptations can use the same adaptation mechanisms, the difference is the context and the confidence with which adaptations are made. An experimental adaptation can be promoted to a target adaptation if it is found to produce a target system configuration with the desired behaviour.

Experimental and target adaptations are illustrated by the hypothetical example in Figure 2.5. The configuration goal for a database system is to minimise the number of processors required, while maintaining a throughput greater than 100 requests per second (rps). The bottom part of the diagram shows a series of experimental adaptations, starting with configuration $T_A$ and adapting the target system by reducing the number of processors to produce configurations $T_B$ and $T_C$. It is found that $T_B$, which delivers a throughput of 110 rps, meets the configuration goal of the minimal number of processors. The set of experimental adaptations used to produce $T_B$ is promoted to a target adaptation, to progress from target system *T* to a new target system *T'*. Sometime later, the database's workload increases causing the throughput to drop to 90 rps (shown in italics in Figure 2.5). This prompts a second set of experimental adaptations, finding a new configuration $T_D$ that meets the configuration goal. These changes are promoted to a target adaptation to progress to a new target system.

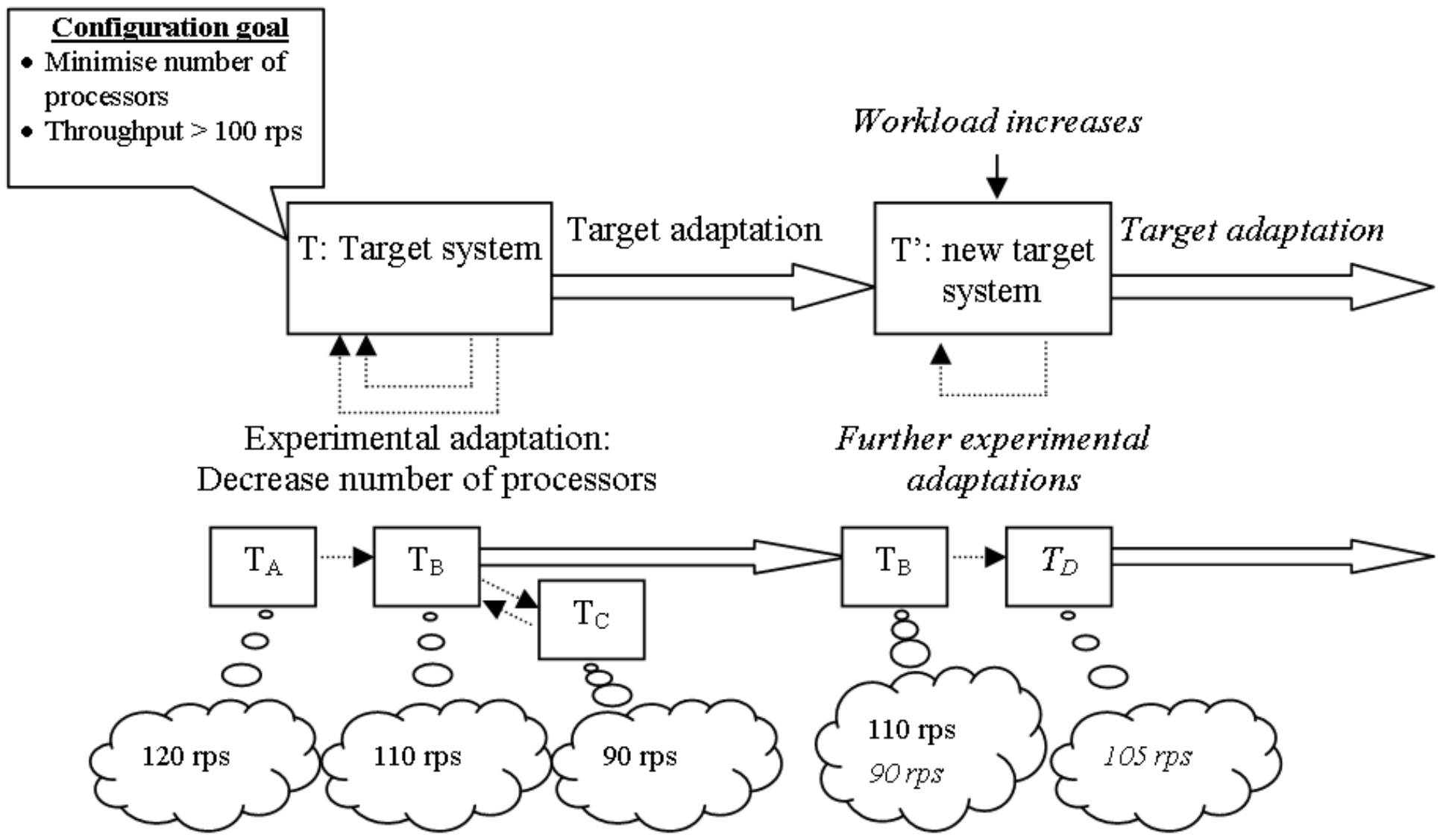

**Figure 2.5: Experimental and target adaptations**

### 2.9.2 Managing adaptation

Techniques for managing the configuration process range from the target system managing itself to an external mechanism managing adaptation.

Self-managing target systems, typified by autonomic systems [63], automatically recognise when something is wrong with their execution and initiate appropriate corrective action to resolve the situation. They can exploit details of their internal structure, but there are drawbacks. Firstly, it is difficult to obtain a global perspective from within a single target system. Secondly, code for observation and adaptation may be intertwined with target system code, making it more difficult to change the policy for configuring the target system and to evolve its code.

Shaw suggests separating the *controller*[6] from the target system being controlled [95]. This can overcome the problems described above, but the external controlling mechanism requires: (1) an understanding of the configuration goal, (2) control over the target system's factors, and (3) some way to deduce the effects of the possible adaptations. Garlan *et al* suggest that external adaptation mechanisms can be studied and reasoned about independently of the target system [56], but the effects of a generic adaptation mechanism are often dependent on the target system in question and on its conditions of use.

[6] The term "controller" comes from the field of control theory.

The CSA project (see section 2.5.1) exploits the advantages of both these approaches by separating policy from mechanism to provide an open implementation [83]. A component's functionality is delivered by a set of mechanisms, and the policy for using these mechanisms can be supplied by other components.

It has been proposed that software architecture should play a central role in planning and coordinating the configuration process [86]. The approach involves maintaining at run-time an explicit architectural model of the target system, which is described using an ADL. An architectural model provides the following benefits:

- It is a context for reasoning about the target system in terms that target system developers and administrators can understand.
- It provides a global perspective for thinking about interactions among target system components, and interaction between the target system and its environment.
- It can include constraints on the target system architecture and behaviour, against which configurations and observed behaviour can be compared.
- Adaptations to the target system's configuration (particularly re-assembly of components) can be proposed and validated at the architectural level, and mapped to adaptation operations on the running target system.

An architecture-based approach assumes that a (constantly updated) architectural model of the target system is available at run-time and that adaptations at the architectural level map to possible adaptations in the implementation.

### 2.9.3 Deciding on appropriate configurations

The *controller* decides how and when to adapt the target system's configuration. *Controllers* are software (or hardware) systems that use observations of the target system's behaviour and conditions to trigger appropriate adaptations, based on a comparison of observed behaviour and the configuration goal. Techniques employed by *controllers* include:

- use of a predictive model (or simulation) of the target system's behaviour to predict the behaviour of various configurations, to estimate which configuration will best meet the configuration goal [22, 66];
- use of a rule-based approach, which responds to observations of the target system and its conditions of use by following pre-defined (though dynamically changeable) adaptation tactics [30, 49].

Listed below are some techniques for producing *controllers*, relying on *a priori* knowledge of the target system's behaviour:

- Experts use detailed knowledge or analytical models of the target system to suggest adaptation tactics.
- Simulations predict the behaviour of various target system configurations, in an attempt to identify a configuration that will meet the configuration goal [22].
- Markov Decision Processes use a simplified discrete state model of the target system to predict behaviour and to decide on beneficial state changes [20]. A cost function predicts the quality of a given state in terms of the variables in the model. Changes in the environment are predicted using pre-defined probabilities of state transitions.
- Control theory provides methodologies for designing *controllers*, based on a model of the target system's behaviour [88].

Making experimental adaptations to a target system, and empirically measuring the various configurations, can help in the development of effective *controllers*. Experiments can aid production of a predictive model of the target system's behaviour and reveal suitable configurations for particular conditions of use (i.e. suggest adaptation tactics).

### 2.9.4 Performance tuning systems

Techniques for performance tuning range from *ad hoc* manual techniques using tools for investigating performance bottlenecks to generic automated performance-tuning tools that support on-the-fly tuning. A selection of generic automated (and semi-automated) performance tuning tools is described below.

The Performance Analysis and Characterisation Environment (PACE), developed at Warwick University, uses modelling techniques to produce a simulation of the target system [22, 66]. PACE provides a toolset for the semi-automated analysis of source code and profiling information to model a target system's computational parts and their parallel execution. However, it is difficult to produce a simulation that exhibits the emergent properties and non-deterministic behaviour of a complex software system. In contrast, ACT can be used to empirically measure such behaviour.

IBM's *AutoTune* [28] uses an artificial neural network to predict the target system's behaviour. This has much in common with the ACT approach: the training set consists of observations of target system configurations run under a variety of conditions. There are two main differences. Firstly, ACT allows feedback from previous observations to be used when deciding on combinations to test. Secondly, statistical techniques for design of experiments and analysis of results give an explicit mathematical model of the target system's behaviour, and confidence levels in the model. In contrast, a model learnt by a neural network is implicit in the weights and is not easily accessible.

The aim of "Autonomic Computing" [63] is to produce an adaptive infrastructure that regulates itself. It is inspired by the metaphor of the autonomic nervous system, which handles crucial but mundane functions automatically, such as increasing heart and breathing rates when required. An autonomic computing system would be self-configuring, self-healing, self-optimising and self-protecting; complexity would be hidden from users by automatically configuring itself according to the users' needs and in accordance with the environment in which it operates. This requires that appropriate configurations be known: again, the use of ACT may help.

Using the *AutoTune agent framework*, Diao *et al* have built an autonomic feedback control system (see section 2.10.1 for a discussion of control theory). Agents perform three functions [43, 44]:

- A *modelling agent* creates a linear model of the target system's behaviour. The case study in [44] models CPU and memory usage of the Apache Web Server as two factors are varied.
- A *controller design agent* uses standard techniques from linear control theory to derive a feedback control algorithm, based on the model of resource usage. The *controller* uses the difference between desired and measured resource usage to determine the next levels of the factors.
- A *run-time controller agent* adapts the target system, based on the control algorithm.

ACT serves a similar role to the modelling phase. The key difference is that search strategies such as Taguchi Methods require testing of fewer combinations to produce a model. Also, the aims of ACT

– to improve comprehension and aid the configuration process – are more general than those of Diao *et al*, whose feedback control algorithm corresponds to a single adaptation tactic.

### 2.9.5 DASADA

"Dynamic Assembly for System Adaptability, Dependability and Assurance" (DASADA) is a DARPA-funded programme [8, 9]. The DASADA approach is based on use of architectural models for reasoning about complex software systems. Adaptability is achieved through on-the-fly re-assembly of the target system's components. The need for change is derived from a comparative analysis of the target system's specification and feedback on its implementation [79].

Probes and gauges provide this feedback by monitoring the executing target system and its environment. Observations are interpreted in the context of an architectural model maintained at run-time. The model is annotated with QoS requirements – violations of these requirements are used to trigger automated re-assembly of components or to request human intervention to perform re-composition. Current work in the DASADA programme assumes that appropriate adaptation tactics are somehow known. Use of ACT could help to determine the effects of re-assembling components and the quality of various configurations.

#### 2.9.5.1 Software Surveyor

The DASADA project "Gauges to Dynamically Deduce Componentware Configurations" [105] aims to model the connectivity and behaviour of applications. The *Software Surveyor* system has been developed for this purpose. It consists of a suite of tools to construct a dynamic, constantly updated model of an evolving, under-specified application. The aim is to answer questions about a target system, such as "how are components connected?", "how does the current configuration compare to other configurations?", "are there unused or unexpected components?" and "how are the components interacting?" Software Surveyor differs from ACT in that it does not make experimental adaptations to explore behaviour. Also, the focus is on modelling the target system at the architectural level, whereas ACT interprets the target system's behaviour in terms of the configuration goal.

Software Surveyor constructs and maintains a model of the target system by combining static and run-time information. It combines information from the specification and software development environment with run-time information about binding decisions, component execution, interactions and resource usage. ACT could be used to gather additional information by empirically measuring a selection of combinations.

#### 2.9.5.2 Kinesthetics eXtreme

The DASADA project "Kinesthetics[7] eXtreme" (KX) [14, 57] aims to produce an infrastructure for the run-time monitoring and adaptation of component-based distributed target systems. KX consists of a probe infrastructure, an event infrastructure and a gauge infrastructure. The project focuses particularly on the probe infrastructures for observing the target system, and the event infrastructure for disseminating information.

The probe infrastructure is summarised in section 2.6.3. One implementation is the *Active Interface Probe Run-Time Infrastructure* [57]. Probe stubs are inserted into the target system by modifying the source code at compile-time. At run-time, the target system administrator can associate callbacks with the before- and after-phases of method calls. Code invoked by the callback may log, augment, override or deny a method's activities.

The event infrastructure disseminates events (i.e. messages) from producers to consumers. Lightweight events are encoded in XML using the *Smart Events Schema* [6] and routed using Siena [106], which requires conversion to and from Siena's attribute-value pairs format. *Gaugents* are heavier mobile software agents that transmit the executable for interpreting the event along with the data.

KX complements ACT in that it provides an infrastructure to collect observations of a target system and to transmit this information to ACT. The contribution of ACT would be to coordinate adaptation of the target system, and to use the observations to infer the behaviour of target system configurations. Indeed, the openness and flexibility of the event infrastructure has informed the design of a new version of ACT, described in section 7.3.

#### 2.9.5.3 Rainbow

The "Rainbow" project for "Architecture-based Adaptation of Complex Systems" [34] aims to support automated target system adaptation at run-time. The approach, first proposed in [86] and discussed in section 2.9.2, involves maintaining at run-time an architectural model of the target system and its QoS requirements. This is encoded using the Acme ADL [55]. Performance-oriented run-time gauges are used to interpret low-level observations of the target system in the context of the architectural model. If measured performance shows an architectural constraint violation (i.e. a QoS violation), a pre-defined adaptation tactic is triggered to reconfigure the target system.

[7] Kinæsthesis is "the sense of muscular effort that accompanies a voluntary motion of the body. Also, the sense or faculty by which such sensations are perceived." [15]

These techniques have been demonstrated with several target systems, such as controlling the transfer of files between a sender and a server by adapting the file compression policy in response to available bandwidth. It has not yet been shown that such techniques scale to complex software systems. ACT differs from this approach in that it does not require an architectural description of the target system or the existence of pre-defined adaptation tactics. It is argued that ACT can be used to discover beneficial adaptation tactics without *a priori* knowledge of the target system.

#### 2.9.5.4 Containment units

The project "Process Guidance and Validation for Dependable On-the Fly System Adaptation" [87] aims to produce adaptable software systems, built from Containment Units (CUs). These are hierarchically composed modules that can self-diagnose the need for change, based on their operational characteristics, and that can make a limited set of changes aimed at meeting these needs. The CUs are defined in Little-JIL, an executable high-level process language with graphical syntax for modelling co-ordination between agents. Little-JIL contains both proactive and reactive control mechanisms, and uses resources for constraining and managing process execution.

This approach has several assumptions and limitations:

- Little-JIL must be used throughout the target system's lifecycle;
- the set of possible adaptations is limited to resource (re)allocation and module replacement, the latter consisting of reassembly of designated components within a CU (using a pre-defined set of available modules and resources);
- all changes occur within existing architectures that cannot themselves change during execution;
- the adaptation tactics are specified *a priori*;

- the set of changes required in the future must somehow be known – if an unexpected contingency arises, off-line human intervention is required to adapt the target system's architecture and incorporate additional CUs.

### 2.9.6 ArchWare

The aim of the ArchWare project is to produce architecture-centric languages, frameworks and tools for engineering evolvable software systems that are compliant to the needs of particular applications [4]. Importantly, both the target system and the process for evolving the target system can change over time.

The main areas of interest are:

- formal architectural style-based languages, e.g. the π-SPACE ADL [33], to describe and analyse evolvable target systems;
- support for evolution.

*Architectural styles* (e.g. a *client-server* style) impose constraints upon the target system's architecture and behaviour. They provide users with high-level abstractions appropriate to a specific domain and encourage reuse of designs.

Target systems developed using the ArchWare infrastructure will support on-the-fly adaptation of components, connectors and their topology. Evolution will be achieved by decomposing the target system into its constituent components, evolving them and then recomposing to form a new target system. This will be done while preserving any state or shared data of the running target system [59]. ACT could be used to explore the effects of possible adaptations, to help discover which configurations would meet the configuration goal under a variety of conditions of use.

### 2.9.7 Reflective middleware

Reflective middleware offers an open implementation, allowing inspection and adaptation of the middleware's components. In general, on-the-fly adaptation is controlled by a rule-based mechanism, which manages the target system's performance by monitoring and adapting the target system according to pre-defined (though dynamically changeable) adaptation tactics [23, 29, 49].

Blair *et al* have developed the *Open ORB* reflective middleware architecture [49]. Its design uses a component-based programming model, where an instance of *Open ORB* is a particular configuration of components. *Open ORB*'s meta-space, which is its support environment, is partitioned into four orthogonal meta-models:

- The *interface meta-model* allows inspection of the external representation of a component, in terms of its (immutable) interfaces.
- The *architectural meta-model* allows inspection and adaptation of *Open ORB*'s architecture.
- The *interception meta-model* allows adaptation of a component's behaviour, through insertion of pre- and post-behaviour using behavioural reflection.
- The *resource meta-model* provides access to inspect and adapt the management of resources.

ACT could run experiments for target systems that use *Open ORB*: adaptation and observation mechanisms exposed by the middleware provide facilities to configure and observe the target system. ACT could determine the effects of adaptations under various conditions of use to help suggest appropriate adaptation tactics.

The *Distributed Systems Group* at Trinity College have produced *K-ORB*: a configurable component model for building adaptable distributed systems, based on a light-weight version of CORBA [48]. It uses the *Iguana* reflective programming model [58] to support dynamic customisation of the middleware. Their *AutoORB* project aims to optimise the middleware system to meet the needs of a particular application, based on analysis of how the application has used the middleware [48]. This aim has much in common with the aim of ACT 2.0: to configure the target system on the fly, and deduce predictive models and adaptation tactics from observations of the target system's execution.

## 2.10 Control theory and process modelling

The process of configuring a target system involves manipulating its *control inputs* (i.e. factors) to affect its outputs (i.e. behaviour). This is clearly a control system, and is related to work in the field of control theory [45, 53] and process modelling [3].

### 2.10.1 Control theory and feedback

Control theory is the mathematical analysis of systems used to achieve a desired state under changing internal and external conditions. Describing ACT as a feedback control mechanism provides a mapping between the domains of software configuration and control theory, and encourages the use of well-established control methodologies. ACT can be viewed as a disturbance-compensated closed-loop control system with command compensation. To explain these terms:

- *Disturbance-compensation* involves using measurements of uncontrolled inputs (e.g. network load) in the control algorithm.
- *Closed-loop* refers to the use of *feedback* (i.e. comparing the measured output with the desired values). In contrast, *open-loop* implies the absence of feedback.
- *Command compensation* exploits knowledge of the process' characteristics, such as a lag before a change has an effect. Disturbance- and command-compensation are examples of *feed-forward* loops.

The algorithm for deciding when and how to adjust the target system's factors is called the *control law*, and is typically derived from a model of the target system's behaviour. A *feedback controller* aims to maintain the desired target system behaviour, where corrective action is based on the *error* (i.e. the difference between desired and observed behaviour) and not on why the error occurred. It is therefore not necessary to know the exact effects of changing the factors, just whether the adjustment makes the response increase or decrease. This allows for simplifying assumptions in the target system model.

There are two main stages to designing and implementing a feedback control system [43, 53]:

- Target system modelling involves producing a predictive model of the target system's behaviour. A continuous model is often used to approximate discrete target systems. Modelling involves:
  - designing an experiment that gives dense and uniform coverage of the input space – called *persistent excitation*, where the target system is continually adapted to display its behaviour for the full range of combinations;
  - running the experiment to collect the data;
  - using *system identification* techniques to produce a model;
  - validating the models by running further experiments.
- The *controller* is designed using standard control algorithms, such as *proportional-integral-derivative* (PID) design.

A common experiment design in control theory is to vary the factors' levels according to discrete sine waves (called *excitation signals*), whose frequencies are "relatively prime" [43]. This tests a wide range of combinations and allows the individual effect of each factor to be determined. A linear model is often considered adequate for capturing the relationship between control inputs and system outputs (i.e. fitness metrics), particularly for a small region of the response surface.

PID *controllers* incorporate three types of control [19]: (1) *proportional control*, where the correction is proportional to the error in the response; (2) *integral control*, where the correction is proportional to the duration for which the error is present; and (3) *derivative control*, where the correction is proportional to the rate of change of the error. The output of the PID *controller* is the sum of these three terms. Tuning the *controller* involves changing the weightings of each term. Proportional control can make the target system unstable, undulating from one side of the desired value to the other. Integral control compensates for this effect, but introduces a lag between the detection of the error and corrective action. Derivative control, also called *rate* or *pre-act*, inhibits rapid changes in the target system outputs to prevent overshoot (i.e. instability from proportional control).

*Adaptive control theory* is the study of *controllers* that automatically redesign themselves as the target system and conditions of use change. The control law is adjusted, based on a model of the target system's behaviour that is updated on-the-fly [73]. This advanced topic of control theory is clearly relevant to configuring target systems that evolve. An aim of future work on ACT 2.0 is to run experiments on-the-fly that explore the target system's behaviour to update, or even generate, a predictive model of its behaviour.

### 2.10.2 Process modelling

There are parallels between tools for configuring software systems and *process support environments* (PSEs). A target system is a type of process: it consists of a partially ordered set of activities in which ag*ents* (i.e. humans and software systems) interact to achieve a common goal. This section discusses PSEs that support modelling, enacting and analysing human-intensive processes. Of particular interest are PSEs that support process evolution.

Techniques for process control in an organisation include continuous regulation with respect to a model, and *ad hoc* decision-making in response to situations [40]. The former involves comparing observations of the process against a plan or template to identify deviations outwith acceptable bounds, and then applying often well-established corrective actions. The "plan" is analogous to a model of the target system's architecture and desired behaviour, while "corrective actions" correspond to pre-defined adaptation tactics. In contrast, corrective actions in "*ad hoc* decision-making" are often unclear, requiring support to aid assessment of possible responses in terms of cost, time and quality. ACT could provide this support for software systems by running experiments to measure the quality of a selection of target system configurations under likely conditions of use. This aids production of a predictive model of the target system's behaviour to establish appropriate corrective actions.

Process modelling includes work on *meta-processes* [104], which guide the observation and evolution of processes. Indeed, ACT can be viewed as a meta-process of the target system: it can coordinate adaptation of the target system's configuration.

#### 2.10.2.1 FEAST

"Feedback, Evolution And Software Technology" (FEAST) aims to investigate software evolution by studying software processes as multi-loop, multi-level feedback systems. The FEAST hypothesis is that software processes in the real world "evolve strong system dynamics and the global stability tendency of other feedback systems. The resultant stabilisation effects are likely to constrain efforts at process improvement" [75]. Localised change has little effect on global process behaviour because the outer feedback loops have a more dominant effect. Significant improvement requires adjustment of feedback loops and therefore an understanding of the process model.

A principle goal of FEAST is the "identification of the drivers of evolution and of the mechanisms that control and direct it, to learn to control these mechanisms and to improve direction, planning and management of product evolution to serve the best interests of the determining stakeholders" [76]. Lehman *et al* look at attributes of processes to capture patterns and trends of the software development process. An example attribute is a count of the number of modules for different releases of a software system, but this reveals little of the emergent properties of the complex software system or development process. His analysis techniques therefore do not transfer readily from the field of process modelling to software configuration.

### 2.10.3 Catastrophe theory

*Catastrophe theory* studies and classifies phenomena characterised by sudden shifts in behaviour arising from small changes in circumstances; a *catastrophe* is a loss of stability in a dynamic system [16]. It is possible that some complex software systems suffer from catastrophes, and that catastrophe theory is pertinent to studying phase changes and non-determinism. Figure 2.6 shows a hypothetical response surface containing a *cusp catastrophe*, which involves two factors. The vertical arrows show sudden shifts in behaviour, and the two layers of surface show that a combination can exhibit multiple behaviours depending on the previous state of the target system.

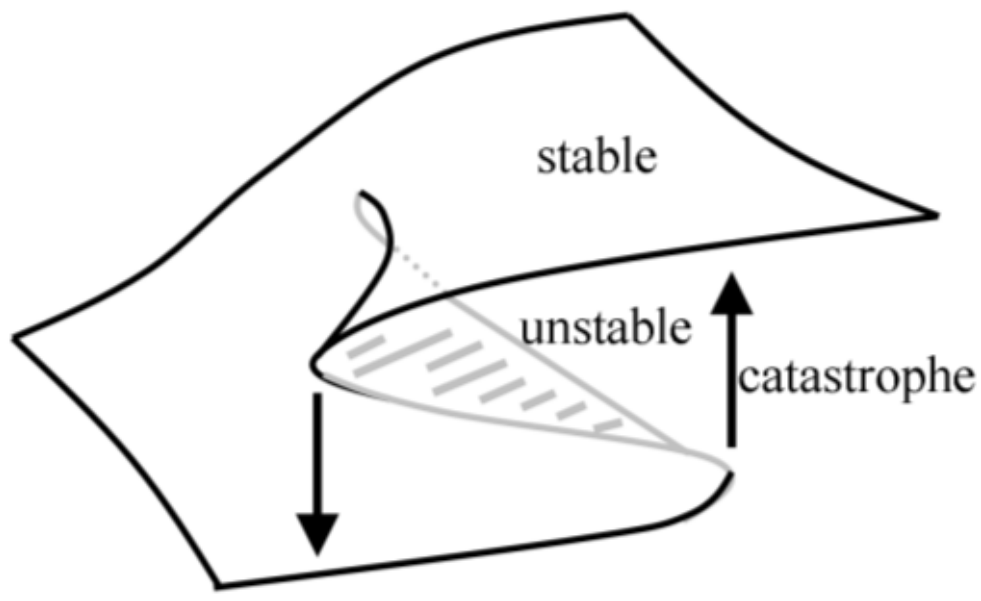

**Figure 2.6: A cusp catastrophe**

Catastrophe theory suggests that, when adapting a target system on-the-fly, the effects of adaptations can depend on the order in which they are applied. However, ACT runs experiments before the target system goes into use, setting it to a known state before every trial. Therefore, the order in which combinations are tested should not influence behaviour.

The factors that form the dimensions of the response surface in Figure 2.6 could be *uncontrolled factors* (i.e. aspects of the target system or conditions of use that are not set explicitly – see section 4.3.6 for further discussion). Variation in the levels of uncontrolled factors could cause a catastrophe, resulting in a sudden and unexplained shift in the target system's behaviour. This could cause high variability in responses from replicated trials.

It is yet to be proven whether software systems do suffer from such catastrophes. Future research topics include investigation of whether catastrophes do occur and incorporation of ideas from catastrophe theory into work on configuring software systems.

## 2.11 Software testing

Running experiments to measure a sequence of combinations has much in common with *software testing*. This section focuses on *system tests*, where the target system is tested as a whole. The aim of software testing is to exercise a target system to identify differences between specification and behaviour, to find faults or to show their absence with some level of confidence. This involves running a sequence of *test cases* and observing the target system's behaviour during each.

A test case specifies a target system configuration to test and the inputs to use (i.e. it specifies a *combination*). The target system either passes or fails a test case – equivalent to a simple *fitness metric*. Software testers generally assume that tests are deterministic (i.e. the same outcome is obtained every time a test case is used). Exhaustive testing is infeasible due to the *curse of dimensionality*, so a technique is required to decide on the set of test cases to use.

Choosing a set of test cases and testing each is analogous to ACT running an experiment for a target system, where a search strategy attempts to find faults. Common practice in software testing is to test the target system's *boundary* and *extreme conditions*, which are unusual conditions near the edges of the target system's functionality. Figure 2.7 shows a hypothetical input space consisting of factors *X* and *Y*, and indicates two boundary conditions (`X=50` and `X>Y`) as dotted lines, which partition the space into regions. Software faults can be divided into *isolated faults* and *region faults* [91]. The former occur for only one specific combination of factor levels. For the latter, all combinations in a region of the input space exhibit the fault. Region faults can be sub-divided as follows:

- *single-mode faults*, where a failure consistently occurs when a single factor has particular levels (e.g. failure whenever `X>50`, which corresponds to the right-hand side of a boundary condition in Figure 2.7);
- *double-mode faults*, where a failure consistently occurs when two factors have particular pairs of levels (e.g. failure whenever `X<Y` and `X<50`);
- *multi-mode faults*, where a failure consistently occurs when multiple factors have particular combinations of levels.

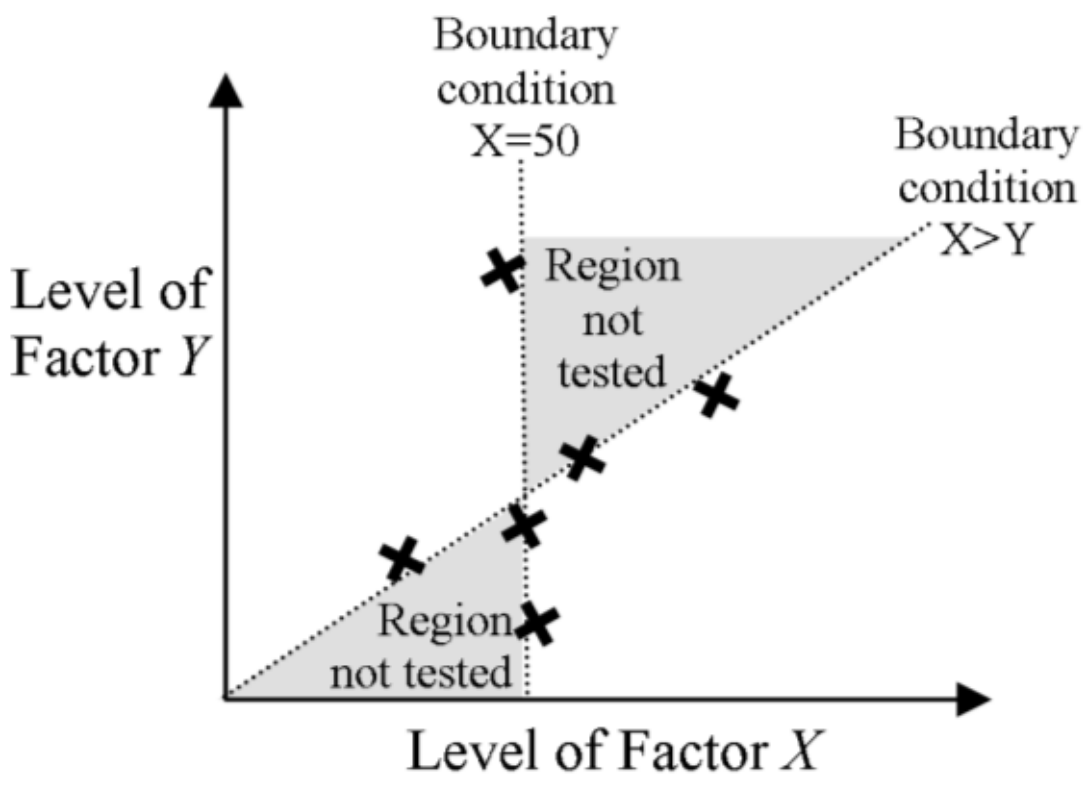

**Figure 2.7: Test cases in a 2D input space**

There should ideally be a test case in every region of the input space, to detect all region faults. To decide on test cases, common practice is to test just above, below and on a boundary condition (e.g. for `X=50`, if `X` must be an integer then set `X` to `49`, `50` and `51`). The crosses in Figure 2.7 show possible test cases that leave two regions of the input space untested. When there are many factors (i.e. many dimensions) and many boundary conditions, it is difficult to ensure that every region is tested.

Generation of test cases is related to Design Of Experiments (see section 2.8). Significant *main effects* correspond to single-mode fault, and significant *interaction effects* correspond to double-mode or multi-mode faults. Phadke proposes use of Taguchi Methods for generating test cases, which he calls *Robust Testing* $^{TM}$ [91]. He describes a systematic approach for selecting combinations of parameter values (i.e. factor levels), given a predefined set of factors and a small set of "interesting" levels for each. Testing these combinations reveals all single-mode faults, all double-mode faults and many multi-mode faults. This application of Taguchi Methods differs from its use in configuring target systems: "quality" is simply pass or fail, and results are not used to predict the performance of untested combinations.

Phadke's approach has a *balance* requirement: for every pair of factors, every pair of values is tested the same number of times. Cohen *et al* observe that the balance requirement is not necessary for the generation of test cases. The weaker requirement that every pair is covered at least once [37] is sufficient for detecting all single-mode and double-mode faults, and greatly reduces the number of test cases required (it grows logarithmically with the number of factors, and quadratically with the number of levels per factor). For example, an input space of 100 factors with two levels each requires 101 test cases using *Robust Testing* $^{TM}$, while an unbalanced test set requires only 10 test cases. The cost of removing the balance requirement is that the individual factor level, or combination of factors' levels, that causes the failure is not identifiable due to *aliasing*. The balance requirement is therefore important for target system tuning and evolution.

## 2.12 Summary

It has been argued by many researchers that software systems should be configurable to meet the needs of all stakeholders under various conditions of use [69, 74, 83]. Identifying when the target system should adapt requires observation (or prediction) of its conditions and/or behaviour.

Oreizy *et al* propose that software architecture should play a central role in planning and coordinating the configuration process [86]. The technique involves maintaining at run-time an explicit architectural model of the target system, which is described using an ADL. This provides a context for reasoning about the target system: observations are compared to constraints on the target system's architecture and behaviour, which trigger adaptation of the target system's configuration.

Deciding how to adapt the target system's configuration requires (some) comprehension of its behaviour. This can be acquired through expert intuition, analytical modelling, simulation or empirical measurement.

Expert intuition, analytical modelling and simulation are infeasible for some complex software systems due to emergent properties and non-deterministic behaviour. Although these techniques are sometimes useful, a higher level of confidence is obtained with results from empirical measurement. This requires observation mechanisms to measure the performance of the running target system, and adaptation mechanisms to configure the target system and conditions of use when measuring a selection of combinations during an experiment.

The experimenter has a number of options when running experiments, which relate to where the experiments are run (e.g. laboratory or customer's site), when they are run, and whether the target system is treated as a black or white box. Adaptations made to the target system's configuration can be categorised as *experimental adaptations* or *target adaptations*: speculative adaptations where the effects are not known in advance, or adaptations known to produce a desirable configuration.

Several projects have involved experimental adaptations and empirical measurement, prior to the target system going into use:

- Vetland and Woodside produced an automated tool for running full factorial experiments [103];

- IBM's *AutoTune* used an artificial neural network that was trained by testing a set of combinations [28];
- Diao *et al* tested combinations to produce a model of the target system, from which they generated a *controller* using techniques from control theory [43];
- Courtois and Woodside tested a set of combinations to produce a model of the target system using *multivariate adaptive regression splines* [39].

The above projects required that a large number of combinations be tested, which is infeasible for some complex software systems. Design Of Experiments, in particular Taguchi Methods, provide techniques for producing fractional factorial designs that involve only a small subset of the possible combinations. These have been used by Phadke [91] for software testing, but the techniques have not previously been applied to configuring software systems.

The statistical basis of Taguchi Methods surmounts a further shortfall in previous research: it provides a statistically rigorous way to design and analyse experiments, such that the confidence level in results can be estimated. It also provides a metric to measure the *robustness* of a combination, given replicated measurements of its performance.

# 3 ACT 1.0

This chapter describes the architecture and use of ACT 1.0, referred to as ACT throughout the chapter. A glossary of the terminology used can be found in Appendix A.

## 3.1 Tool architecture

Figure 3.1 is a UML component diagram showing the main components of ACT and the dependencies between them. They are described below, starting from the right side of the diagram.

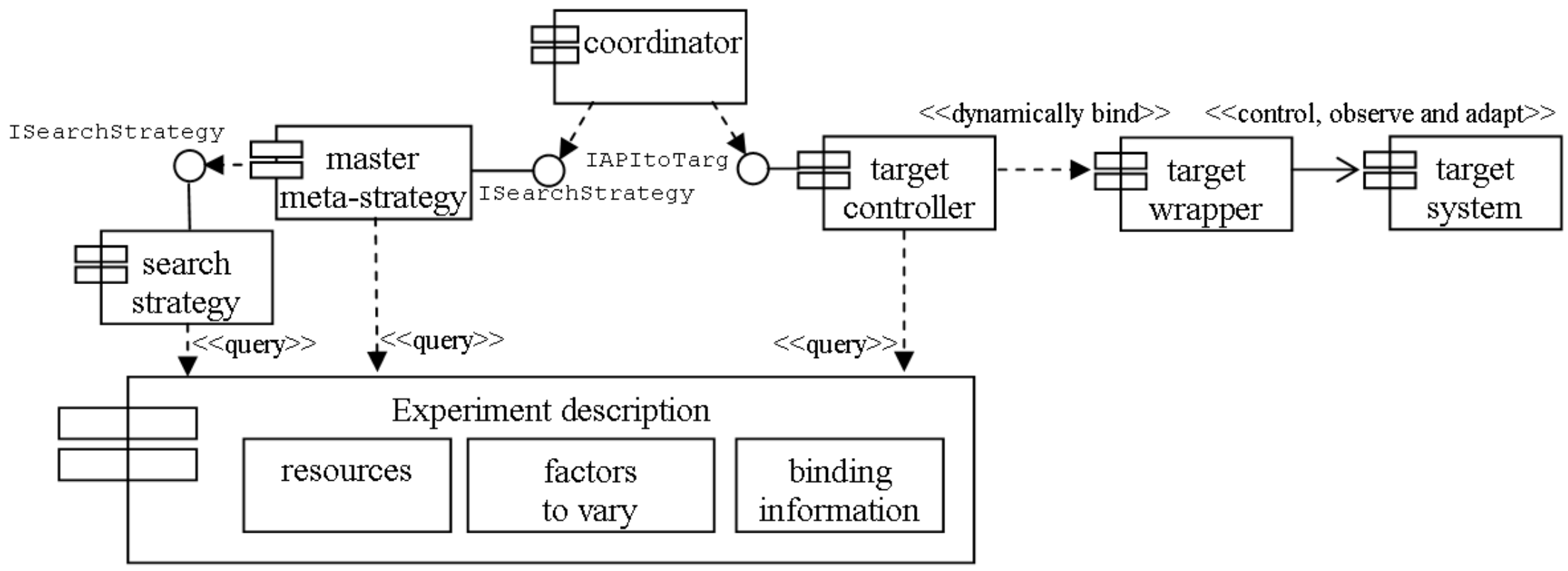

**Figure 3.1: Component diagram showing structure of ACT**

The *target system* is the software system to be configured. The *target wrapper* component is associated with the *target system*, and contains functions to control the target system during experiments.

The *target controller* component dynamically loads the functions in the *target wrapper* and uses them to interact with the target system. The *target controller* implements the `IAPItoTarg` interface, which includes methods to run a trial and to configure the target system.

The *master meta-strategy* component dynamically binds to and uses *search strategy* components. They implement the `ISearchStrategy` interface, which includes a method to get the next combination to test and a method to record the results of a trial as feedback to the *search strategy*.

The *coordinator* is responsible for using the *search strategy* and *target controller* components to run a series of trials. It binds to the *master meta-strategy* component and uses it as a proxy, through the `ISearchStrategy` interface, to repeatedly get the next combination to test from a *search strategy*. It uses the *target controller*, through the `IAPItoTarg` interface, to adapt the target system and conditions of use accordingly and to run trials.

The *core* of ACT consists of the *coordinator*, *target controller* and *master meta-strategy*. The experimenter can choose from a library of search strategies included in the ACT distribution.

The input to ACT, denoted *experiment description* in Figure 3.1, is a description of:

- the resources available (used by *search strategy* components);
- the factors to vary, including the legal levels for each (used by *search strategy* components);
- the location of each function in the *target wrapper*, giving its name and the path of a Dynamically Linked Library (used by the *target controller*).

## 3.2 Human roles

There are a number of human roles inherent in the use of ACT:

- The *ACT implementer* is the author of ACT; the programmer who produces the core of ACT.
- *Search strategy implementers* write new search strategy components for ACT. The *ACT implementer* has written some generic search strategies. Third party developers could also write search strategies, and *target system administrators* could write search strategies tailored to a specific domain or target system.
- *Target system developers* are responsible for the target system's implementation; they are experts in the details of the target system's operation.
- Use of the target system is the responsibility of the *target system administrator* who provides suitable functions for configuring, running, observing and evaluating the behaviour of the target system.
- The *experimenter* is the user of ACT and is thus responsible for configuring and invoking ACT for a particular target system.
- The *customer* is the person who will use the target system in a real-world situation, for whom the configuration process is performed.

## 3.3 ACT set-up

To set up ACT for use with a target system, and to start an experiment, requires the following steps:

- The *target wrapper*'s functions are written by the target system administrator.

- The *experiment description* is written by the experimenter.
- The experimenter executes ACT, supplying the following command line arguments:
    - A file containing the *experiment description*.
    - A file containing a description of the environment in which the experiment is run. This is not used by ACT, but is recorded for reproducibility. Any format is suitable, provided future experimenters can understand its contents and use the information to recreate the test environment.
    - The output directory in which to record the results.

## 3.4 Experiment description

The *experiment description* is encoded in XML format, and contains the following information (see Appendix C for an example):

- It describes the factors that can be varied, divided into two sets: the *configurable aspects* of the target system, and the *usage aspects* that comprise aspects of the conditions of use that can be configured. For each factor, the *experiment description* gives:
    - a name, used by ACT for meaningful output;
    - the type of the levels to which the factor can be set, restricted to *int*, *float* and *string* for simplicity;
    - the set of legal levels to which the factor can be set (see below);
    - the predicted length of time required by the *adaptation function* to change the factor's level, which may be used by the search strategy to guide the choice and order of combinations to test;
    - the location of the *adaptation function* for changing this factor's level, giving the path of the DLL and the name of the function.
- It lists the names of the *fitness metrics* measured during each trial, used by ACT for meaningful output when the results files are generated.
- It specifies the locations of the *target wrapper*'s functions, giving for each the path of the DLL and the name of the function. Section 3.5 describes these functions.
- It lists the resources available.

- It gives miscellaneous information for ACT's components, including:
    - an upper limit on the time allowed per trial;
    - the maximum number of consecutive attempts to test a combination before it is abandoned;
    - the *search strategy* component to use.

A set of legal levels for a factor can be described using either an *enumeration* or a *range*. An enumeration consists of a list of levels. A range (for *int* or *float* levels) specifies a *lower bound*, an *upper bound*, a *legal granularity* and an optional *sample granularity*. The *legal granularity* specifies the acceptable step size for incrementing and decrementing the factor's level. The *sample granularity*, which should be a multiple of the *legal granularity*, suggests a step size to use when changing the level. It recommends to the search strategy the number of levels at which to test the factor. It also provides a mechanism for the experimenter to indicate the sensitivity of the factor (i.e. recommend a small *sample granularity* if it is believed that a small change in level has a large effect, or vice versa).

### 3.5 Target wrapper

Interaction with the target system is through functions in the *target wrapper*, which consists of a collection of dynamically linked libraries (DLLs). It contains the functions described below (see Appendix D for an example):

- There is an *adaptation function* for each factor to be varied. Each *adaptation function* takes as an argument the new level for the factor, represented as an *int*, *float* or *string*.
- The *validation function* checks that a proposed target system configuration is legal (e.g. that it does not violate target system invariants). It takes a description of a proposed target system configuration (specifying a level for each factor) and returns *true* if the configuration is valid and *false* otherwise.
- The *run function* runs and measures a single trial of the target system. It takes no arguments and returns values for the fitness metrics in a *result object*, which implements the `IResultObj` interface described below.

- The *recovery function* restores the target system to a stable state in the event of failure, allowing the experiment to continue. It takes a description of the current configuration, to which the target system should be restored.
- The *new result object function* instantiates and returns a new *result object*, given a set of values for the fitness metrics. This function is required for the continuation of an interrupted experiment, where previous trials' results (stored in the output files) are re-instantiated as *result objects*.

*Result objects* implement the `IResultObj` interface, shown by the C++ code in Figure 3.2. ACT can use this interface to process results independently of the target system concerned. In a typical implementation, the methods `getNumFitnessMetrics` and `getFitnessMetric` return the number of fitness metrics and the value for the `i`$^{th}$ fitness metric respectively. The order of the fitness metrics is assumed to be the same as in the *experiment description*'s list of fitness metrics. The ordering method (the *less-than operator*) implements the aggregating function discussed in section 2.2. It compares two *result objects* to determine which best meets the configuration goal: the greater the better. A *search strategy* can use this to interpret the results of previous trials, to guide the choice of combinations to test.

```
class IResultObj {
public:
   virtual bool operator<( const IResultObj& s ) const = 0;
   virtual int getNumFitnessMetrics() const = 0;
   virtual Value *getFitnessMetric( int i ) const = 0;
   virtual float eval() const = 0;
   virtual ~IResultObj() {};
};
```

**Figure 3.2: The `IResultObj` interface**

## 3.6 Target controller

The *target controller* acts as an intermediary between the *target wrapper* and the other components of ACT; it dynamically loads the functions in the *target wrapper* to interact with the target system. The *target controller* implements the `IAPItoTarg` interface, shown in Figure 3.3. A typical implementation is described below:

- The `adaptTarg` method sets the `i`$^{th}$ factor of the target system's configurable aspects to the level `v`. Factors are numbered (from zero) in the order in which they are listed in the *experiment description*. The `adaptTarg` method calls the appropriate *adaptation function* in the *target wrapper*.

- The `adaptCond` method sets the $i^{th}$ factor of the usage aspects to the level `v` by calling the appropriate *adaptation function* in the *target wrapper*. The *target controller* thus maintains a logical distinction between factors of the target system and factors of the conditions of use – the search strategy decides how these factors are used when generating a sequence of combinations to test.
- The `validateConfig` method checks that a target system configuration, `conf`, is valid, returning *true* if it is and *false* otherwise. This maps directly to a call to the *validation function* in the *target wrapper*.
- The `recover` method restores the target system to a stable state in the event of failure by calling the *recovery function* in the *target wrapper*. The `recover` method returns a boolean value to indicate whether the combination should be re-tested (*true*) or abandoned (*false*). It returns *false* if the number of consecutive recovery attempts for the current combination is equal to the maximum number of recovery attempts specified in the *experiment description*; otherwise *true* is returned.
- The `run` method calls the *run function* in the *target wrapper*, and measures the duration of the function call. It implements a timeout mechanism that terminates the *run function* if the time taken exceeds a threshold value, defined in the *experiment description*. The `run` method also catches any run-time exceptions thrown by the *run function* and verifies that a *result object* has been returned (i.e. result is not `null`). In the event of error, an exception is thrown indicating failure. If the run is successful, the *result object* is returned.

```
class IAPItoTarg {
public:
   virtual void adaptTarg( const unsigned int i, const Value& v ) = 0;
   virtual void adaptCond( const unsigned int i, const Value& v ) = 0;
   virtual bool validateConfig( const FactorLevels& conf ) = 0;
   virtual bool recover( const FactorLevels& conf ) = 0;
   virtual IResultObj *run() = 0;
   virtual ~IAPItoTarg() {}
};
```

**Figure 3.3: The `IAPItoTarg` interface**

## 3.7 Search strategy

A *search strategy* component generates a sequence of combinations to test, potentially using results from previous trials as feedback to guide its choice. How and why these combinations are chosen by a search strategy is a policy issue, determined by the *search strategy implementer*. The choice of search strategy is made by the experimenter.

A typical *search strategy* manages the resources available for the search (listed in the *experiment description*), such as the total time available for testing. *Search strategy* components implement the `ISearchStrategy` interface, shown in Figure 3.4. A typical implementation is described below:

- The `getCombination` method returns the target system configuration to test and the conditions to use during the next trial. The result is an instance of the `Combination` class, which contains a `FactorLevels` object describing the target system's configuration and a `FactorLevels` object describing the conditions of use (giving a level for each factor).
- The `recordResult` method takes as arguments a description of the combination tested during a trial and the *result object* obtained. This provides a feedback loop, allowing the search strategy to use the results of previous trials when deciding on the next combination.
- The `isFinished` method returns *true* if the search is complete and *false* otherwise. Completion means that no further combinations are to be tested (e.g. no more time available). When the search is complete, the result of subsequent calls to the `getCombination` method is undefined.

```
class ISearchStrategy {
public:
   virtual const Combination *getCombination() = 0;
   virtual void recordResult( const Combination *c, const IResultObj *results ) = 0;
   virtual bool isFinished() = 0;
   virtual ~ISearchStrategy() {};
};
```

**Figure 3.4: The `ISearchStrategy` interface**

A *meta-strategy* component is a special search strategy that dynamically binds to and uses other *search strategy* components; it can configure *search strategy* components and dynamically switch between strategies. To configure a *search strategy* component requires strategy-specific knowledge. The search strategy implementer has freedom to use any scheme desired for communication between a *meta-strategy* and *search strategy*. For example, a *search strategy* could implement an interface that exposes its configurable aspects, through which a suitable *meta-strategy* (i.e. one that knows about the interface) could configure it.

The *master meta-strategy* component provides a consistent mechanism for the controller to query *search strategy* components. It is also responsible for updating the database of *result objects* (see section 3.9.1 for a description of this database).

The experimenter specifies in the *experiment description* a choice of *search strategy* by giving the name of a DLL that contains a *search strategy* component (if absent, ACT uses the default *grid sampling* strategy, described in chapter 4). At start-time, the *master meta-strategy* dynamically loads this DLL and calls a function in it named `instantiate`, which returns an instance of the *search strategy* component. The `instantiate` function takes as arguments a handle to the *experiment description* and a handle to the database of *result objects* obtained to-date (which is initially empty).

The *search strategy* component loaded by the *master meta-strategy* can itself be a *meta-strategy* that chooses and configures other search strategies for use.

## 3.8 Coordinator

The *coordinator* component manages the experiment: it runs a set of trials that test a sequence of combinations. For the duration of the experiment, the *coordinator* binds to a *master meta-strategy* component (referred to here as the *search strategy*) and to a *target controller* component, which it accesses through the `ISearchStrategy` interface and `IAPItoTarg` interface respectively.

The *coordinator* implements a simple loop, shown in Figure 3.5 to Figure 3.7 in pseudo-code and simplified by removing error-handling.

Figure 3.5 shows the main control loop that repeats until the *search strategy* reports that the experiment is finished. The *coordinator* queries the *search strategy* to get the next combination to test (line 3), then checks whether the target system configuration is valid by querying the *target controller* (line 4). If the configuration is valid, the *coordinator* configures the target system and conditions of use (line 5, shown in Figure 3.6), and then runs the target system to measure its behaviour (line 6, shown in Figure 3.7). Otherwise, an *error result object* is generated (line 8). The result of the trial is then fed back to the search strategy (line 10).

```
   /**
    * Run trials for a variety of combinations.
   */
1  void Coordinator::runExperiment() {
```

```
2     while( not searchStrategy.isFinished() ) {
3        next combination = searchStrategy.getCombination()
4        if( targetController.validateConfig(next target system configuration) ) {
5           changeCombination( next combination )
6           result = runTrial()
7        } else {
8           result = new error result( "Invalid configuration" )
9        }
10       searchStrategy.recordResult( result )
11    }
12 }
```

**Figure 3.5: Main control loop**

Figure 3.6 shows how the *coordinator* configures the target system and conditions of use. Each factor whose level is to change (compared to its level under the current combination) is set by calling the `adaptCond` or `adaptTarg` method of the *target controller*. This is done in two loops: first for the conditions of use (lines 14 to 18), and second for the configurable aspects of the target system (lines 19 to 23).

```
   /**
   * Set the target system configuration and conditions of use.
   */
13 void Controller::changeCombination( const Combination *info ) {
14    for( each factor of conditions of use ) {
15       if( new level != current level ) {
16          targetController.adaptCond( factor's index, level )
17       }
18    }
19    for( each factor of target system ) {
20       if( new level != current level ) {
21          targetController.adaptTarg( factor's index, level )
22       }
23    }
24 }
```

**Figure 3.6: Changing the target system's configuration and conditions of use**

Figure 3.7 shows how the *coordinator* uses the *target controller* to empirically measure a combination's behaviour. The *coordinator* repeatedly attempts to run the target system by calling the `run` method of the *target controller* (line 29). In the event of `run`'s failure (caught at line 31), the *coordinator* calls the *target controller*'s `recover` method. This loop repeats until either `run` is successful or `recover` returns *false*. The latter indicates that further runs of the target system with that combination should not be attempted. The result of the trial (returned on line 36) is either the *result object* returned by `run` or an *error result object*.

```
   /**
   * Run a trial for the current combination.
   */
25 IResultObj Controller::runTrial() {
26    success = false
27    do {
28       try {
29          result = targetController.run()
30          success = true
31       } catch( Exception e ) {
```

```
32          tryAgain = targetController.recover(current target system configuration)
33          result = new error result( "Run failed" )
34       }
35    } while( not success and tryAgain );
36    return result
37 }
```

**Figure 3.7: Running a trial**

## 3.9 Recording and reporting results

Results of the trials are stored in a transient database in main memory, and are written incrementally to a collection of XML output files for stable storage.

### 3.9.1 Results database

A *results database* contains the set of results in main memory. Each record of the database corresponds to a trial, storing the combination tested and the *result object* obtained. Search strategies can use the results to guide their subsequent choice of combinations to test. The *results database* is created incrementally as the experiment progresses. It can also be recreated from a set of XML output files.

The `exportTabSeparated` method of the *results database* outputs the database's results to a tab-separated text file, suitable for data analysis in a spreadsheet or statistics application (e.g. Minitab ™). Each record of the database is output on a single line, giving a level for each factor and a value for each fitness metric.

### 3.9.2 Output files

Results are written incrementally to a collection of XML output files for stable storage, located in a directory specified as a command line argument to ACT. The files passed as input to ACT are also stored to aid reproducibility. The following files are generated:

- The file `experimentDescription.xml` contains the *experiment description*, supplied as input to ACT.
- The file `envDescription.txt` contains a description of the environment in which the experiment was conducted, supplied as input to ACT.
- The file `configs.xml` stores the target system configurations tested. A configuration is described by giving a level for each factor of the target system.

- The file `conditions.xml` stores the conditions used during the trials. A condition of use is described by giving a level for each factor of the environment and workload.
- The file `resultObjs.xml` stores the *result objects* from the trials. A *result object* is described by giving a value for each of the fitness metrics.
- The file `links.xml` contains for each trial a reference to the target system configuration, a condition of use, and a *result object*. These references are represented using *xlinks* (i.e. multi-directional links).
- The file `results.xml` contains references to the files described above, giving a single entry point from which to parse the results.

## 3.10 Running an experiment

The UML activity diagram in Figure 3.8 shows the activities involved in using ACT to run an experiment, starting from the initial invocation of ACT.

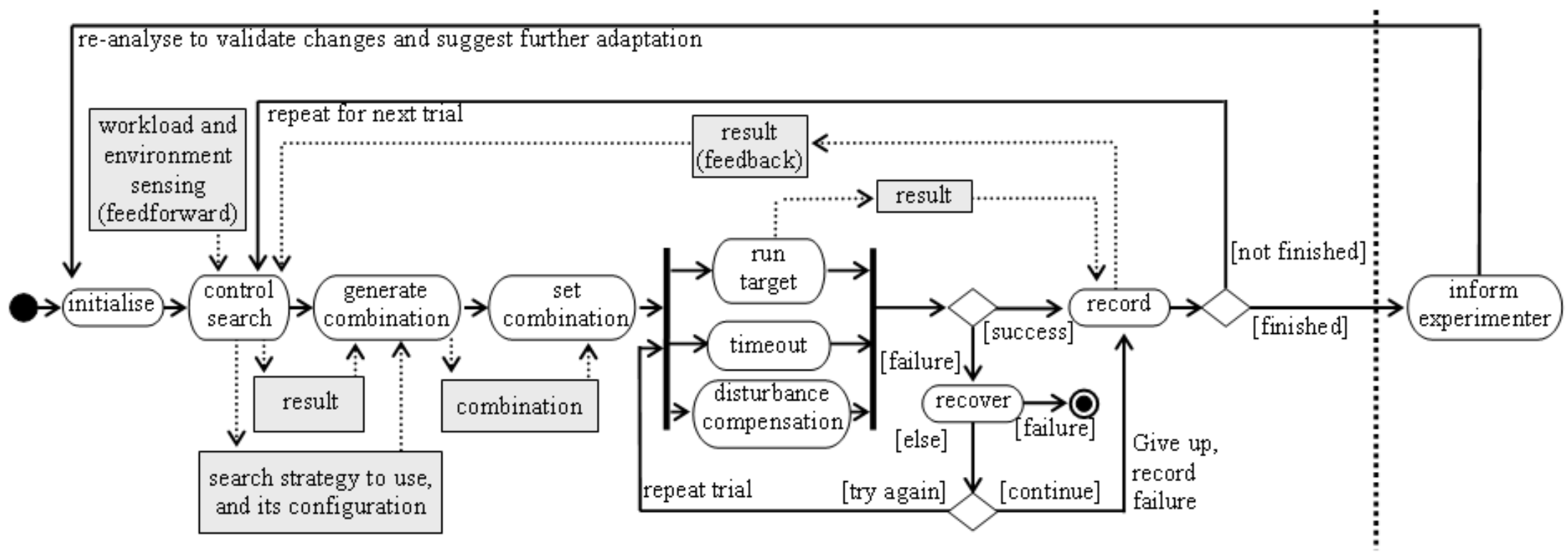

**Figure 3.8: Activity diagram of the configuration process**

During the *initialise* activity, ACT dynamically loads the functions in the *target wrapper*. The *master meta-strategy* component binds to a *search strategy* component, which can itself be a *meta-strategy* component.

During the *control search* activity, the *meta-strategy* component (if any) chooses, binds to and configures an appropriate *search strategy* for use. This may be guided by *feedback* information (i.e. results from previous trials) and *feed-forward* information (e.g. observations of conditions of use and resources available). For example, the number of replications may depend on the variability of observed behaviour in previous trials and the time available for the experiment.

During the *generate combination* activity, the *search strategy* currently in use identifies a configuration and condition to use for the next trial. During the *set combination* activity, the *target controller* uses the *adaptation functions* in the *target wrapper* to adapt the target system's configuration and conditions of use.

During the *run target* activity, ACT runs the target system by invoking the *target wrapper*'s *run function*. This sets the target system to a consistent state and runs it, observing its behaviour and returning a *result object*. Failure is detected by monitoring run-time errors, and by a *timeout* mechanism that puts an upper bound on the length of time allowed per trial.

The *disturbance compensation* activity is optional and can be done in parallel with running the target system. It involves measurement of uncontrolled factors (e.g. network load) to detect any disturbance caused by external sources.

If the trial is successful, ACT performs the *record* activity to feed back the result to the *meta-strategy* and *search strategy*. The result is added to the *results database* and appended to the XML output files. If ACT detects a failure, it performs the *recover* activity to invoke the *target wrapper*'s *recovery function* to restore the system to a stable state and optionally to collect diagnostic information. If the *recover* activity results in a *failure* condition, ACT terminates. Otherwise, there is a decision to either *try again* to test the combination (with an upper bound on the number of consecutive attempts) or to *continue*. For the latter, the *record* activity involves recording failure.

The *search strategy* is then queried to decide whether the experiment is *not finished* or *finished*. The experiment continues, testing further combinations, until the *finished* condition is met.

The last activity, *inform experimenter*, involves presenting the results to the experimenter. Changes made up until this point are experimental adaptations. With guidance from the experimenter, the results could be used to estimate which configuration would best meet the configuration goal. Adapting the target system to this configuration would be a target adaptation.

Target system administrators could also use results to guide target system usage, to prevent or encourage particular behaviour. Target system developers could use experiment results to guide development of future versions. For example, they could add facilities for making on-the-fly observations that trigger adaptation to a configuration believed to behave well under current or predicted conditions. ACT could then try further experimental adaptations: to validate the new configuration, cope with new conditions of use and meet subsequent configuration goals.

The methodology described for configuring a target system is recursive: it can be applied to various levels of a target system, configuration goal, and ACT itself:

- A "systems of systems" can be configured by configuring individual sub-systems, and by changing the architecture of the target system as a whole.
- A configuration goal can be met by first satisfying a sub-goal, e.g. determine the behaviour of a set of configurations, and then using the result of the sub-goal to attain the primary goal.
- ACT itself can be configured, to improve its use with a given target system, by using another instance of ACT. Configuring ACT involves choosing and configuring a *search strategy* component.

## 3.11 Conclusions

ACT 1.0, written in C++, provides a generic infrastructure for running automated experiments. It is generic in that it can explore the behaviour of a wide variety of target systems using a variety of search strategies. This is achieved by encapsulating the target-specific code behind a set of functions in the *target wrapper* and by implementing *search strategy* components as pluggable DLLs. This allows ACT to be used with any target system for which appropriate functions can be written and allows new search strategies to be developed and easily bound to ACT 1.0.

The core of ACT 1.0 is kept simple by:

- delegating to a *search strategy* component the potentially complicated task of choosing combinations to test;
- assuming that experiments are always run before the target system goes into use;
- delegating to the *target wrapper*'s *run function* the task of running and measuring the target system, which encapsulates the probes and gauges used to measure and evaluate performance.

Section 7.3 discusses future work on a new version of ACT, which will support on-the-fly adaptation of the target system's configuration. It will also make explicit the probe and gauge components, and will include *advice* components that will encode expert knowledge of the target system's behaviour.

# 4 Exploring target system behaviour

Exploring a target system's behaviour involves testing a selection of configurations under a variety of conditions. Deciding on the sequence of combinations to test is the task of a *search strategy*. There are many possible search strategies that aim to:

- find a configuration that meets the configuration goal under a particular condition;
- explore the target system's behaviour to find characteristics of interest;
- help construct a predictive model of the target system that can estimate behaviour for untested configurations under given conditions.

These aims are realisable for at least some target systems:

- Some combinations may be found that meet the configuration goal better than the default – an indication of success. This is true even for simple search strategies, such as randomly choosing combinations to test or using *grid sampling*, where the input space is divided into a grid and each point on the grid is tested in turn (forming a full factorial design).
- Examples of identifiable trends and interesting features on the response surface include:
    - conditions that produce highly variable (or highly consistent) behaviour, or that cause failure;
    - points in the input space at which continuing to increase/decrease a factor's level starts having the opposite effect on behaviour.
- Experiments designed with statistical rigour, such as when using Taguchi Methods, can produce predictive models of a target system's behaviour.

## 4.1 Meta-strategies

A meta-strategy is a special kind of search strategy that dynamically binds to and uses other search strategy components. Search strategies may be thought of as mechanisms for deciding on a sequence of combinations to test, while a meta-strategy provides the policy for choosing and configuring search strategies for use. Policy decisions can be based on the information desired from the experiment, resources available, results obtained to-date and knowledge of the target system's behaviour.

Use of a meta-strategy makes the selection and tailoring of a search strategy explicit, compared to use of autonomous search strategies that are self-configuring. The amount of possible “tailoring” depends on the particular search strategy.

## 4.2 Use of feedback

Some search strategies use feedback from previous trials to guide the choice of combinations to test. Examples include the use of *iterative improvement algorithms* [92], which start with some combination and move around the response surface in search of the optimal. These are standard algorithms from the field of artificial intelligence, but their use in applying experimental adaptations to configure complex software systems is novel.

*Gradient descent*[8] is an iterative improvement algorithm that finds a path from an initial configuration to a local minimum by following a “downward” slope. Testing the neighbouring combinations in each direction allows the gradient of the slopes to be calculated, and the neighbour in the steepest direction to be chosen each time. Alternatively, neighbouring combinations may be tested in turn until a better combination is found, prompting a move to this point without need to test the other neighbours (this still guarantees a “downward” motion). Gradient descent suffers from three drawbacks relating to the shape of the response surface:

- There is a risk of being trapped at a *local minimum* (as opposed to the global minimum) because the algorithm terminates when a combination performs better than all neighbouring combinations.
- The search strategy will conduct a random walk when on a *plateau*, which is a region of the response surface that is essentially flat. It will take a long time to leave the plateaux, or will terminate at a minimum on the plateaux.
- A *valley* with steeply sloping sides and a gently sloping base can be hard to follow. It is easy to descend the sides of the valley but, if there is no series of adjacent combinations that follows the valley floor, the search can oscillate from side to side and make little progress.

---

[8] This description assumes that the response variable is to be minimised; the algorithm is called *hill climbing* when the response is to be maximised.

*Simulated annealing* is an iterative improvement algorithm that can avoid the drawbacks listed above, and that can be used in combination with gradient descent. The simulated annealing algorithm operates as follows: given an initial combination, a second combination is generated within the vicinity of the first by randomly adjusting the levels of the factors. If this second combination gives improved behaviour, the search moves to this new point in the space. Otherwise, the probability that the search will move to the new point is calculated using the "badness" of the move and the *temperature* of the search. Temperature is a measure of the "energy" of the search, and decreases as the search progresses: high temperatures lead to larger changes in combination and make "bad" moves more likely. This allows escape from local minima early in the search as it can move to a worse point while looking for the global minimum. As the temperature nears zero, the choice of combination stabilises at a minimum because only good moves are made.

The starting point for an iterative improvement algorithm's search could be a random point in the input space, a combination suggested by the experimenter (i.e. their informed guess) or the best found when using another search strategy. The last is an example of a simple *meta-strategy*, where the search strategy used is switched during the experiment.

Some search strategies use *feed-forward* information to guide the choice of combinations to test. For example, the cost of adapting the target system's configuration may influence the order in which combinations are tested, e.g. cheap adaptations should be made more often than expensive adaptations.

## 4.3 Design Of Experiments (DOE)

Some search strategies use a structured statistical approach for the design of experiments and analysis of results. Using notation from the field of DOE, a *design matrix* depicts the combinations to test during an experiment. Each row represents a combination and each column corresponds to a factor. The numbers in the matrix specify the *coded levels*, which can be mapped to factors' uncoded levels. Any set of combinations can be depicted. For example, Figure 4.1 depicts a full factorial design for three factors, each with two levels. ACT automatically runs experiments, taking as input a design matrix and the values to which the *coded levels* correspond.

| | | Factor A | Factor B | Factor C |
|---|---|---|---|---|
| Combinations | **1** | 1 | 1 | 1 |
| | **2** | 1 | 1 | 2 |
| | **3** | 1 | 2 | 1 |
| | **4** | 1 | 2 | 2 |
| | **5** | 2 | 1 | 1 |
| | **6** | 2 | 1 | 2 |
| | **7** | 2 | 2 | 1 |
| | **8** | 2 | 2 | 2 |

**Figure 4.1: Design matrix**

*Taguchi Methods* standardise the statistical techniques of DOE, and provide a method for creating fractional factorial designs using *orthogonal arrays* [101]. Experiments that use these designs can identify the effects of a large number of factors, and selected interactions, by testing only a small number of combinations[9].

The experiments are conducted in two phases. The *first phase* involves running a fractional factorial experiment to produce a mathematical model of the system's behaviour for the region of the response surface investigated[10]. The model is used to predict a combination near the optimal in the investigated region. The *second phase* tests combinations in a small region around the predicted optimal to produce a more accurate model of the response surface in that region. This model is used to more accurately predict the optimal combination.

According to Taguchi, good performance implies *robustness*: consistently high performance with low variability, even when *uncontrolled factors* vary. The levels of uncontrolled factors (of the system and its conditions of use) are not set explicitly for cost or technical reasons. For example, the disk access speed is hard to control as it depends on the ordering of read requests and initial position of the read heads. By replicating trials, different values for the uncontrolled factors are likely to be encountered. This is important for identifying robust configurations of a target system: *robust* implies that the target system will perform well irrespective of the values of uncontrolled factors.

The main benefit of Taguchi Methods is a reduction in the number of combinations to test compared to full factorial designs. For example, 16 factors with two levels for each gives $2^{16}$ (i.e. 65,536) possible combinations. A fractional factorial design, produced using Taguchi Methods, requires the testing of only 32 combinations to determine the effect of each factor and of 15 two-factor interactions. A second benefit is the ability to identify robust target system configurations.

[9] The combination of Taguchi Methods and ACT yields the semi-automated *TACT process*.

[10] The validity of such models is discussed in section 6.2.

Taguchi Methods make several assumptions about the target system[11]:

- The experimenter knows which factors to vary, appropriate levels to test for each, and which interactions are of interest.
- Interaction effects involving three or more factors are seldom significant.
- Main effects are more significant than interaction effects: if the effect of a factor is aliased with the effect of a two-factor interaction, the observed effect is wholly credited to the factor.
- If an interaction effect, say between factors *A* and *B*, is not investigated, the effect of *A* is the same for all levels of *B*, and vice versa.
- A model produced from the fractional factorial experiment will accurately predict a combination that is near the optimal in the tested region of the response surface.

### 4.3.1 First phase experiment

Taguchi's technique for creating fractional factorial designs is called *parameter design*. It can be done either manually or with the assistance of a statistics package such as Minitab ™. The steps are:

- choose factors to vary and the levels for each, and choose interactions to investigate;
- choose an appropriate orthogonal array;
- allocate factors to the columns of the orthogonal array to produce a *design matrix*.

The experimenter first chooses the set of factors to vary and the levels to test for each. Factors of the target system should be chosen that are believed to have a significant influence on behaviour. The experimenter may also choose factors of the workload and environment that vary during normal customer usage but that can be controlled explicitly during the experiment[12]. A selection of interaction effects can be investigated, in particular two-factor interactions chosen by the experimenter.

[11] The validity of these assumptions for complex software systems is discussed in section 6.2.

[12] These are referred to as *noise factors*, discussed in section 4.3.6.

The number of levels per factor should be restricted to at most four: more than four levels make experiment designs complicated due to the orthogonal arrays available [94]. Standard practice is to choose a low, medium and high level [107]. Expert knowledge can help decide on the range of levels (i.e. the region of the response surface to investigate) and on particular levels of interest.

An orthogonal array (OA) is a special kind of matrix, which stems from Euler's *Latin squares*. Figure 4.2 shows the $L_8(2^7)$ orthogonal array, meaning it has eight rows and has seven columns that each have two coded levels. OAs have the following properties:

- Every pair of columns includes every combination of coded levels an equal number of times. For example, columns *C1* and *C2* in Figure 4.2 have the pairs: 1,1; 1,1; 1,2; 1,2; 2,1; 2,1; 2,2 and 2,2. This is important for statistical analysis: the effect of a factor is calculated using the target system's response for different levels of the factor, as the average of its effect when the other factors are set to each of their levels.
- Some columns represent (i.e. are aliased with) the interactions between other columns. For example, column *C3* is aliased with the interaction between columns *C1* and *C2*: it has a "1" when columns *C1* and *C2* have the same coded level and a "2" when they differ. If factors *A* and *B* are allocated to columns *C1* and *C2*, then the two-factor interaction between *A* and *B*, denoted *AB*, is represented by column *C3*. If factor *C* is allocated to column *C3*, then the effect of *AB* is aliased with the effect of factor *C*: the effects of *AB* and *C* cannot be separated during statistical analysis of the results. If no factor is allocated to column *C3*, then the effect of *AB* is *clear* [107]: it is not aliased with the effect of any other factor or two-factor interaction.

| | ***C1*** | ***C2*** | ***C3*** | ***C4*** | ***C5*** | ***C6*** | ***C7*** |
|---|---|---|---|---|---|---|---|
| **1** | 1 | 1 | 1 | 1 | 1 | 1 | 1 |
| **2** | 1 | 1 | 1 | 2 | 2 | 2 | 2 |
| **3** | 1 | 2 | 2 | 1 | 1 | 2 | 2 |
| **4** | 1 | 2 | 2 | 2 | 2 | 1 | 1 |
| **5** | 2 | 1 | 2 | 1 | 2 | 1 | 2 |
| **6** | 2 | 1 | 2 | 2 | 1 | 2 | 1 |
| **7** | 2 | 2 | 1 | 1 | 2 | 2 | 1 |
| **8** | 2 | 2 | 1 | 2 | 1 | 1 | 2 |

**Figure 4.2: The $L_8(2^7)$ orthogonal array**

The experimenter can choose an appropriate OA from published tabulated sets of OAs [101]. The choice depends on the number of factors and number of levels for each, and the interactions of interest. The OA must have at least as many columns as there are factors and interactions of interest:

each factor is allocated to a column, and each interaction of interest should be allocated to an appropriate column so that its effect is clear[13].

*Linear graphs* show the aliasing among columns in an OA, and may be used when deciding on the allocation of factors to columns. Each vertex and edge in a linear graph corresponds to a column of the OA. Each edge shows that the corresponding column is aliased with the two-factor interaction for the columns corresponding to the connected vertices. There are many possible linear graphs for each OA. For example, both linear graphs in Figure 4.3 represent the OA in Figure 4.2. The first graph shows that column *C3* is aliased with the interaction between *C1* and *C2*, that *C5* is aliased with the interaction between *C1* and *C4*, and that *C6* is aliased with the interaction between *C2* and *C4*.

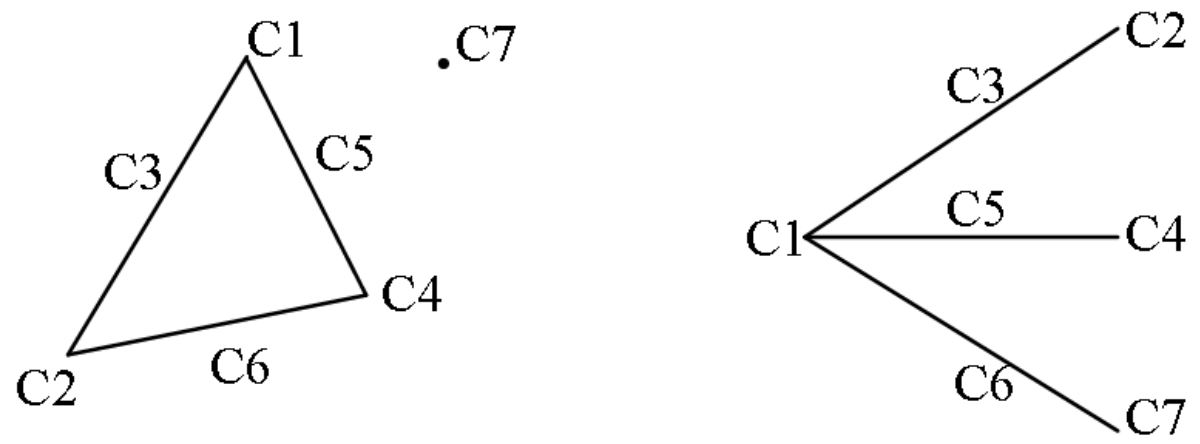

**Figure 4.3: Linear graphs**

A linear graph of the OA should be chosen that allows an allocation of factors to columns such that all interesting interaction effects are clear. If no linear graph is suitable, a larger OA is required.

Consider an example experiment with five factors (*A*, *B*, *C*, *D* and *E*) that each have two levels, and where the interaction effect *AB* is of interest. This requires an OA with at least six columns (for the five factors and the one two-factor interaction) so the $L_8(2^7)$ orthogonal array in Figure 4.2 may be suitable. The experimenter should allocate factors *A* and *B* to a connected pair of vertices, and allocate no factor to the connecting edge. As shown in Figure 4.4, the experimenter could use the first graph of Figure 4.3 by allocating factor *A* to column *C1*, factor *B* to column *C2* and leaving column *C3* unused. If factor *C* is allocated to column *C4*, factor *D* to column *C7*, and factor *E* to *C6* (i.e. leaving *C5* unused), then the two-factor interaction *AC* will also be clear. The design matrix, indicating the combinations to test for this experiment, is shown in Figure 4.5.

[13] Minitab ™ automates the process of allocating factors to columns of an OA. To aid understanding of Taguchi Methods, it is described how this can be done by hand.

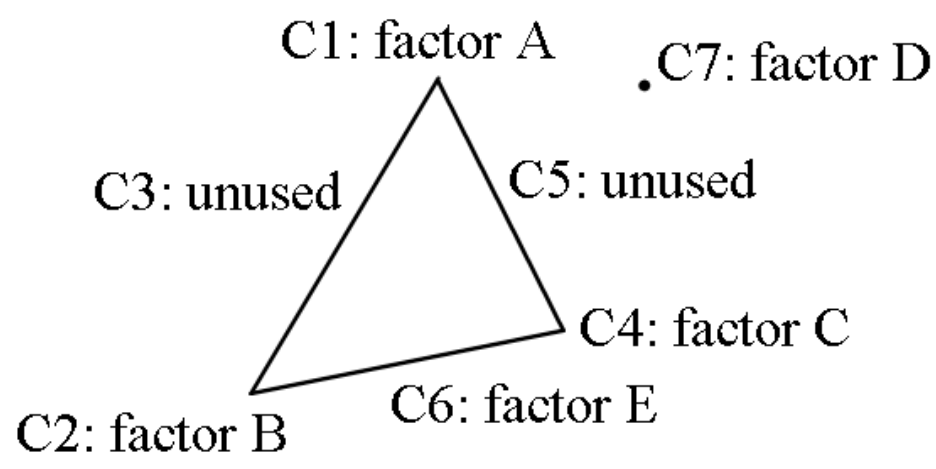

**Figure 4.4: Linear graph showing allocation of factors**

| | *Factor A* | *Factor B* | *Factor C* | *Factor D* | *Factor E* |
|---|---|---|---|---|---|
| **1** | 1 | 1 | 1 | 1 | 1 |
| **2** | 1 | 1 | 2 | 2 | 2 |
| **3** | 1 | 2 | 1 | 2 | 2 |
| **4** | 1 | 2 | 2 | 1 | 1 |
| **5** | 2 | 1 | 1 | 2 | 1 |
| **6** | 2 | 1 | 2 | 1 | 2 |
| **7** | 2 | 2 | 1 | 1 | 2 |
| **8** | 2 | 2 | 2 | 2 | 1 |

**Figure 4.5: Design matrix**

It is important to run the trials with combinations in a random order. When replicating trials, every combination should be tested once, then every combination should be tested again in a different order, and so on. Randomising the order prevents testing consecutively all combinations with a factor at a particular level. This reduces the impact on the experiment's results caused by variation in uncontrolled factors that affect a subset of the trials. For example, it counters the problem of performance degradation over time due to a failing network card that drops an increasing number of packets. If factor *A* was set to a low level for the first half of the trials and a high level for the second half, the effect of the failing network card could be credited to changing factor *A*'s level.

### 4.3.2 Signal to noise ratio (SNR)

Comparing target system configurations for a given condition is simple if there is a single metric that summarises the quality of each combination, given a set of responses from replicated trials. Here, quality is measured in terms of robustness using the *signal to noise ratio* (SNR) metric [101]. This has the following properties:

- Consistently high responses give a high overall score.
- An exceptionally high response, which is much higher than the other replicated responses, will only increase the overall score slightly. In contrast, such a response will greatly affect the *mean*.

- Low responses, which indicate that the configuration goal was not met in those trials, are heavily punished to give a low overall score.

SNR is based on a *loss function* that approximates the loss (i.e. cost) resulting from the target system failing to meet the configuration goal. Figure 4.6 shows a suggested loss function[14] for the case of maximising the response variable, where *y* is an observed response [101]. This loss function is used in the SNR formula, shown in Figure 4.7, which is based on the average of the loss for a set of responses. In more detail: *n* denotes the number of responses and $y_i$ denotes the $i^{th}$ response. By summing $1/y_i^2$, even just one low value of $y_i$ will give a much larger total, heavily punishing poor responses. In contrast, a large value of $y_i$ only increases the total a small amount. The result is negated so that optimising robustness is a problem of maximising SNR.

$$Loss = \frac{1}{y^2}$$

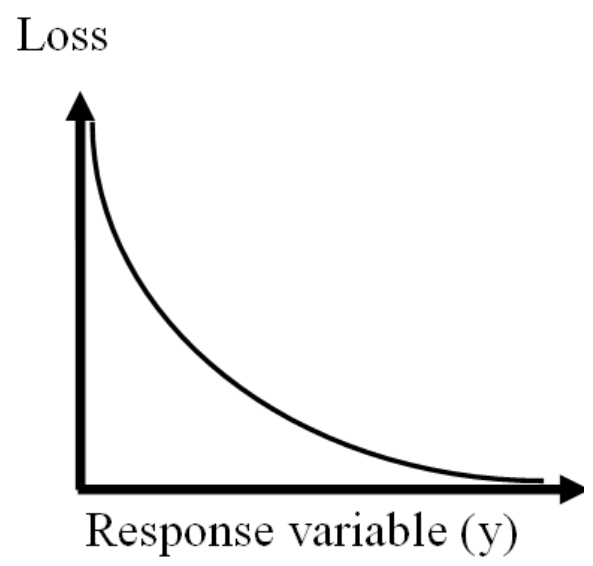

**Figure 4.6: Loss due to a low value**

$$SNR = -10\log_{10}\frac{\sum_{i=1}^{n}\frac{1}{y_i^2}}{n}$$

**Figure 4.7: Signal to noise ratio[15]**

### 4.3.3 Techniques for analysing results

> "He uses statistics as a drunken man uses lamp-posts – for support rather than for illumination."
>
> *Andrew Lang*

---

[14] This is the loss function suggested by Taguchi. Other loss functions could be used instead.

[15] The purpose of taking the logarithm is to make the SNR approximately normally distributed. It uses base 10, and multiplies by 10, because that is common practice in the field of engineering – it makes no difference to the rank ordering of combinations.

Running an experiment generates a set of data, where each data point is the observed value of the response variable for a combination. When SNR is used to combine replicated observations of a combination's response, each data point is an SNR value. Subsequent trials that produce additional observations of a combination's response can be combined to give a second SNR value for that combination (i.e. a second data point), and so on. Statistical analysis can use the data set to produce a predictive model of the target system.

Figure 4.8, taken from [98], shows the input space for a hypothetical experiment, and is used in the following descriptions of main effects and interaction effects. There are three factors (*A*, *B* and *C*) represented by the *x*, *y* and *z*-axes respectively. Each factor has two levels (0 and 1), shown in bold. The observed value of the response variable for each of the 8 (i.e. $2^3$) combinations is marked beside the corresponding point[16], forming the corners of a cube.

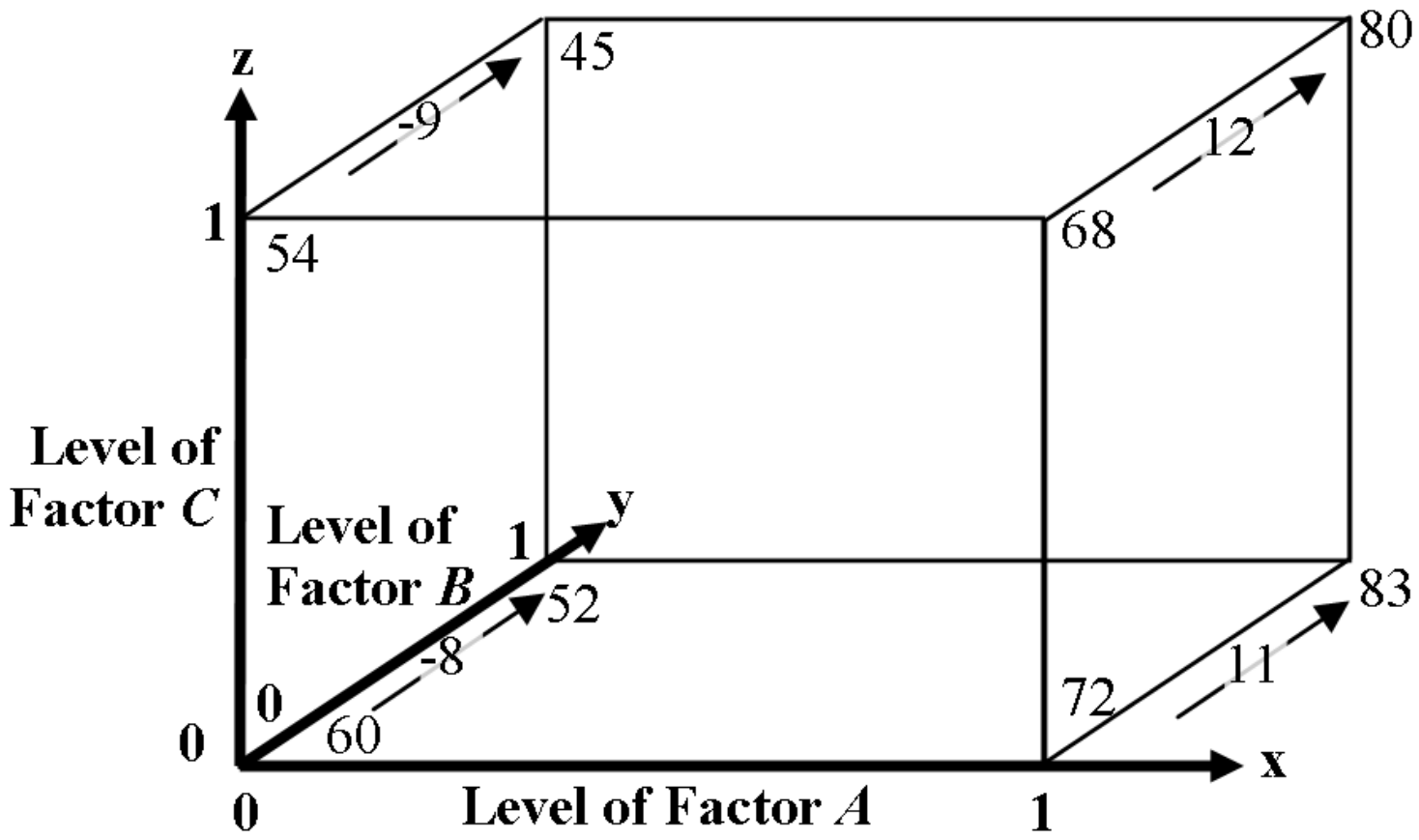

**Figure 4.8: Example input space and results**

4.3.3.1 Main effects

The *main effect* of a factor is the effect on the response variable's value caused by adjusting the level of that factor in isolation. If the factor's level has an effect on the response variable's value, the main effect is *significant*. Multiple linear regression, described in section 4.3.3.3, provides a technique for calculating the magnitude of the effect.

The main effect can be investigated by comparing observations of the response variable at different factor levels. Consider levels 0 and 1 of factor *A* in Figure 4.8: observations of the response variable's value for each are {45,52,54,60} and {68,72,80,83} respectively, and the means are 52.75 and 75.75. This suggests that increasing factor *A*'s level gives an improvement in the value of the response variable.

[16] The measured responses could be plotted on a *response surface* in a fourth dimension, but this is hard to depict graphically.

*Main effects plots* can be used to depict main effects. Figure 4.9 shows the main effects for Figure 4.8. Each box represents the main effect of a factor. Intuitively, the steeper the line the bigger the main effect. For example, the line in the left box suggests that the main effect of factor *A* is big. In contrast, the line in the middle box suggests that the main effect of factor *B* is small.

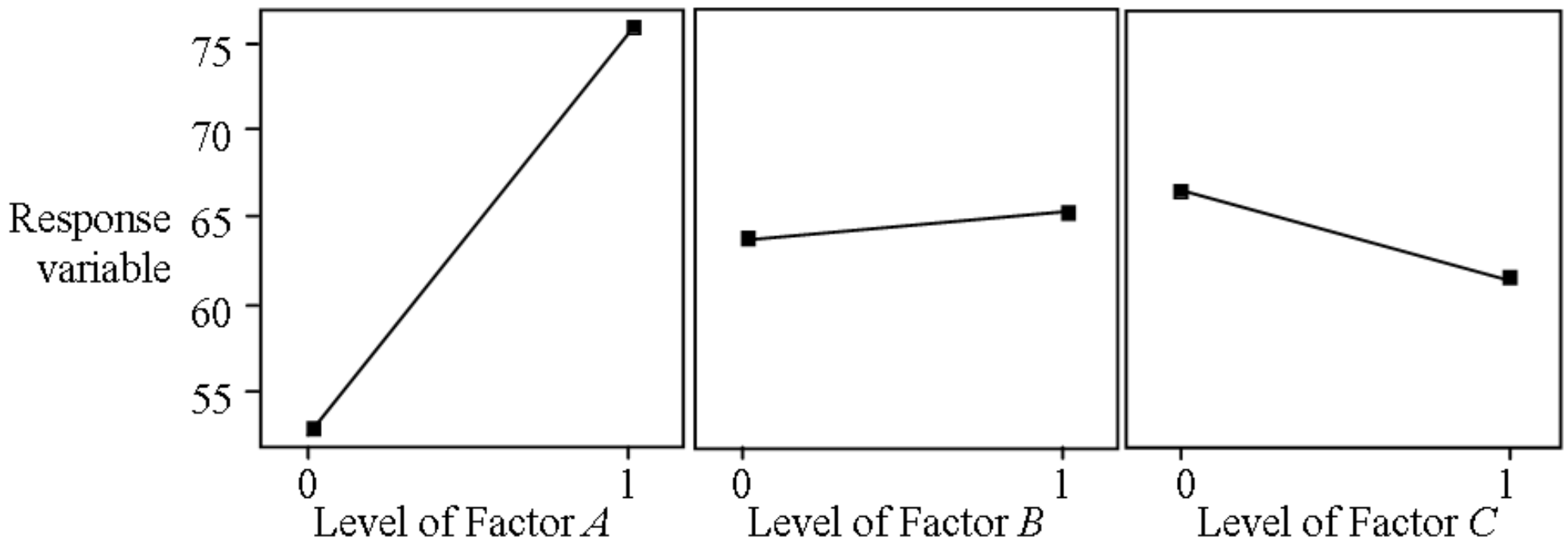

**Figure 4.9: Main effects plot**

A *population* consists of the set of all possible values of the response variable, for combinations with a given factor at a particular level. The main effect of a factor is significant if two populations, for two different levels of the factor, have different means[17]. Each observation of the response variable's value is a *sample* from one of these populations. The combinations tested may be a subset of the possible combinations at that level of the factor. For example, Figure 4.10 shows that there are many legal levels for factors *B* and *C*, each point on the grid being a possible combination. In addition, measuring a combination a second time may give a different value of the response variable. Therefore, the mean of a set of samples may not be the same as the mean of the population from which they were drawn. When comparing the means of two populations, it is insufficient to simply compare the means of the samples.

[17] The difference in the population's means could be due to an interaction effect. Multiple linear regression provides a technique to separate main effects and interaction effects in analysis of results.

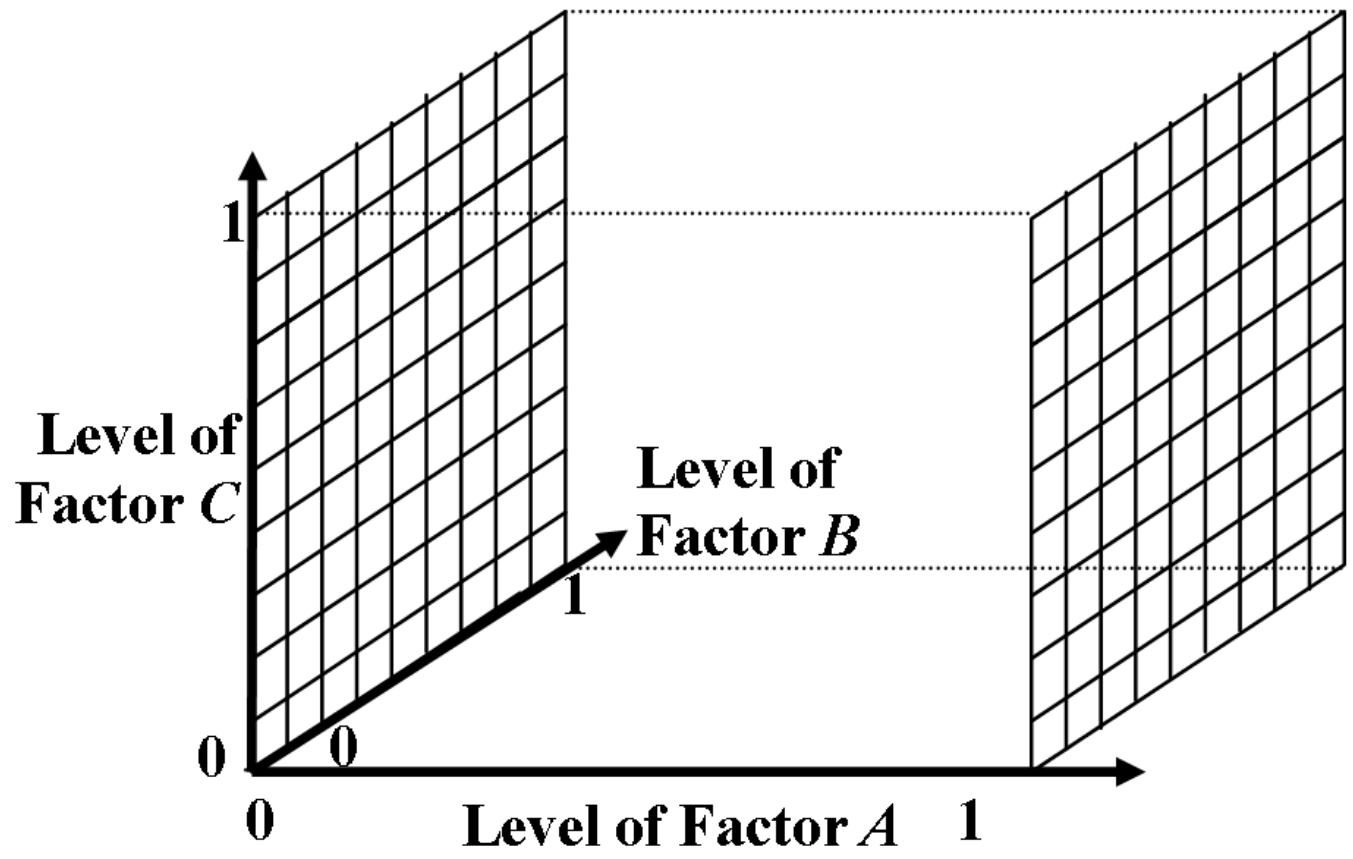

**Figure 4.10: Input space showing legal combinations for factor *A* at levels 0 and 1**

Determining whether a main effect is significant requires testing whether the means of the populations are the same (the *null hypothesis*) or different (the *alternative hypothesis*). This is commonly done at the 5% level (i.e. 95% confidence in the effect being significant when rejecting the null hypothesis).

ANOVA (*analysis of variance*) can be used to test if two groups of samples come from populations with different means. If the variation between the two groups of samples is sufficiently greater than the variation within each group, it suggests that the populations have different means. ANOVA requires the populations to be normally distributed with equal standard deviation. If the assumptions are violated, or if the group sizes are too small, the conclusions of the test may be incorrect[18].

Consider levels 0 and 1 of factor *A*, which give two groups of samples: $\{45,52,54,60\}$ and $\{68,72,80,83\}$. ANOVA gives a p-value of 0.003, so the null hypothesis is rejected at the 5% level (because $0.003 < 0.05$) to conclude that factor *A*'s effect on the response variable is significant.

4.3.3.2 Interaction effects

An *interaction effect* between factors, say *A* and *B* (denoted *AB*), is the effect that factor *A*'s level has on factor *B*'s main effect, and vice versa. If factor *B*'s effect on the response variable depends on the level of factor *A*, the interaction effect *AB* is *significant*. Multiple linear regression, described in section 4.3.3.3.1, provides a technique for calculating the magnitude of the effect.

[18] Tests to validate that the assumptions hold include the *Anderson-Darling test* [24] to check for normal distribution, and *Levene's test* [13] to check whether the standard deviations are equal.

*Interaction plots* can be used to depict two-factor interaction effects. Figure 4.11 shows the interaction plots for Figure 4.8. Each box represents a two-factor interaction, and contains a set of lines that show the effect of varying one factor when the second factor is fixed at a given level. Intuitively, the less parallel the lines the bigger the interaction effect. For example, the top-middle box shows the effect of varying factor *B* when factor *A* is fixed at levels 0 and 1. These lines are not parallel, which suggests that the effect of *AB* is big. Logically, the effect of *BA* (middle-left box) is also big. In contrast, the effect of factor *A* depends little on the level of factor *C*, which suggests that the effect of *AC* is small.

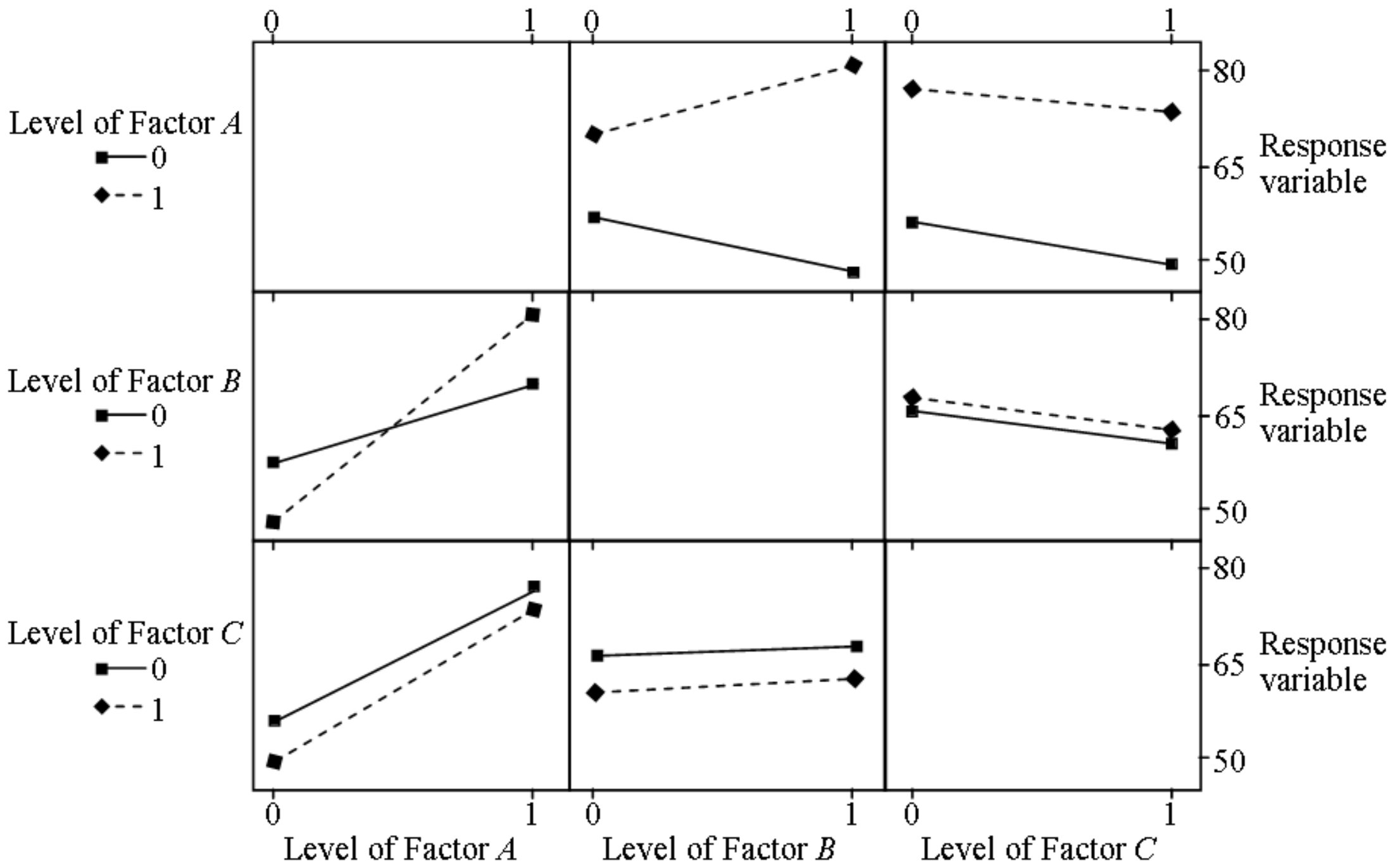

**Figure 4.11: Interaction effects plot**

Interaction effects can be described in terms of the shape of the response surface. An *interaction effect* is a relationship between cross-sections of the response surface[19], obtained by fixing a given factor at various levels. A big interaction effect causes the slope of the cross-sections to differ. This is illustrated by the response surface in Figure 4.12 for a hypothetical experiment: the surface's slope is different when *A*=0 from when *A*=1.

[19] In an n-dimensional input space, cross-sections are n-dimensional surfaces where the *n* dimensions consist of n-1 factors plus the response variable.

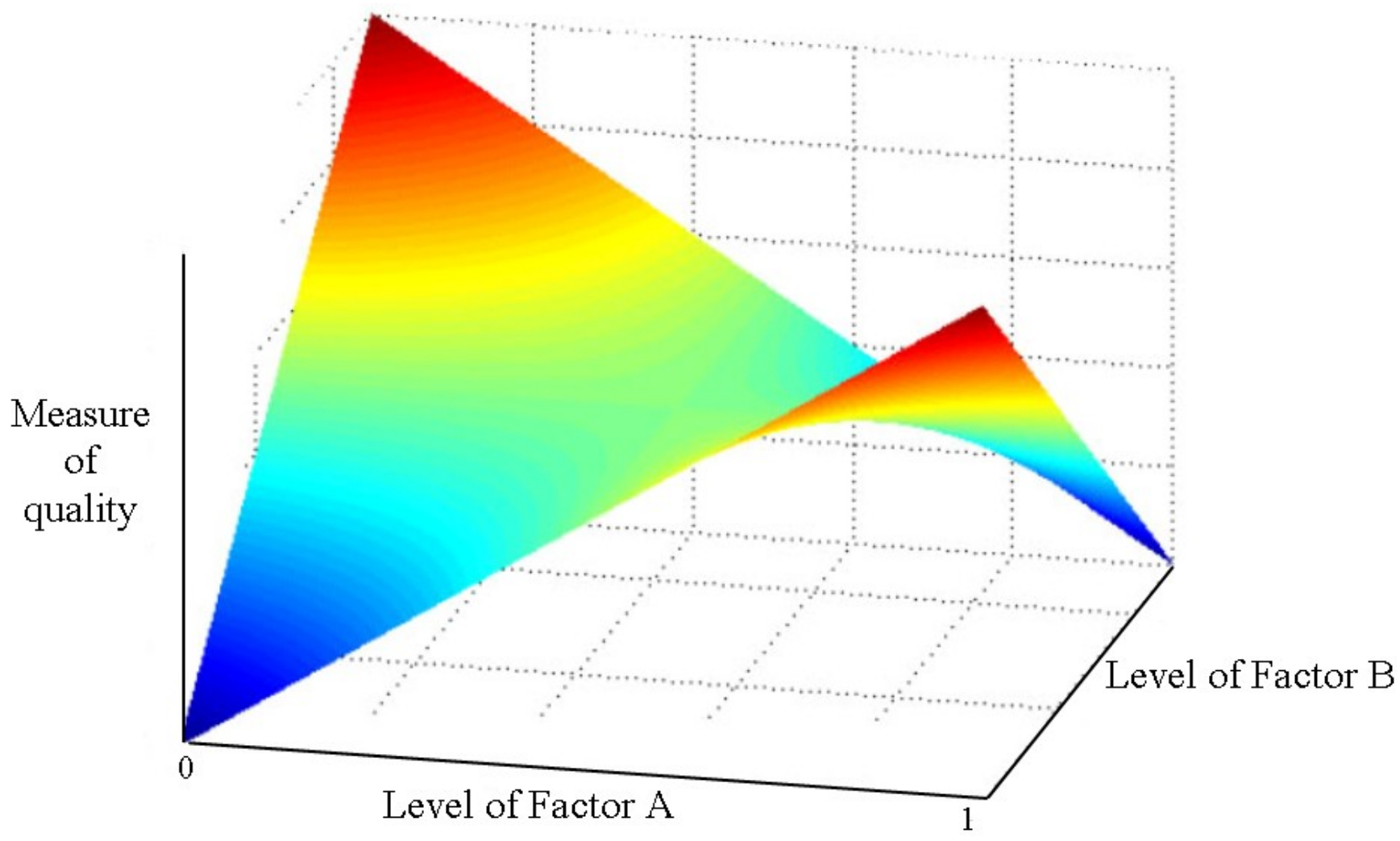

**Figure 4.12: Example interaction effect**

Consider the interaction effect between factors *A* and *B* in Figure 4.8. It can be investigated by comparing the response surface at the left four points (*A*=0) and the right four points (*A*=1). When *A*=0, changing *B* from 0 to 1 decreases the response variable's value by either 8 or 9 (depending on the level of *C*). When *A*=1, changing *B* from 0 to 1 increases the response variable's value by 11 or 12, significantly different from when *A*=0. This suggests that there is a significant two-factor interaction between *A* and *B*.

4.3.3.3 Modelling

Data from experiments can be used to develop and/or calibrate a model of the target system's behaviour. The purpose of a model is to predict the behaviour of untested combinations. Modelling involves determining a curve (i.e. a function) that fits the data. Its goodness of fit can be measured by the closeness of fit to the data set, and the accuracy for predicting the response of untested combinations.

Simple models assume that relationships between factors and the response variable are simple (e.g. linear or quadratic). Use of a complex model (e.g. a high-degree polynomial) can give a closer fit to the data set but can result in *over-fitting*, which decreases the accuracy of predictions. It is therefore not always desirable to have a perfect fit to the data set. Consider the hypothetical example in Figure 4.13. The dotted line is a polynomial of order seven, which gives a close fit to the data set. It predicts that the minimum value of the response variable is -4.2 (at *A*=1.3), but there is no evidence to support this prediction. Use of a simple model can avoid over-fitting. It is arguably sensible[20] to use a simple model when such a model gives an approximate fit to the data set or when the data set is small (e.g. when testing only two or three levels per factor).

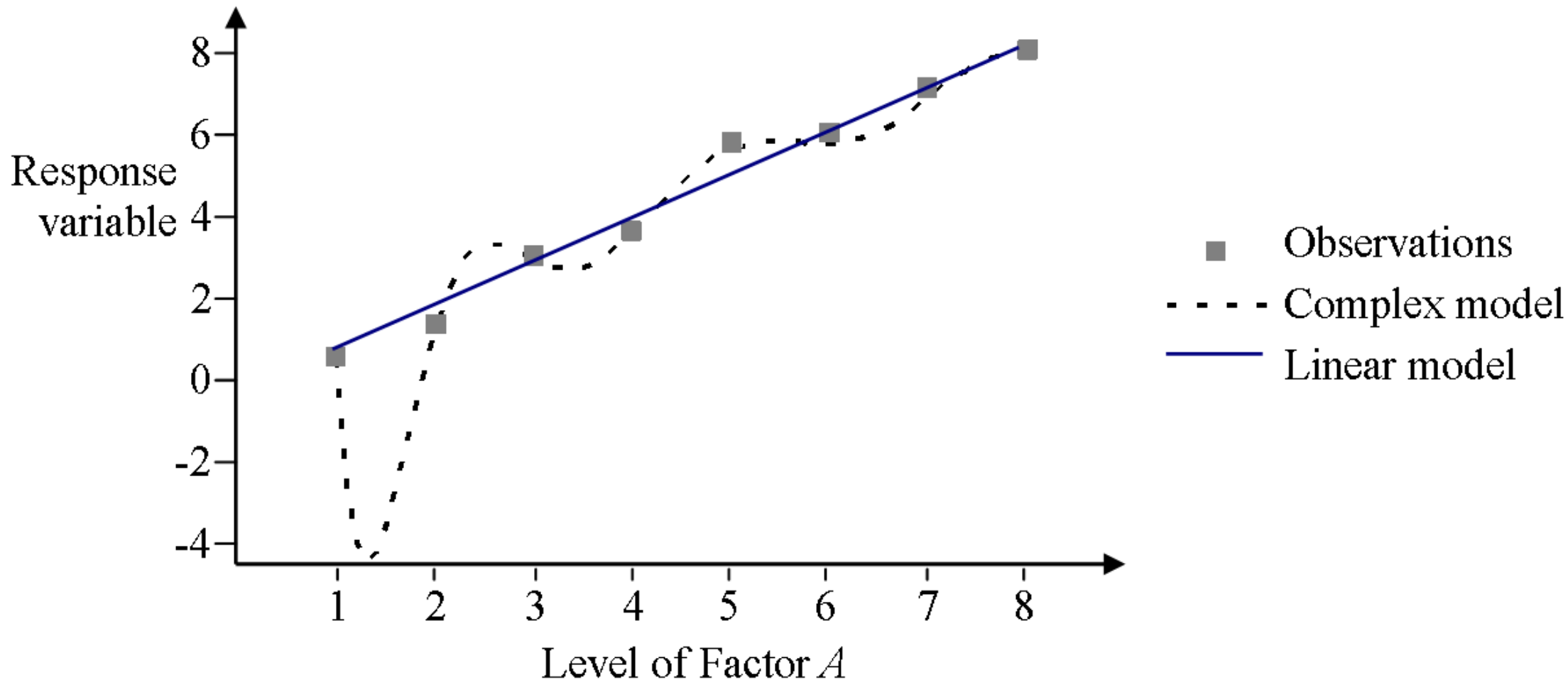

**Figure 4.13: Example of over-fitting**

There are a number of techniques for producing a model of a software system, given observations of the response variable for a selection of combinations. *Multiple linear regression* is useful when relationships between factors and the response variable are simple, or are assumed to be simple such as when the data set is small. The case study in section 5.2.4 uses multiple linear regression. Other modelling techniques include *splines* and *artificial neural networks*, which are useful for modelling discontinuous and non-linear response surfaces when the data set is large.

4.3.3.3.1 Multiple linear regression

*Multiple linear regression* [47, 64] is a statistical technique for fitting a linear model to a data set. Figure 4.14 shows the general form of the model, where: $y$ is the value of the response variable, $x_1$, $x_2$, …, $x_n$ are variables representing the levels of the factors, and $\beta_0$, $\beta_1$, …, $\beta_n$ are the coefficients. Each term gives an estimate of the effect of a factor or interaction, the coefficient indicating the magnitude of that effect. Interactions are modelled by terms such as $\beta_{12}x_1x_2$.

$$y = \beta_0 + \beta_1 x_1 + \beta_2 x_2 + \beta_{12} x_1 x_2 + \ldots + \beta_n x_n$$

**Figure 4.14: Multiple linear regression model**

[20] This follows the principle of *Occam's razor*, which states that, if two theories explain something equally well, the simpler of the two is better.

The errors in the model's predictions are called the *residuals* (i.e. the differences between predicted and observed values of the response variable). Multiple measurements of the response variable's value for a combination can give different residuals, as illustrated in Figure 4.14 for a hypothetical system. The line shows the model, while the crosses show measurements. The residual for a measurement is the distance from the corresponding cross to the line. Observations for a particular combination are drawn from the population of all possible values of the response variable for that combination. It is assumed that the residuals of each population are normally distributed about zero and that all populations have the same standard deviation. This implies that the model is as accurate for predicting large values as for small values in terms of absolute error.

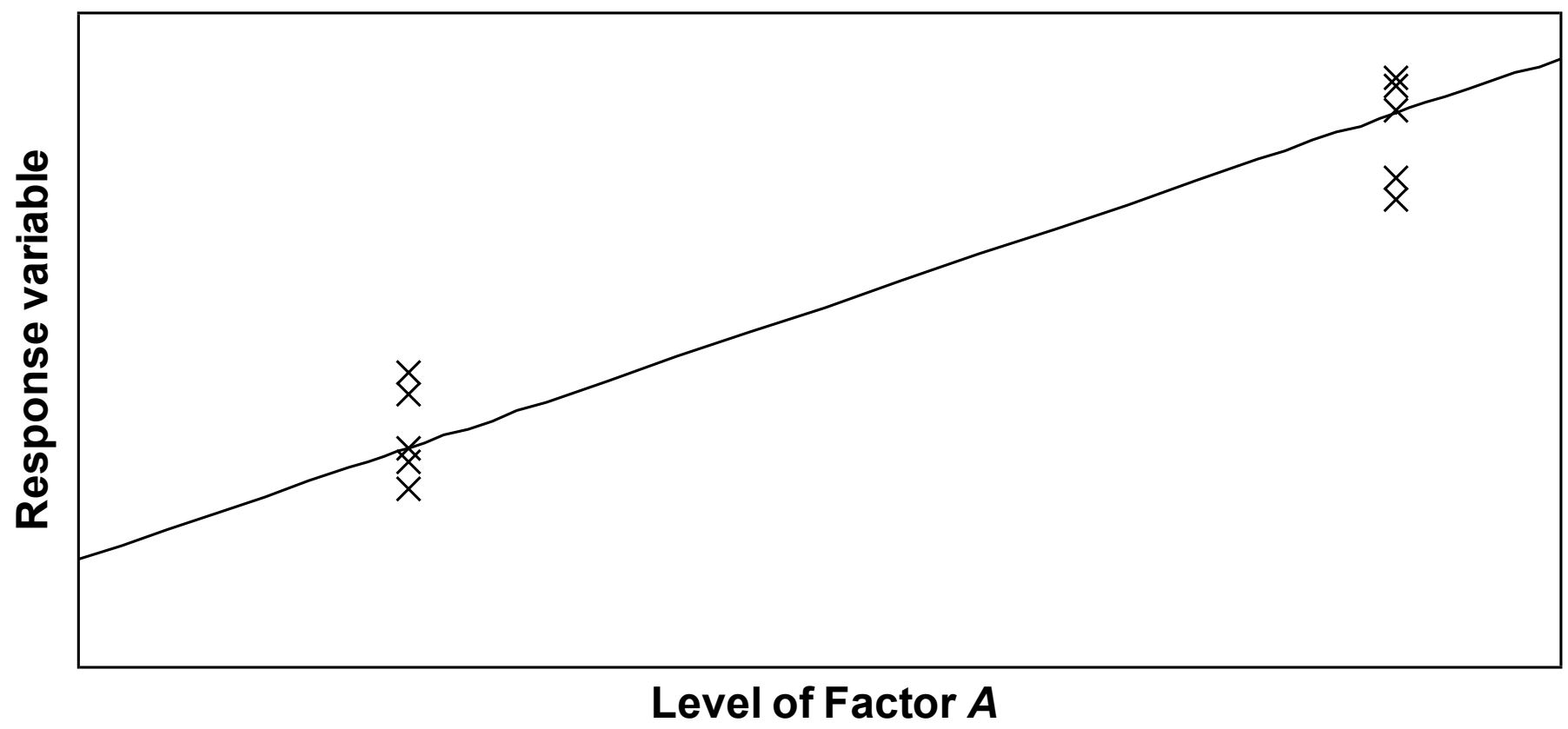

**Figure 4.15: Example residuals**

Effects of interactions that are not clear, due to aliasing, cannot be inferred during analysis. Multiple linear regression should be used to estimate only those effects explicitly investigated, and to estimate the confidence in the significance of these effects.

Terms for which there is no statistical evidence that their effects are significant should be removed from the model. This helps to prevent over-fitting. *Backward elimination* [47] can be used: each non-significant term is removed in turn, starting with interaction effects. A new regression model is fitted after removing each term, and this cycle repeats until the model only contains terms with significant effects.

An estimate of the significance of an effect can be a *false positive* or a *false negative*. The consequence of a false positive is that the magnitude of a non-significant effect is over-estimated and is left in the regression model. The probability of a "significant" effect being a false positive is 5% when working at the 5% level. The consequence of a false negative is that the magnitude of a significant effect is underestimated and is removed from the regression model. The probability of a false negative can be reduced by increasing the size of the data set or by making false positives more likely.

Linear regression models are used to predict which combination maximises the response variable. The assumptions of multiple linear regression include the following (their validity is discussed in section 6.2):

- The response surface is linear. Transformations can be applied if necessary to model some non-linearities (e.g. a quadratic term $\beta_{11}x_1^2$ is the square of the first factor's level). However, multiple linear regression will not work well if the response surface is either discontinuous or is a polynomial of a high degree.
- If an interaction effect, say *AB*, is not modelled, the effect of factor *A* is the same for all levels of factor *B*.

4.3.3.3.2 Parametric representation

*Parametric representations*, such as *splines*, are useful for modelling non-linear and discontinuous response surfaces. They are well suited to producing models that give a close fit to a large dataset. A parametric representation of an *n*-dimensional surface uses a *piecewise polynomial surface* (i.e. a surface made up of multiple segments). Each segment is defined using *n* polynomial functions in a parameter *t*. The most common functions used are polynomials of order three, which are of the form:

$$x(t) = a_x t^3 + b_x t^2 + c_x t + d_x$$

$$y(t) = a_y t^3 + b_y t^2 + c_y t + d_y$$

$$z(t) = a_z t^3 + b_z t^2 + c_z t + d_z \quad 0 \le t \le 1$$

…

where *x*, *y*, *z*, etc are dimensions of the space containing the surface. The value of *t* is never plotted: each value of *t* gives a point in each segment of the surface. Informally, one can think of *t* as a measure of time: for the case of a 2D curve, it is a measure of time as *n* pens move to draw the *n* segments of the line.

The choice of coefficients (i.e. values for *a*, *b*, *c* and *d*) for each segment is defined by constraints on *control points* (i.e. data points that the surface passes through or near to), *endpoints* (i.e. edges of a segment, which are special cases of control points) and the smoothness of joins between segments.

Each cubic surface has four coefficients per dimension, allowing four constraints to be met for each dimension.

*Splines* provide a technique for choosing values for the coefficients. *Natural cubic splines* produce a model that interpolates (i.e. passes through) all data points and for which the joins between segments are smooth to the second derivative. That is, adjacent segments meet, have the same gradient at the point where they meet, and the gradient of each segment is changing at the same rate when they meet. Many types of splines are discussed in [52].

###### 4.3.3.3.3 Multivariate Adaptive Regression Splines

*Multivariate adaptive regression splines* is a non-parametric technique, which does not assume that the data complies with an *a priori* functional form. The result is a connected set of local linear regressions [39]. As with other splines, it requires a much larger dataset to produce a model of a response surface than multiple linear regression and suffers from a lack of theory for calculating confidence intervals. This makes its use unsuitable for modelling some complex software systems.

###### 4.3.3.3.4 Artificial neural networks

Artificial neural networks can “learn” a model, which is then implicit in the network structure and weights between nodes [28]. It is therefore hard to examine the model of a response surface – e.g. determine the effects of factors, and detect phase changes and local maxima – to gain insight into the target system’s behaviour.

### 4.3.4 Validating the model

A model may be used either to predict the rank ordering of a set of combinations, or to predict the value of the response variable for particular combinations. Rank ordering is arguably more important than the exact value of the response variable because configuring a target system involves choosing between combinations. Indeed, rank ordering is less sensitive to variation in the conditions of use, as some changes (e.g. increased network load) may affect the value of the response variable for all combinations equally.

It is important to investigate the accuracy of predictions for combinations not used to produce the model[21]:

- **Validate rank ordering**. The predicted rank ordering of combinations should be compared to that obtained when they are empirically measured. This can be done using the *rank correlation coefficient*, τ (tau), which takes a value between -1 and 1. These indicate perfect disagreement (i.e. reverse ranking) and perfect agreement respectively [65]. Given *n* combinations, there are $\frac{n(n-1)}{2}$ pairs of combinations that can be compared: τ is proportional to the fraction of the pairs of combinations that are in the same order for both rankings.
- **Validate predictions of the response variable**. Empirical measurements of combinations should be compared to the model's predictions.

### 4.3.5 Second phase experiment

The experiment in the second phase of Taguchi Methods produces a finer grained model of the response surface around the predicted optimal combination; the aim is to more accurately predict the optimal combination [107]. An experiment is run to investigate the predicted highest peak on the response surface by testing the predicted optimal combination and combinations that are adjacent in the input space. The size of the region to test is influenced by the shape of the response surface, as predicted by the model from the first phase: the sharper the predicted peak, the smaller the region. The second phase experiment produces a model, which estimates all the main effects (using linear and quadratic terms) and a selection of two-factor interactions. The combination with the maximum response according to this model is taken to be the combination that best meets the configuration goal.

One possible experiment design is the *central composite design* [107], which tests the three sets of points shown in Figure 4.16:

- *Corner points* (squares in Figure 4.16) form a factorial design (with two levels for each of the *n* factors) centred at the predicted optimal combination. The design can be a fractional or full factorial design, and is used to estimate the main effects and interaction effects.

[21] In neural network terminology, combinations used to produce the model form the *training set*, and combinations used to validate the model form the *validation set*.

- *Star points* (crosses in Figure 4.16) are combinations along each axis in the space, increasing and decreasing each factor's level while keeping the other factors fixed. This gives 2*n* combinations and is used to estimate *curvature effects* (i.e. non-linearities), which often occur near a peak on the response surface.
- The *centre point* (circle in Figure 4.16) is the predicted optimal combination. The value of the response variable is measured multiple times (e.g. giving multiple values of the SNR) to give information about variance.

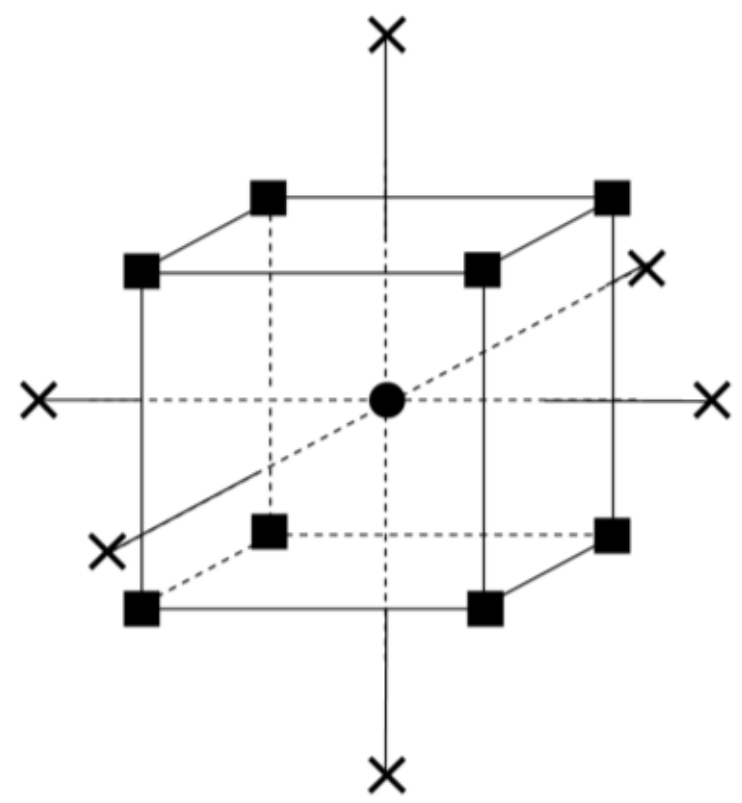

**Figure 4.16: 3D input space showing a central composite design**

The models from the first and second phases may be different. Consider the simple hypothetical example in Figure 4.17, which shows a response surface, measurements, and predictive models. The second phase experiment investigates the predicted peak at *A*=12 (giving three crosses at 7, 12 and 17). Its model is different from the first model (i.e. response increases linearly with factor *A*), and more accurately predicts the optimal combination.

Given a linear model such as that in Figure 4.17, the experimenter should investigate whether the linear relationship holds for higher levels of *A*, and if/when the response peaks. This is done by testing higher levels of factor *A*.

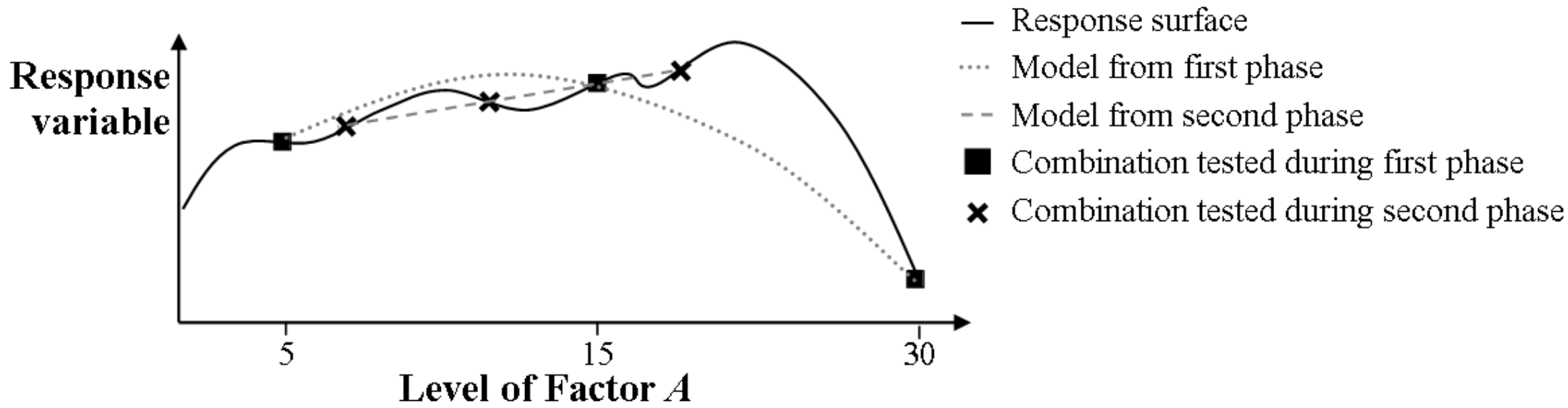

**Figure 4.17: Example second phase experiment**

### 4.3.6 Robust design

It is sometimes desirable to identify target system configurations that perform well under a variety of conditions, instead of a specific configuration for a particular condition of use. This is because conditions can vary dynamically and:

- it may be too expensive to reconfigure the target system on-the-fly;
- conditions may be too expensive to measure;
- conditions may change too quickly to configure the target system for the current conditions.

The aim of Taguchi's *robust design* [107] is to identify target system configurations that are insensitive to variation in *noise factors*. These are factors whose levels are not set explicitly when the target system is in use (for cost or technical reasons), but that can be controlled during the experiment. For example, the load on a network cannot be chosen when a target system is in use, but various levels of network load can be simulated during an experiment. Similarly, a database system cannot control the number of read and write requests issued by clients when in use, but various workloads can be used during an experiment.

Robust design involves exploiting the interaction effects between noise factors and *control factors* (i.e. configurable aspects of the target system) to find target system configurations that are insensitive to variation in the noise factors. Control-by-noise interaction plots, such as the hypothetical example in Figure 4.18, show the effect an interaction has on the response variable. Height of the line indicates goodness of response, and flatness of the line indicates insensitivity to the noise factor. There is a trade-off between these two characteristics: choice of level depends on the expected variation in the noise factor's level and the importance of insensitivity to this variation. In Figure 4.18, factor *A* is a noise factor and factor *B* is a control factor. Factor *B*'s level should be chosen such that the response is consistently high for all levels of factor *A*. Because the solid line is flatter than the dotted line and is higher for two of the three levels of factor *A*, level 1 may be an appropriate choice for factor *B*.

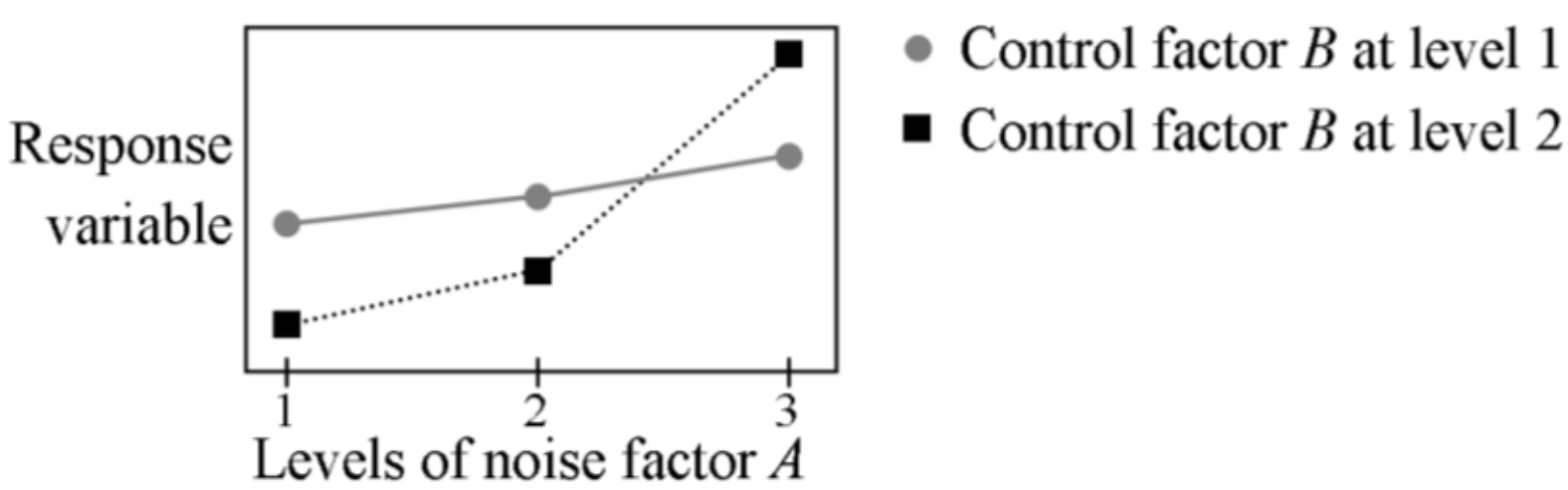

**Figure 4.18: Control-by-noise interaction plot**

## 4.4 Conclusions

There is a wide variety of potential search strategies for exploring a target system's behaviour. Choice of search strategy depends on:

- the aim of the experiment (e.g. to find a good configuration for a given condition, to discover specific characteristics of the target system, or to produce a predictive model of the target system's behaviour);
- the set of available adaptation mechanisms and the time required to configure the target system;
- resources available (e.g. time available);
- information available *a priori* about the target system's behaviour.

One promising search strategy involves the use of Taguchi Methods, which allows for statistical analysis of results. A predictive model can be produced, e.g. using multiple linear regression, that estimates the effects of factors and selected interactions. A second phase experiment can then produce a more accurate model of the response surface around the predicted optimal combination. The next chapter uses an industrial case study to investigate the applicability of Taguchi Methods for configuring a complex software system.

# 5 Case studies

This chapter describes two industrial case studies, which used the Data Connection Ltd (DCL) products DC-MailServer and DC-Directory.

Both products are configurable to support a broad customer base with a wide variety of potential usage patterns, hardware platforms, operating systems and network topologies. DCL currently undertakes performance analysis and tuning by hand, which relies heavily upon costly expertise and only permits testing of a few configurations for a given installation due to time constraints. In the following experiments, for both DC-MailServer and DC-Directory, it took over thirty minutes to configure and run the target system. Measurements were taken during a “hot run”, letting the performance stabilise to reflect customers’ continuous usage, and then measuring performance over a reasonable time.

The experimental base at St Andrews was a 64 node Beowulf cluster running RedHat 7.1 and connected through a 100Mb/s Ethernet switch. Each node consisted of a Pentium II 450MHz processor with 384MB of RAM and a 6.4GB hard disk.

The first case study (using DC-MailServer) illustrates some of the problems inherent in measuring complex software systems. The second (using DC-Directory) demonstrates how Taguchi Methods can be used to model and configure software systems.

## 5.1 DC-MailServer

The aim of the first case study was to explore the behaviour of DC-MailServer and to investigate the effects of a selection of configurable aspects on throughput of e-mail messages.

DC-MailServer [11] is a back-end mail server product from Data Connection Ltd (DCL). It is designed to be scalable, allowing deployment of multiple instances of each component in a server farm (e.g. in a Beowulf cluster).

DC-MailServer exposes hundreds of configurable aspects, including caching policies, concurrency policies and communication policies. A description of its configuration is stored in an instance of DC-Directory [10], an LDAP and X.500 directory server. A scripting language (TXDS) is used to query and change the directory data, changes affecting DC-MailServer’s configuration when it next restarts. Additional scripts start, stop and check the status of DC-MailServer.

### 5.1.1 Experimental infrastructure

The synthetic workload used to drive DC-MailServer was generated by the Microsoft Exchange Stress and Performance tool, ESP [7]. This simulated simultaneous access of many users sending and retrieving mail, using SMTP and POP3 protocols respectively. The workload could be configured by setting the number of users, the size of e-mail messages sent, the order of commands and the delay between them.

For these experiments, DC-MailServer was configured to output a count of sent and retrieved e-mail messages, writing the results to DC-Directory at five minute intervals. DCL supplied scripts to get the measurements from DC-Directory and calculate the throughput.

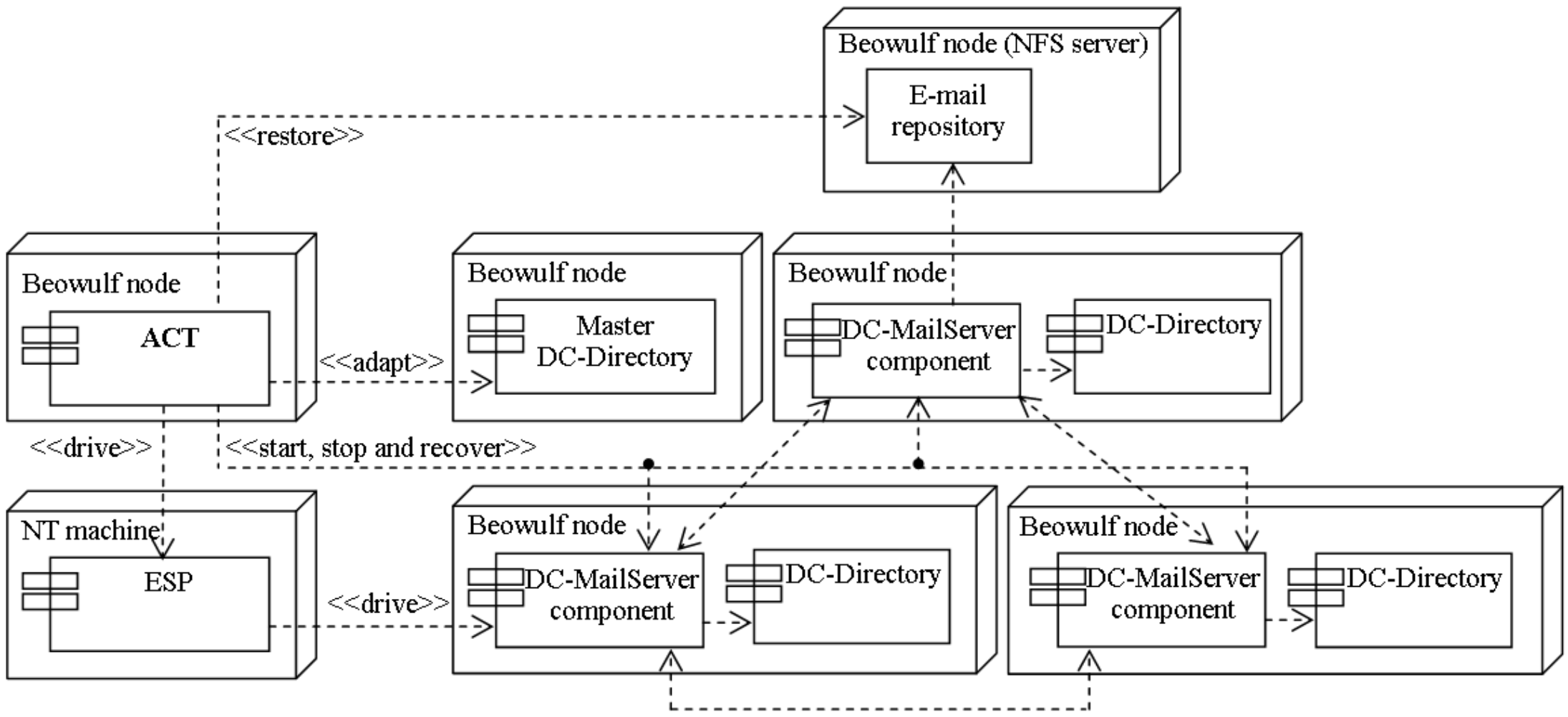

**Figure 5.1: Deployment diagram of experimental infrastructure for DC-MailServer**

The UML deployment diagram in Figure 5.1 shows the machines and components involved in running the experiments:

- A simple deployment of DC-MailServer was used, distributed over three nodes. Each node ran an instance of a particular component type.
- A node was used as a central file server to store the e-mails.
- A node ran the *master DC-Directory* component, which was responsible for managing a description of DC-MailServer's configuration. This information was replicated across each node of the DC-MailServer deployment. For simplicity, interactions among DC-Directory components are omitted from the diagram.

- ESP ran on an NT machine, connected directly to the Beowulf switch. It interacted with DC-MailServer by connecting to a pre-determined node on a known port.
- ACT ran on a separate node. It used remote shell invocations to call appropriate scripts on each of the other nodes.

For each trial, ACT:

- restored DC-MailServer to a consistent state by re-initialising the repository of e-mails and clearing the queues of messages awaiting processing;
- started DC-MailServer on each node in the deployment and started ESP to drive the system, waited 200 seconds for DC-MailServer to warm-up, and then measured throughput over a five minute period[22];
- stopped DC-MailServer and ESP.

In the event of failure, ACT checked the state of DC-MailServer, and used the point of failure and number of consecutive failures to guide the choice of recovery response. Diagnostics were collected and the target system's processes terminated, ready for restart. ACT also checked the state of the environment to ensure that file systems were usable (i.e. mounted correctly).

To run the experiments required no alterations to the target system itself. The *target wrapper* consisted of 260 lines of C++ code and a further 500 lines of shell scripts. This was less than 1% of the target system's size.

### 5.1.2 Variability in behaviour

The aim of this experiment was to investigate variability in DC-MailServer's behaviour. The default configuration was tested using a workload of thirty users retrieving mail and one user sending mail. Throughput was measured in terms of the number of e-mail messages retrieved and sent, referred to as *fetch* and *rcpt* (short for *recipient* in the SMTP protocol) respectively. Measurements were replicated 214 times to investigate variability in throughput experienced when using this single combination.

[22] These times were chosen based on advice from the software manufacturer.

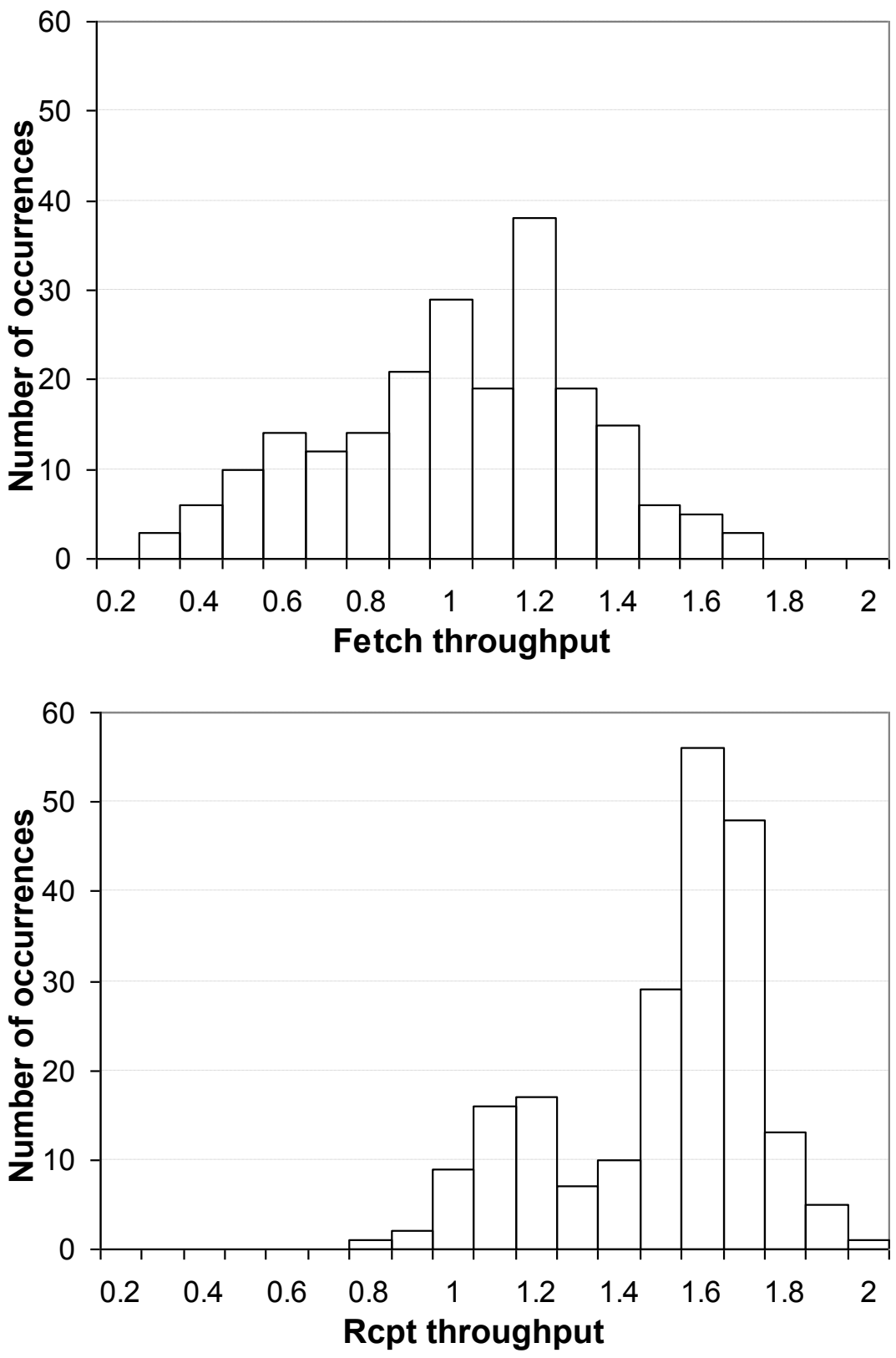

**Figure 5.2: Histograms of measured throughput[23]**

The histograms in Figure 5.2 show the number of occurrences for various ranges of throughput values. There is statistical evidence at the 5% level that the populations from which the *fetch* and *rcpt* results were drawn are not normally distributed (an *Anderson-Darling test*[24] gave p-values of 0.005 and 0.000 respectively, which are both less than 0.05).

Given a set of measurements, different metrics capture different characteristics of their distribution – which metric is appropriate depends on the distribution of the data and on the characteristics of interest. In general, *mean* and *standard deviation* are only appropriate if the data is normally distributed [35] and are therefore inappropriate for describing DC-MailServer's throughput. In contrast, SNR can be used to measure DC-MailServer's quality in terms of both how high throughput is and its consistency.

[23] All measurements of throughput for DC-MailServer were normalised with respect to the median of the *fetch* throughput, calculated using the results in Figure 5.2. This did not affect the shape of the graphs or the conclusions.

[24] The Anderson-Darling test [24] is a standard technique for testing if a set of samples come from a population with a normal distribution.

The variability in throughput (i.e. the spread of values observed) raised two issues:

- Given observations (from a small number of trials) that suggest configuration *A* has a higher throughput than configuration *B*, the confidence level is low that configuration *A* will perform better in future trials.
- Configurations observed to meet the configuration goal during an experiment may fail to do so when used by a customer.

### 5.1.3 Varying workload

The aim of this experiment was to investigate the effect on DC-MailServer's behaviour of varying the workload. It investigated the effect on throughput of the number of users that simultaneously retrieved mail using the POP3 protocol. The number of users sending mail was fixed at one. Figure 5.3 shows the observations of *fetch* and *rcpt* throughput. Figure 5.4 shows the SNR for these responses.

The graphs show that, for low numbers of clients, *fetch* throughput increased and *rcpt* throughput decreased with the number of users retrieving mail. Both throughput measures then levelled off for higher numbers of clients. These results suggest that threads servicing POP3 and SMTP requests compete (e.g. for CPU time and disk usage). The level sections of the graphs suggest that DC-MailServer could sustain at least 485 users retrieving mail without deterioration in throughput (compared to say 225 users retrieving mail).

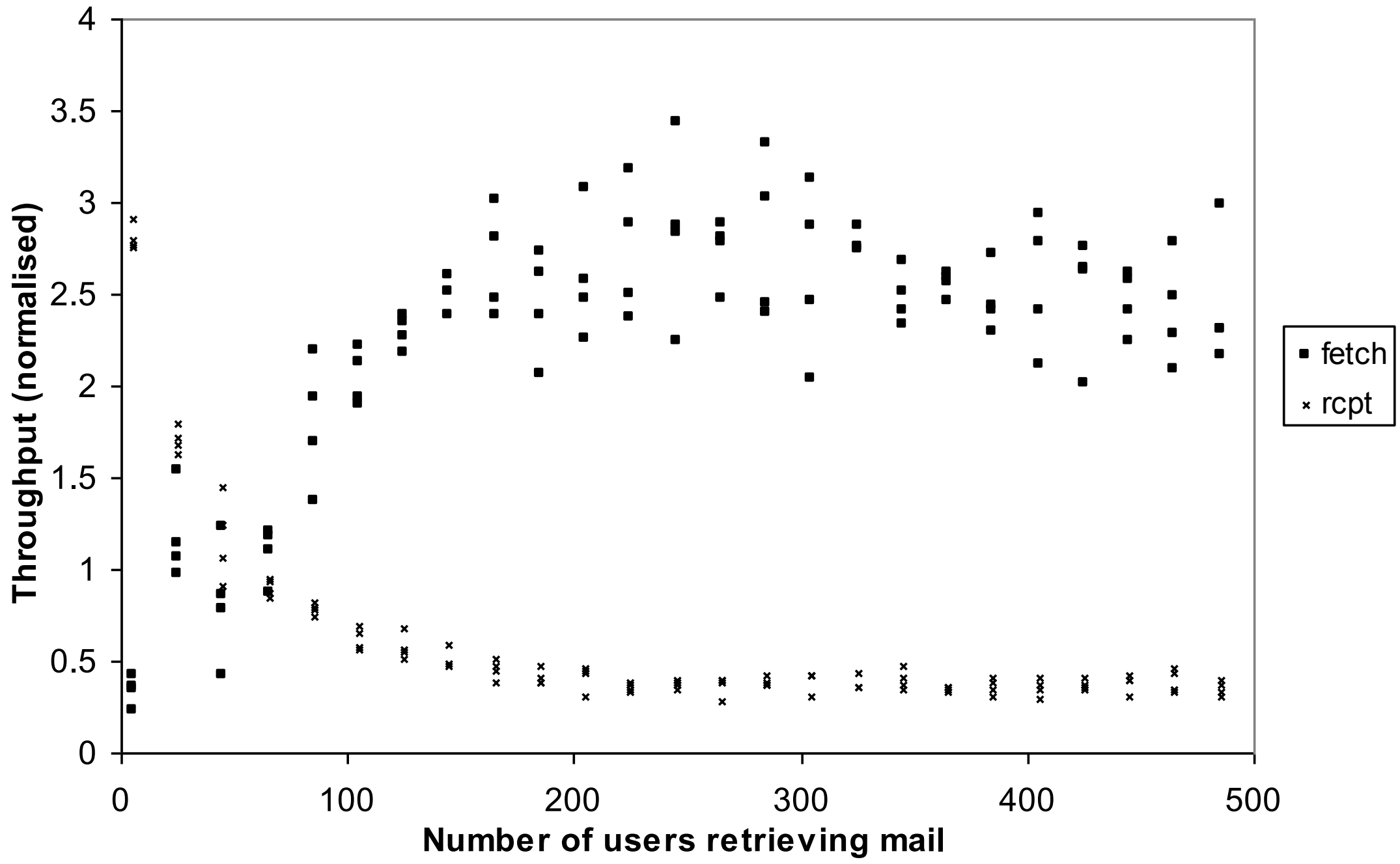

**Figure 5.3: Effect on throughput of the number of users retrieving mail**

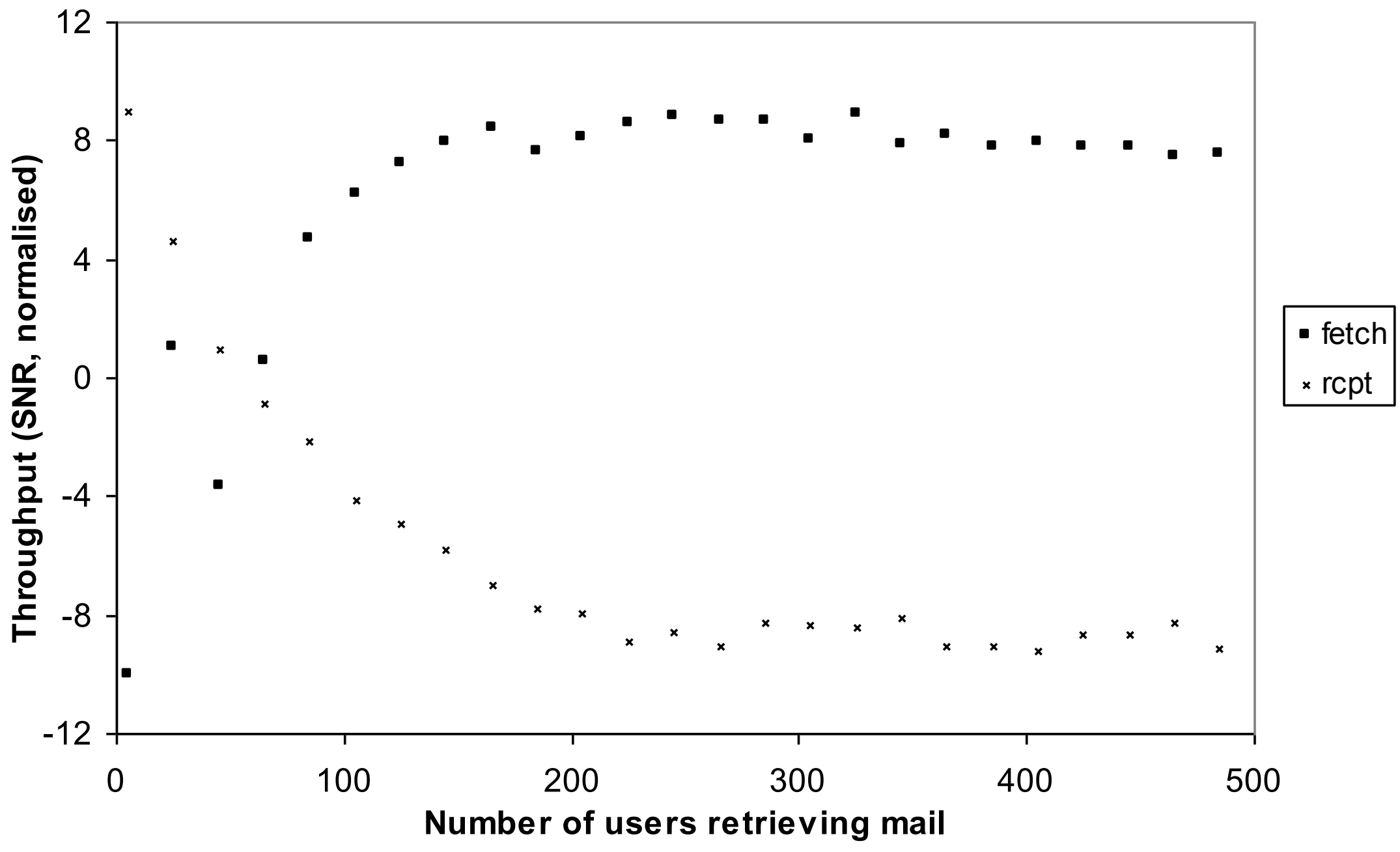

**Figure 5.4: Effect on throughput (SNR) of the number of users retrieving mail**[25]

[25] The SNR metric provides a good view of each combination's robustness: it is easier to compare combinations and see trends in Figure 5.4 than in Figure 5.3. However, it hides some information so is not always an appropriate representation for presenting results.

Observing the effect of varying the workload illustrates an alternative use of ACT. Previous discussion focused on finding a configuration of the target system that met a configuration goal, but ACT could also tune the workload to suit the target system. For example, an upper bound could be found for the acceptable number of users simultaneously retrieving mail. When this number was reached, a *throttle back* mechanism could be activated to reduce the number of additional clients connecting to DC-MailServer (e.g. by introducing a delay in responding to *connect* requests). The policy for choosing an upper bound would depend on the trade-off between *fetch* and *rcpt* throughput, and on the importance of response time compared to throughput.

### 5.1.4 Exploring effects of configurable aspects

Attempts to measure a wide range of DC-MailServer configurations proved problematic for two reasons: the large number of possible combinations meant that only a small part of the input space could be tested, while variability in performance led to a low confidence in results. Attempts to replicate interesting behaviour observed during experiments (i.e. combinations with very high or low throughput, compared to neighbouring combinations) failed.

Experiments exploring the input space of DC-MailServer revealed that:

- ACT can be used to run experiments that measure the performance of target system configurations. Inability to replicate observations was not the fault of ACT: it was due to variability in the target system's behaviour. This was verified by replicating trials independently of ACT and by monitoring the state of the experimental infrastructure.
- The set of factors under experimental control and the set of target system attributes observed were insufficient for replicable performance measurement. Uncontrolled factors in the environment and in the target system had a significant effect on throughput. This assumes that the target system did not deliberately behave non-deterministically; it assumes that, if all factors were controlled, the behaviour would be the same every time.
- Variability in performance is a serious issue. The acceptable variability depends on the degree of consistency demanded by the customer.

Approaches for coping with variability in performance include:

- increasing the number of replicated measurements to determine the distribution of results for each combination tested;

- identifying, and attempting to compensate for, the causes of variability by reducing the number of uncontrolled/unobserved factors – additional probes could further monitor the behaviour and state of both the target system and its conditions of use;
- designing experiments using Taguchi Methods (see section 4.3) to search for target system configurations that are robust (i.e. that give consistently high performance and are insensitive to the effects of uncontrolled factors).

## 5.2 DC-Directory

Given the results for DC-MailServer, the next case study investigated a simpler target system. DC-Directory [10] is an LDAP and X.500 directory server from Data Connection Ltd (DCL). It exposes hundreds of configurable aspects in a textual file. DCL supply scripts to make changes to the file, which affect DC-Directory's configuration when it next restarts. Additional scripts start, stop and check the status of DC-Directory.

### 5.2.1 Experimental infrastructure

The synthetic workload used to drive DC-Directory in the experiments was generated by DirectoryMark [12], an LDAP server benchmarking tool. It simulated multiple clients sending sequences of LDAP requests, and reported the time taken for these requests to be processed. The workload could be configured by setting the size of the directory information base, the number of clients, the type of requests and the number of requests per client. A directory of 100,000 entries was used, with 10 clients each sending 10,000 addressing (i.e. lookup) requests.

The configuration goal was to maximise DC-Directory's *throughput* (in terms of requests serviced per second), which was measured by DirectoryMark.

The UML deployment diagram in Figure 5.5 shows the machines and components involved in running the experiments. For simplicity, DC-Directory was run on a single node. DirectoryMark ran on an NT machine, connected directly to the Beowulf switch. ACT ran on a separate node, and used remote shell invocations to call appropriate scripts that controlled DC-Directory and DirectoryMark.

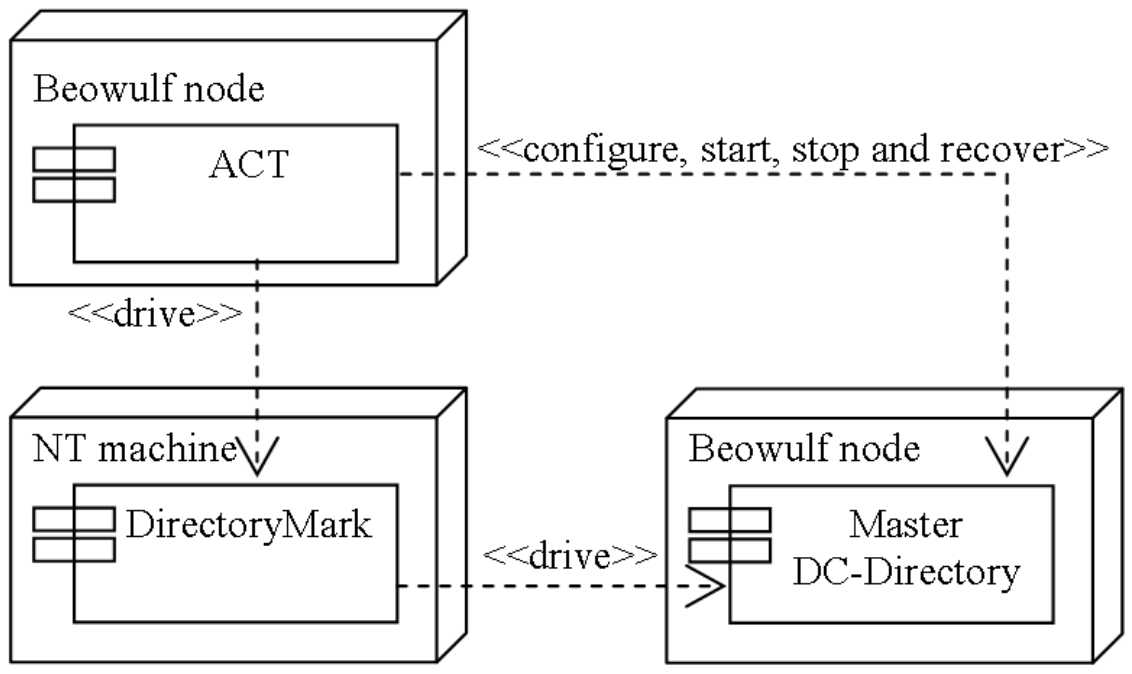

**Figure 5.5: Deployment diagram showing experimental infrastructure for DC-Directory**

For each trial, ACT ran DC-Directory and obtained a measure of its throughput. In the event of failure, the *recovery function* was invoked to report where the failure occurred and to collect diagnostic information.

To automatically run, configure and measure DC-Directory required no alterations to the target system itself. The *target wrapper* consisted of 345 lines of C++ code and a further 400 lines of shell scripts. This was less than 1% of the target system's size.

### 5.2.2 Normalising results

All measurements of throughput (i.e. all responses) for DC-Directory were normalised with respect to the median response of the default configuration. This did not affect the shape of the graphs or the conclusions.

Calculations of SNR used the normalised responses. This had the effect of decreasing all SNR values by a constant. Figure 5.6 explains this algebraically, where α is the normalising term.

$$SNR = -10\log_{10}\frac{\sum_{i=1}^{n}\frac{1}{\left(\frac{y_i}{\alpha}\right)^2}}{n} = -10\log_{10}\frac{\sum_{i=1}^{n}\frac{\alpha^2}{y_i^2}}{n}$$

$$= -10\log_{10}\left(\alpha^2 \times \frac{\sum_{i=1}^{n}\frac{1}{y_i^2}}{n}\right) = -10\log_{10}(\alpha^2) - 10\log_{10}\frac{\sum_{i=1}^{n}\frac{1}{y_i^2}}{n}$$

**Figure 5.6: Normalised SNR**

Consider a combination that consistently gives the same response as the default's median response. It will have a normalised SNR of 0 because all normalised values of the response will be 1, and $\log_{10}(1)$ is 0. However, responses for the default configuration gave a normalised SNR of -0.004. This is because there was variability in the responses, and SNR punishes low values more than it rewards high values. For example, the hypothetical responses {1,1,1,1} give an SNR of 0, while the hypothetical responses {0.97,0.99,1.01,1.03} give an SNR of -0.007.

### 5.2.3 Importance of replicating observations

Based on advice from Data Connection Ltd (DCL), an experiment was conducted to investigate the effects of varying DC-Directory's `TNE`. This factor indicates the typical number of entries in the database. DC-Directory uses `TNE`'s level to decide on the amount of memory required: the higher the `TNE`, the larger the cache of database records stored in main memory.

At the beginning of each trial, DC-Directory was warmed using DirectoryMark to simulate ten clients sending addressing requests for ten minutes. Trials were replicated four times for each level of `TNE`, and throughput measured during each trial.

Figure 5.7 shows the results of the experiment. The graph can be partitioned into three regions, indicated by the dotted lines: `TNE ≤ 60,000`, `70,000 ≤ TNE ≤ 160,000`, and `TNE ≥ 170,000`. In the first and third regions, throughput was consistent. In the middle region, throughput was generally more variable[26]. This illustrates the importance of replicating trials.

[26] Throughput was consistent when TNE was 80,000 – no explanation of this behaviour is available.

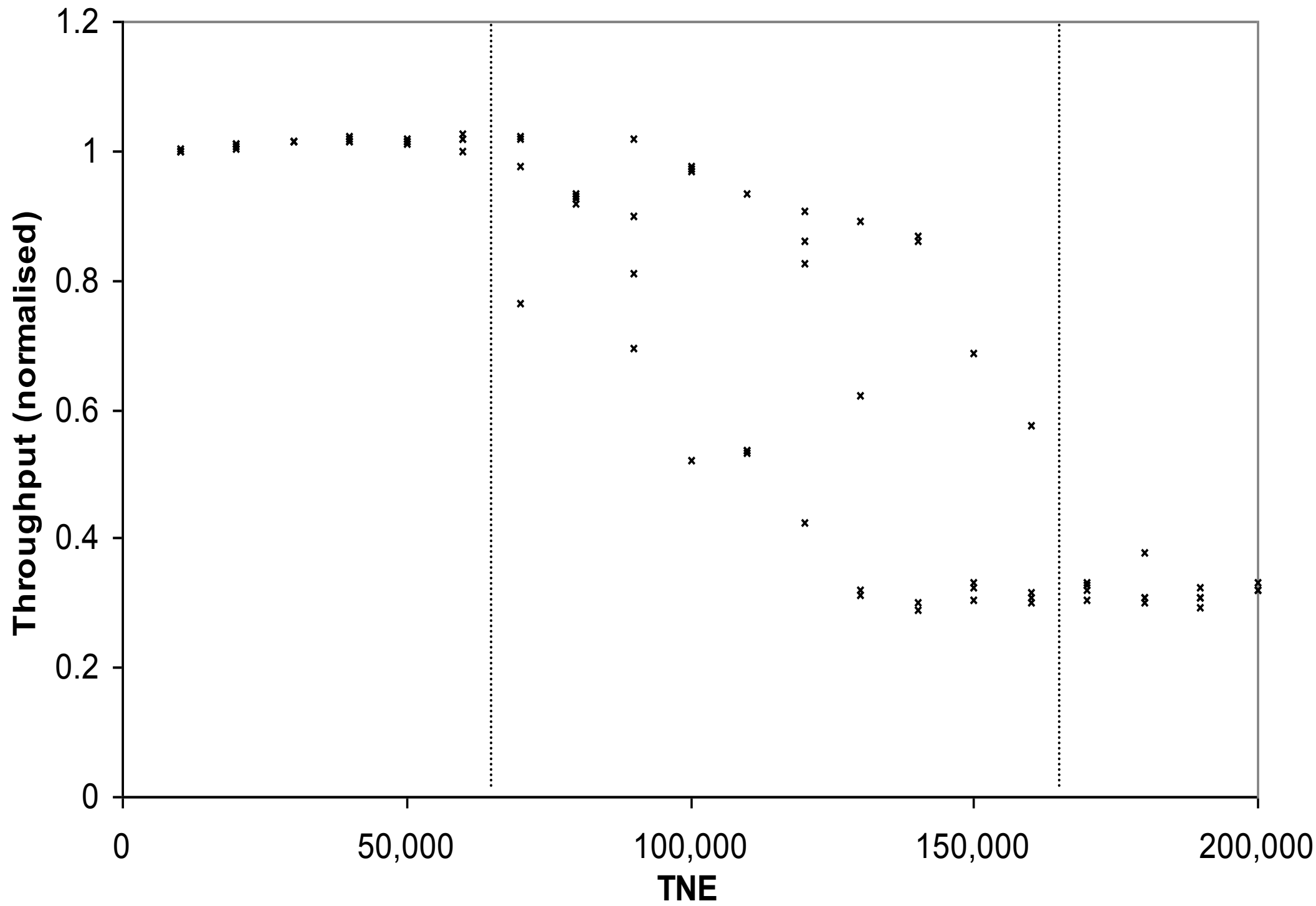

**Figure 5.7: Observations of DC-Directory's throughput for a selection of cache sizes**

Additional metrics gathered during each trial included CPU and memory usage. A correlation was observed between the variability in DC-Directory's throughput and swap space usage (i.e. amount of memory stored on disk rather than in RAM)[27]. Figure 5.8 shows the amount of swap space used, measured 60 seconds into each trial. Again, the graph can be partitioned into three regions: swap space usage was consistent when `TNE` was low or high, but varied in the middle region of the graph. However, the correlation is not perfect: swap space usage varied when `TNE` was 60,000, yet throughput was consistent (the range of observed responses was less than 3% of the median).

[27] It is assumed that all usage of swap space was due to DC-Directory.

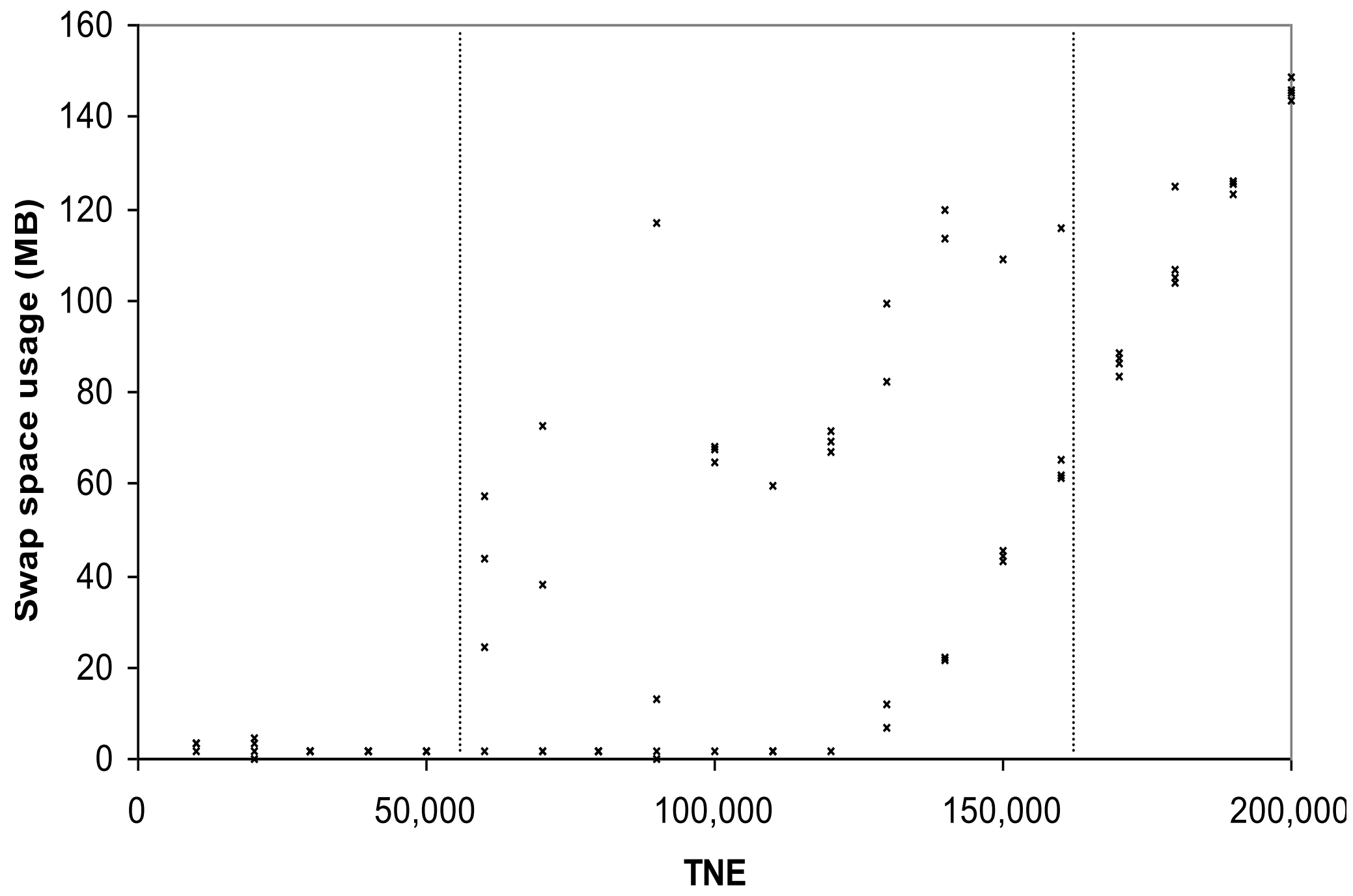

**Figure 5.8: DC-Directory's swap space usage for a selection of cache sizes**

An experiment was conducted to investigate the effects of TNE on throughput after a longer warm-up period. The warm-up consisted of ten clients sending addressing requests for ten minutes plus a subsequent 10,000 lookup operations per client. Figure 5.9 shows the results, using 0.9 as the origin of the graph.

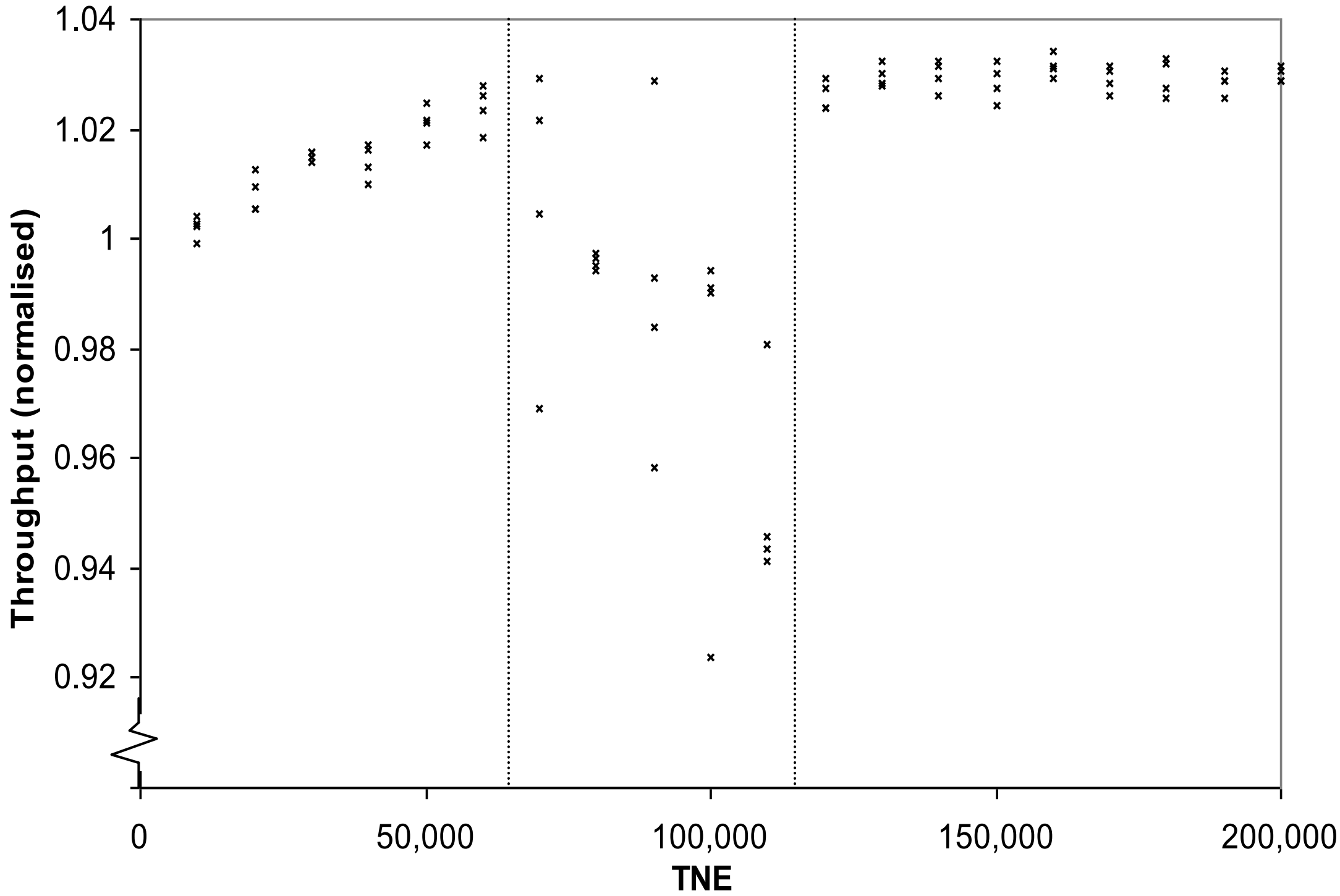

**Figure 5.9: Observations of DC-Directory's throughput after a longer warm-up period**

This graph can be partitioned into three regions: `TNE` levels of 60,000 or less gave consistently high throughput, as did `TNE` levels of 120,000 or greater. Levels in the middle region generally gave larger variability in throughput.

The effect of `TNE` on throughput depended on the length of the warm-up, as can be seen from the graph of SNR in Figure 5.10. This indicates that DC-Directory's behaviour changes over its run-time. The appropriate warm-up to use during an experiment depends on the expected usage pattern of the customer.

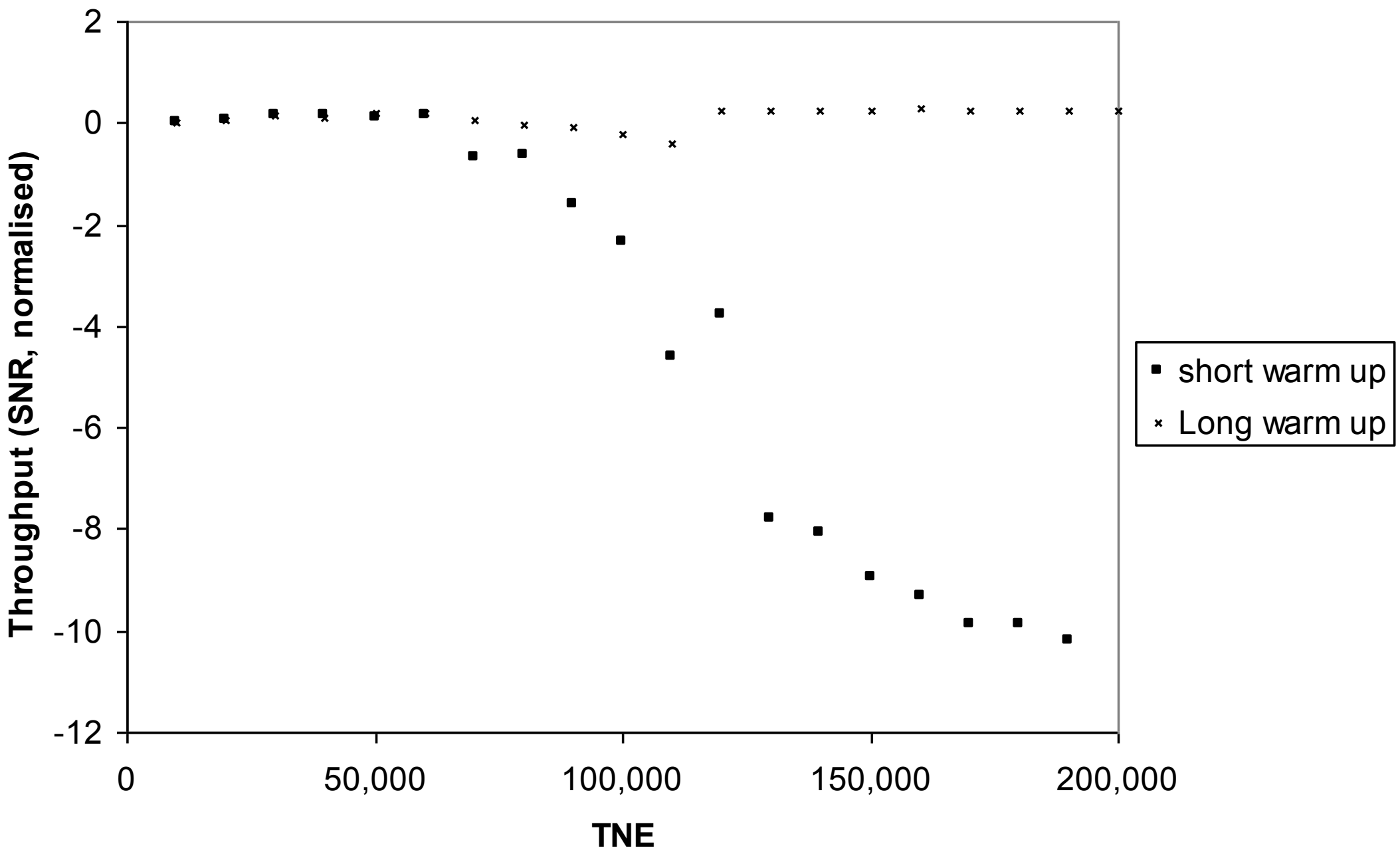

**Figure 5.10: DC-Directory's quality after various warm-ups**

### 5.2.4 Use of Taguchi Methods

The following experiment demonstrates the use of Taguchi Methods for investigating the behaviour of DC-Directory. For simplicity just four factors were varied, but Taguchi Methods allow the design of experiments that explore the effects of dozens of factors.

The aim was to determine the effects of the following factors on DC-Directory's throughput:

- `TNE`, which indicates the expected typical number of entries in the database;
- `MaxLDAP`, which is a constraint on the maximum number of simultaneous LDAP users that can connect to DC-Directory;
- `LDAPnum`, which corresponds to the number of threads servicing the queue of LDAP requests;

- `DispNum`, which corresponds to the number of threads that search the database.

5.2.4.1 First phase experiment

The first phase experiment investigated the main effect of each factor, plus a sample of the interaction effects. These were `TNE.MaxLDAP` (i.e. the interaction effect between `TNE` and `MaxLDAP`), `MaxLDAP.LDAPnum` and `MaxLDAP.DispNum`. Figure 5.11 shows the levels tested for each factor, the default levels used in a fresh installation of DC-Directory, and the constraints on legal configurations defined by the software manufacturer.

| Factor | Number of threads | | | | |
|---|---|---|---|---|---|
| | Level 1 | Level 2 | Level 3 | Default | Constraints |
| TNE | 10,000 | 30,000 | 50,000 | 10,000 | Integer; TNE > 0 |
| MaxLDAP | 100 | 1001 | 2000 | 1001 | Integer; MaxLDAP > 0 |
| LDAPnum | 1 | 2 | 4 | 2 | Integer; LDAPnum > 0 |
| DispNum | 2 | 5 | 8 | 5 | Integer; DispNum > 1 |

**Figure 5.11: Factors**

Minitab ™ was used to help design the experiment. It revealed that the most suitable orthogonal array was $L_{27}(3^{13})$, shown in Appendix E, with 27 combinations. This is a third of the size of a full factorial design, which would have required 81 (i.e. $3^4$) combinations.

DC-Directory was warmed using DirectoryMark, which simulated ten clients that each sent addressing requests for ten minutes plus a subsequent 10,000 lookup operations per client. Trials were replicated four times[28] to measure the robustness of each combination. The experiment took 100 hours.

Appendix E contains a table of the experiment results showing, for each combination, the four measurements of throughput and the SNR (i.e. quality). Figure 5.12 shows the main effect of each factor on quality (see section 4.3.3.1 for a description of *main effects plots*). For example, it shows that increasing `MaxLDAP` had a detrimental effect on quality.

[28] Professor Harry Staines of Abertay University recommended a minimum of four replications.

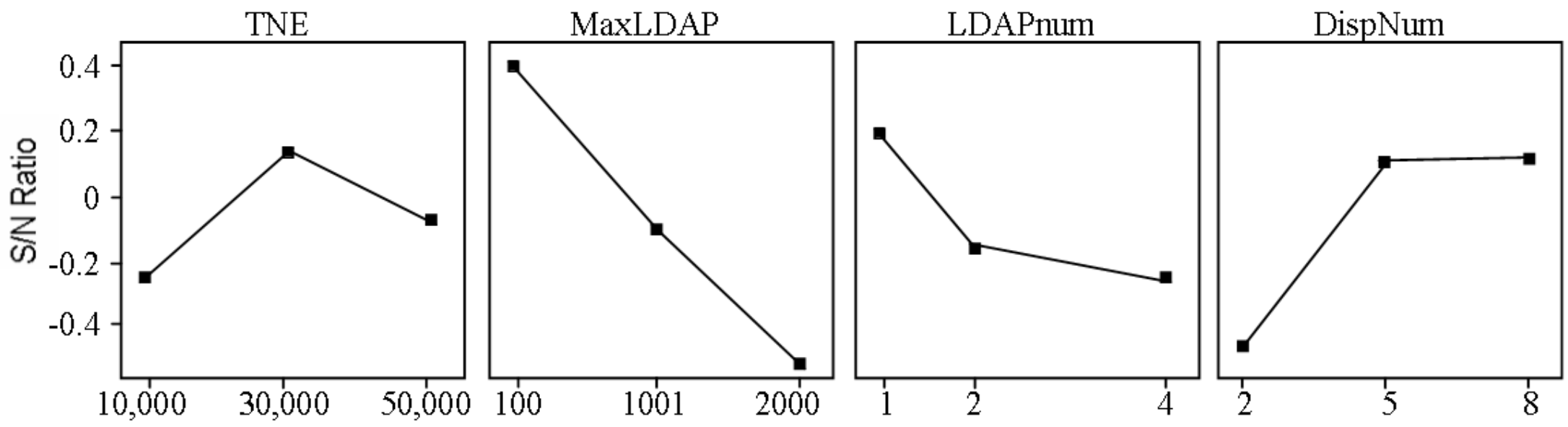

**Figure 5.12: Main effects plot**

Figure 5.13 depicts the interaction effects that could be inferred from the experiment results (see section 4.3.3.2 for a description of *interaction plots*). It suggests that the effect of `TNE.MaxLDAP` was small, while the effects of `MaxLDAP.LDAPnum` and `MaxLDAP.DispNum` were big.

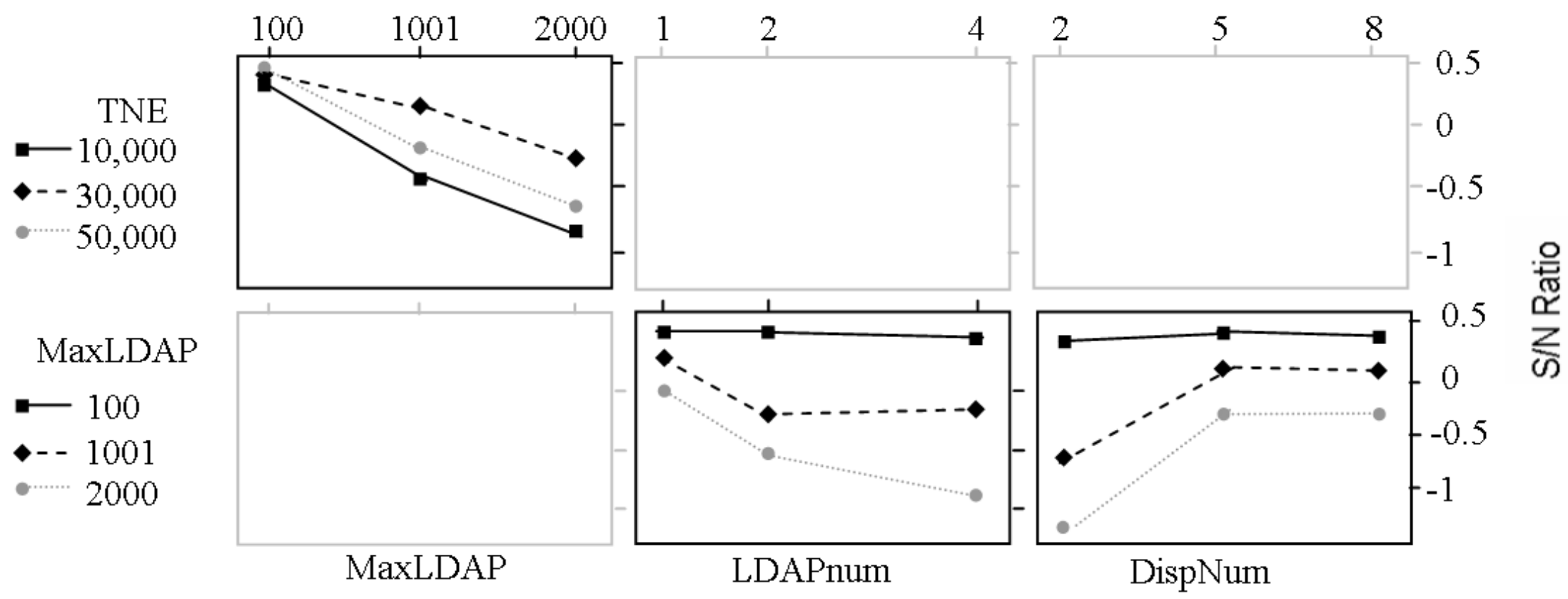

**Figure 5.13: Interaction plot**

Minitab [TM] was used to produce a multiple linear regression model of the target system's quality, shown in Figure 5.14. Linear and quadratic terms were used for each main effect because three levels were tested for each factor. For some terms, there was no statistical evidence at the 5% level that their effects were significant (as reported by Minitab [TM]). Using *backward elimination* to remove non-significant terms (see section 4.3.3.3.1) gave the regression model in Figure 5.15.

$$
\begin{aligned}
SNR = & -0.0345 + 0.513x_{TNE} - 0.733x_{MaxLDAP} - 0.523x_{LDAPnum} + 0.343x_{DispNum} \\
& - 0.0782x_{TNE}^2 + 0.076x_{MaxLDAP}^2 - 0.103x_{LDAPnum}^2 - 0.0323x_{DispNum}^2 \\
& + 0.0020x_{TNE}x_{MaxLDAP} - 0.138x_{MaxLDAP}x_{LDAPnum} + 0.0758x_{MaxLDAP}x_{DispNum}
\end{aligned}
$$

**Figure 5.14: First phase model, before backward elimination[29]**

[29] The levels of `TNE` were divided by 10,000, and `MaxLDAP` by 1000 to keep the coefficients manageable.

$$SNR = -0.892 + 0.515x_{TNE} - 0.573x_{MaxLDAP} + 0.343x_{DispNum} - 0.0782x_{TNE}^2 - 0.0323x_{DispNum}^2 - 0.135x_{MaxLDAP}x_{LDAPnum} + 0.0758x_{MaxLDAP}x_{DispNum}$$

**Figure 5.15: First phase model of DC-Directory's quality**[30]

According to this model, the combination (within the region of the input space tested) of highest quality would be that shown in Figure 5.16. The predicted optimal level of 5.43 for `DispNum` has been truncated to 5, since it must be an integer.

| Factor | Level |
|---|---|
| `TNE` | 32,930 |
| `MaxLDAP` | 100 |
| `LDAPnum` | 1 |
| `DispNum` | 5 |

**Figure 5.16: Predicted optimal combination**

5.2.4.2 Second phase experiment

A *central composite design* (see section 4.3.4) was used to test combinations in a region of the response surface around the predicted optimal combination. The aim was to produce a model that could more accurately predict whether there was a peak, and where it lay. The *corner points*, centred at the predicted optimal combination, were at `TNE ±5000`, `MaxLDAP ±45` and `DispNum ±1`. These levels were chosen based on the predicted sharpness of the peak. The level of `LDAPnum` was fixed at 1 because levels below 1 are invalid, and the design must be symmetric about the centre. This decision can be justified using the model in Figure 5.15, which predicts that quality decreases as `LDAPnum` increases so its optimal level is its smallest legal level (i.e. 1). Appendix F contains the experiment design and results.

Multiple linear regression was used to model the response surface, giving the model in Figure 5.17. *Backward elimination* was used to remove non-significant terms, giving the model in Figure 5.18.

$$SNR = 0.465 - 0.083x_{TNE} + 0.203x_{MaxLDAP} - 0.0207x_{DispNum} + 0.0259x_{TNE}^2 - 0.0810x_{MaxLDAP}^2 + 0.00119x_{DispNum}^2 - 0.0389x_{TNE}x_{MaxLDAP} - 0.0025x_{TNE}x_{DispNum} + 0.0139x_{MaxLDAP}x_{DispNum}$$

**Figure 5.17: Second phase model, before backward elimination**

[30] Standard diagnostic tests showed that the assumptions of multiple linear regression were not violated.

$$SNR = 0.0266 + 0.0363x_{TNE} + 0.152x_{MaxLDAP} - 0.0851x_{MaxLDAP}^2$$

**Figure 5.18: Second phase model of DC-Directory's quality**

Figure 5.18 predicted that, for the region of the response surface tested, quality would increase linearly with `TNE`. To identify the optimal combination therefore required investigation of higher levels of `TNE`. A further experiment was conducted to produce a model for higher levels of `TNE`. Because there were no significant interaction effects in Figure 5.18, the main effect of `TNE` was investigated in isolation of factors `MaxLDAP`, `LDAPnum` and `DispNum`. These were fixed at 100, 1 and 5 respectively, which allowed results from the above experiment to be used in producing the model (i.e. results for the centre point and two of the star points). This gave the model in Figure 5.19, which predicted that quality would peak when `TNE` is 58,720. The predicted optimal combination is that shown in Figure 5.20.

$$SNR = 0.0266 + 0.0752x_{TNE} - 0.00641x_{TNE}^2$$

**Figure 5.19: Model of `TNE`'s effect on DC-Directory's quality**

| Factor | Level |
|---|---|
| `TNE` | 58,720 |
| `MaxLDAP` | 89 |
| `LDAPnum` | 1 |
| `DispNum` | 5 |

**Figure 5.20: Predicted optimal combination from follow-up experiment**

5.2.4.3 Results of validating the model

The aim of this experiment was to test the accuracy of predictions made by the first phase model in Figure 5.15, in terms of the rank ordering of combinations. It used a validation set consisting of 16 previously untested combinations[31]. Their predicted rank ordering was compared to the observed rank ordering using the *rank correlation coefficient* (see section 4.3.6). It yielded a correlation coefficient of 0.62, which means that the predicted ranking gave the same order as the observed ranking for 81% of the pairs of combinations.

[31] The validation set consisted of a full factorial design with two levels per factor. Results revealed that significant two-factor interaction effects at the 5% level were `MaxLDAP.LDAPnum`, `MaxLDAP.DispNum` and `LDAPnum.DispNum`. There was no statistical evidence that `TNE.MaxLDAP`, `TNE.LDAPnum` or `TNE.DispNum` were significant.

Accurate prediction of which combination is optimal is more important than the rank ordering of other combinations. It is important that the predicted optimal does indeed perform well. Of the 16 combinations in the validation set, the observed best was ranked 6$^{th}$ by the first phase model in Figure 5.15 – a poor prediction. However, the combination ranked 1$^{st}$ by the model was observed to be 2$^{nd}$ best. Finding a near-optimal combination (e.g. 2$^{nd}$ best) may be sufficient for meeting the configuration goal.

The predicted maximum investigated during the second phase experiment was not as high as the first phase model in Figure 5.15 predicted. Indeed, the predicted optimal from the second phase experiment was not as good as the combination in Figure 5.21, which was tested during the first phase experiment. This suggests that assumptions made in designing the first phase experiment and analysing the results were violated. This is discussed in the following two sections.

| Factor | Level |
|---|---|
| TNE | 50,000 |
| MaxLDAP | 100 |
| LDAPnum | 1 |
| DispNum | 8 |

**Figure 5.21: Combination with highest observed quality**

5.2.4.4 Consequences of ignoring significant effects

Validation of the first phase model in Figure 5.15 revealed that the model's predictions were sometimes inaccurate. The aim of this follow-up experiment was to test the hypothesis that inaccuracies in predictions were (at least partly) due to ignoring a significant two-factor interaction effect.

The interaction effects investigated in the first phase experiment described in section 5.2.4.1 were TNE.MaxLDAP, MaxLDAP.LDAPnum and MaxLDAP.DispNum. However, results from the validation experiment in section 5.2.4.3 revealed that, for this region of the response surface, the significant interaction effects were MaxLDAP.LDAPnum, MaxLDAP.DispNum and LDAPnum.DispNum. The predictive model should therefore have included these effects and the main effects of TNE, MaxLDAP, LDAPnum and DispNum.

This follow-up experiment investigated these effects. It was designed using Taguchi Methods and was similar to the first phase experiment described in section 5.2.4.1. The levels tested were as shown in Figure 5.11. As before, DC-Directory was warmed by ten clients that each sent addressing requests for ten minutes plus a subsequent 10,000 lookup operations per client. Trials were replicated four times.

The resultant models, before and after *backward elimination*, are shown in Figure 5.22 and Figure 5.23 respectively. The combination predicted to be of optimal quality according to this model is shown in Figure 5.24.

$$\begin{aligned} SNR = 0.436 &+ 0.181x_{TNE} - 0.726x_{MaxLDAP} - 0.689x_{LDAPnum} + 0.217x_{DispNum} \\ &- 0.0156x_{TNE}^2 + 0.076x_{MaxLDAP}^2 + 0.0956x_{LDAPnum}^2 - 0.0308x_{DispNum}^2 \\ &- 0.151x_{MaxLDAP}x_{LDAPnum} + 0.0829x_{MaxLDAP}x_{DispNum} + 0.0431x_{LDAPnum}x_{DispNum} \end{aligned}$$

**Figure 5.22: Model from follow-up experiment, before backward elimination**

$$\begin{aligned} SNR = 0.018 &+ 0.0869x_{TNE} - 0.565x_{MaxLDAP} - 0.197x_{LDAPnum} + 0.217x_{DispNum} \\ &- 0.0308x_{DispNum}^2 - 0.151x_{MaxLDAP}x_{LDAPnum} + 0.0829x_{MaxLDAP}x_{DispNum} \\ &+ 0.0431x_{LDAPnum}x_{DispNum} \end{aligned}$$

**Figure 5.23: Model of DC-Directory from follow-up experiment**

| Factor | Level |
|---|---|
| `TNE` | 50,000 |
| `MaxLDAP` | 100 |
| `LDAPnum` | 4 |
| `DispNum` | 6 |

**Figure 5.24: Predicted optimal combination**

Using the validation set from 5.2.4.3, the *rank correlation coefficient* for this model's predictions was 0.68 (i.e. 84% of the pairs of combinations have the same predicted and observed ranking). This improved on the previous first phase model in Figure 5.15, which had a rank correlation coefficient of 0.62. This suggests that inclusion of all significant two-factor interaction effects is a necessary condition for accurate interpolation. However, it is not a sufficient condition: for a simple model to accurately describe the response surface, the relationship between factors and the response variable must be simple.

#### 5.2.4.5 Relationships between factors and the response variable

In the first phase experiments of sections 5.2.4.1 and 5.2.4.4, it was assumed that a linear or quadratic model could accurately describe the main effect of a factor. The follow-up experiments described here tested the validity of this assumption.

In each of the following experiments a single factor was varied, while the other three factors were fixed at their middle levels (as listed in Figure 5.11); testing more than three levels per factor gave a more detailed view of the factors' effects. As before, DC-Directory was warmed by ten clients that each sent addressing requests for ten minutes, plus a subsequent 10,000 lookup operations per client. Trials were replicated four times.

The graphs in Figure 5.25 to Figure 5.28 show the main effects of `TNE`, `MaxLDAP`, `LDAPnum` and `DispNum` respectively. Graphs on the left show the replicated measurements of throughput, while graphs on the right show the SNR calculated using these values.

Lines in the SNR graphs show the SNR predicted by the first phase model in Figure 5.23 – data points in the graphs below were not used to produce the model. The solid lines show predictions for combinations in the tested region of the response surface, while the dotted lines show extrapolation from the model. A straight line indicates that, when producing the model, there was no statistical evidence at the 5% level that the quadratic term in the main effect was significant.

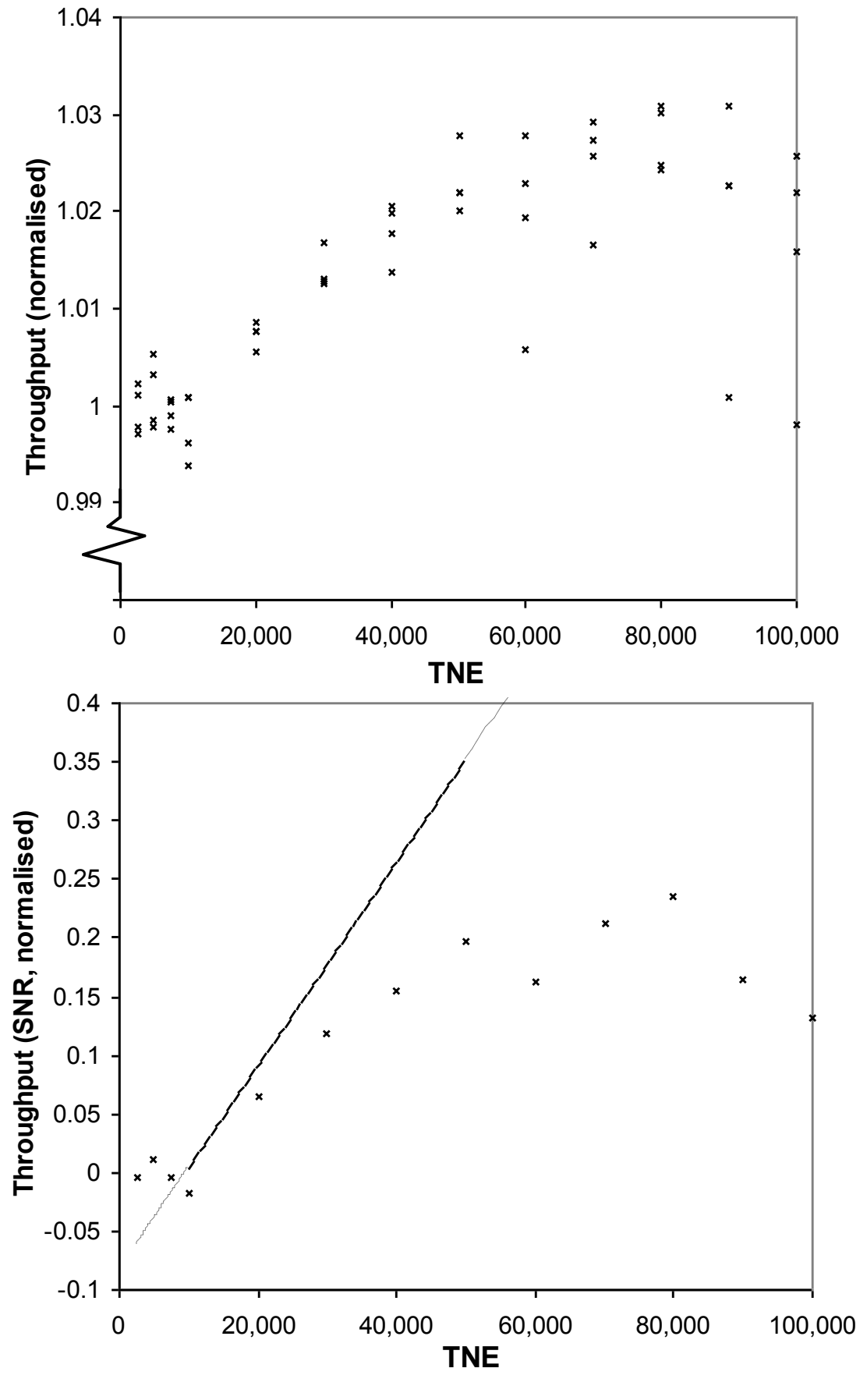

**Figure 5.25: Main effect of `TNE`**

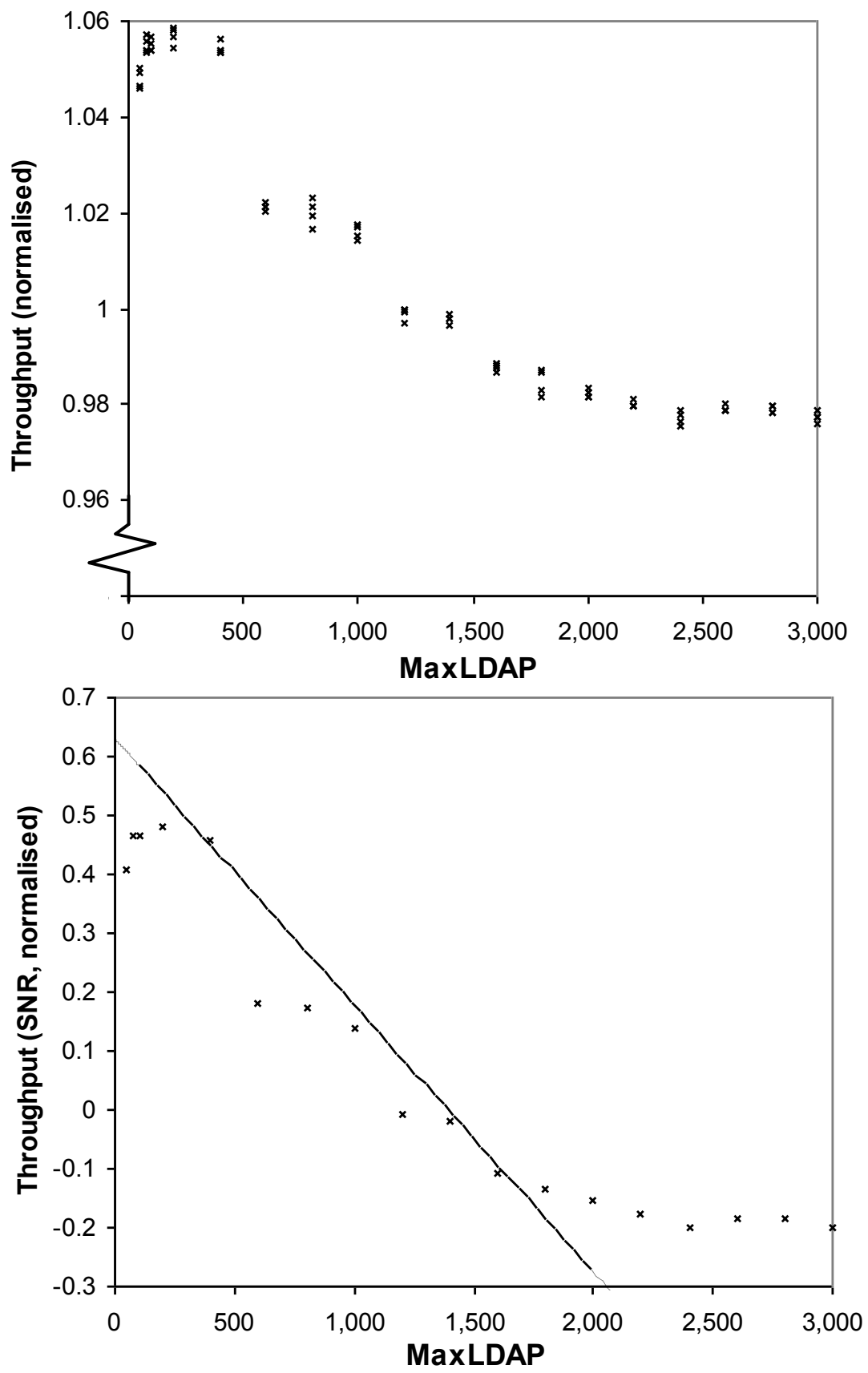

**Figure 5.26: Main effect of `MaxLDAP`**

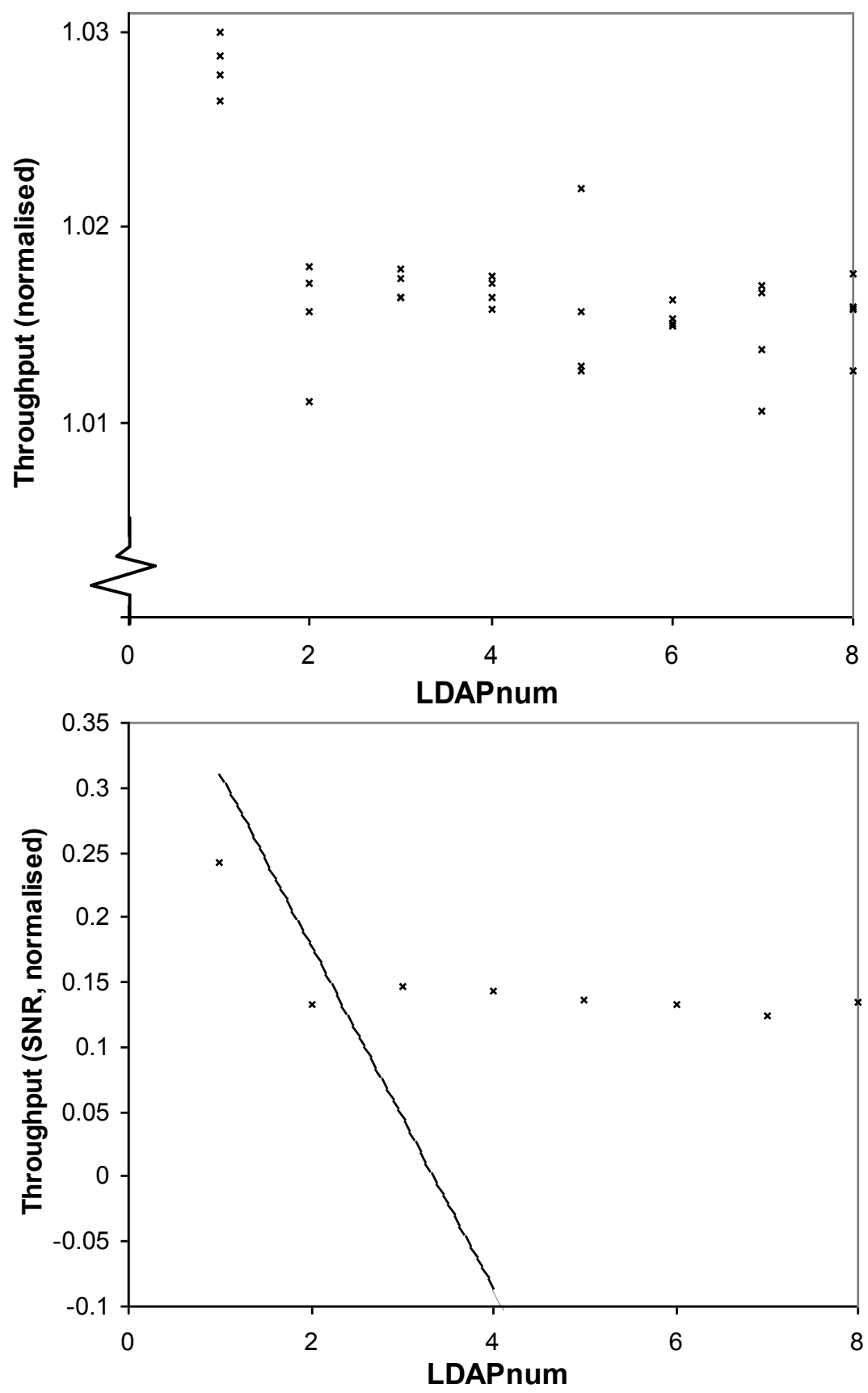

**Figure 5.27: Main effect of `LDAPnum`**

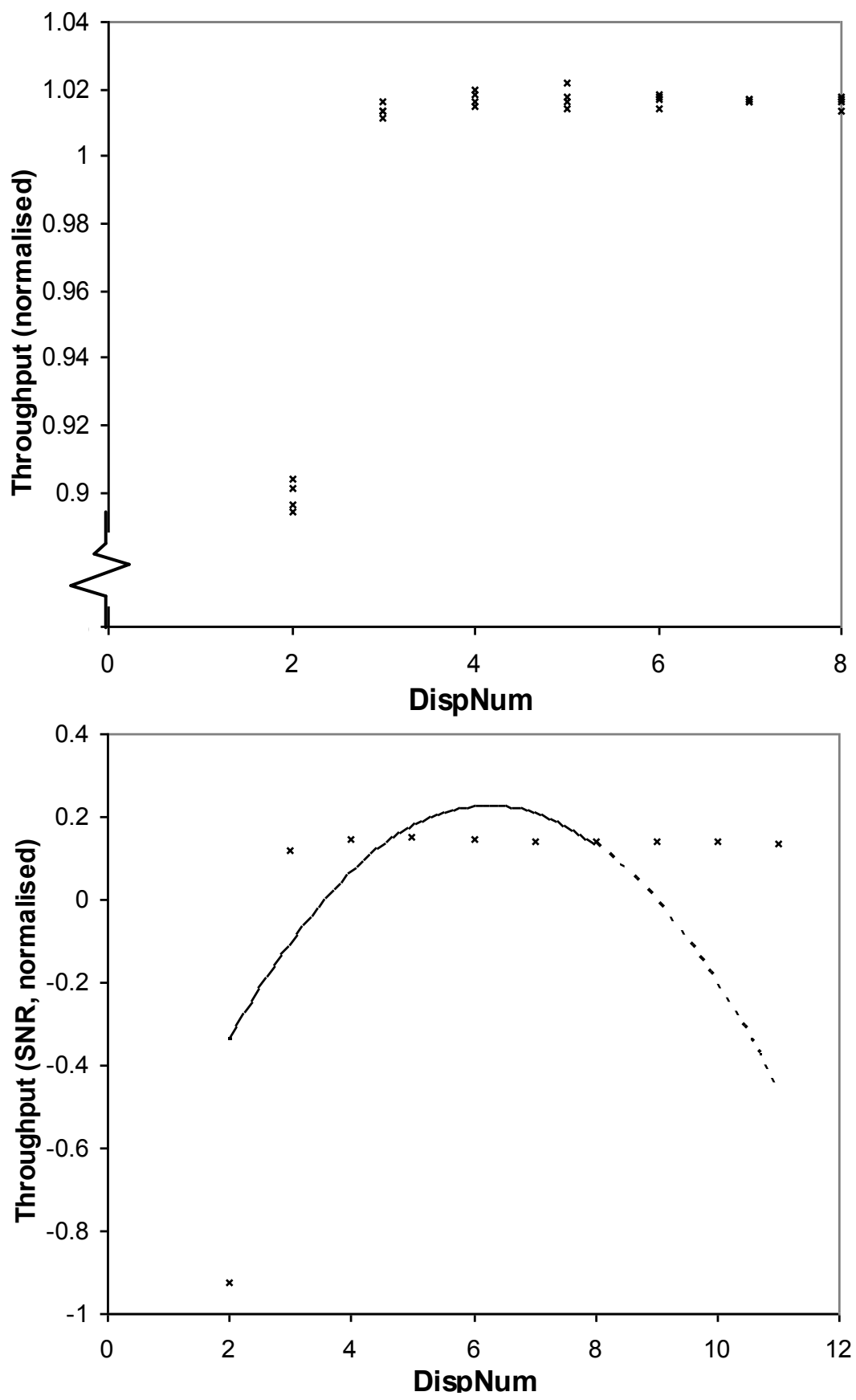

**Figure 5.28: Main effect of `DispNum`**

There were differences between the model's predictions and the observed SNR shown in the graphs above. This was due to simplifying assumptions made when producing the model: it was assumed that the main effects were linear or quadratic. It was also assumed that some interaction effects were zero (e.g. those involving three or more factors) and could therefore be ignored. Some of these interaction effects were aliased with main effects: if one or more of these interaction effects were non-zero, it would have affected the estimate of the main effect.

The graphs show that the main effect cannot always be accurately modelled by a quadratic polynomial. For example, the main effect of `DispNum` (when the other factors are set to their middle levels) would have been more accurately modelled by a piecewise curve: the effect of increasing `DispNum`'s level from 2 to 3 was different from the effect of increasing `DispNum`'s level in the range 3 to 12.

The main purpose of models from first phase experiments is to predict a combination that is near the optimal in the tested region of the response surface. This does not require that the model accurately interpolates the response variable's values for other combinations. The maxima for the solid lines in Figure 5.25 to Figure 5.28 are close to the highest observed SNR (e.g. the solid line in Figure 5.27 correctly predicts that the highest SNR will be when `LDAPnum` has level 1, despite the line being a poor fit to the data). This suggests that the model in Figure 5.23 is sufficient for estimating the optimal levels of `TNE`, `MaxLDAP`, `LDAPnum` and `DispNum`, based on their main effects.

Investigation of lower levels of `MaxLDAP` (i.e. less than 50) revealed a phase change, shown in Figure 5.29. There was a five fold improvement in throughput when `MaxLDAP` had the level 10, compared to throughput when `MaxLDAP` had a level between 14 and 3000. This result shows the importance of choosing the correct range of levels to test when designing an experiment. It also demonstrates that extrapolating from a model can be inaccurate: the model in Figure 5.23 does not predict a phase change. Predictions should only be made for combinations within the region tested.

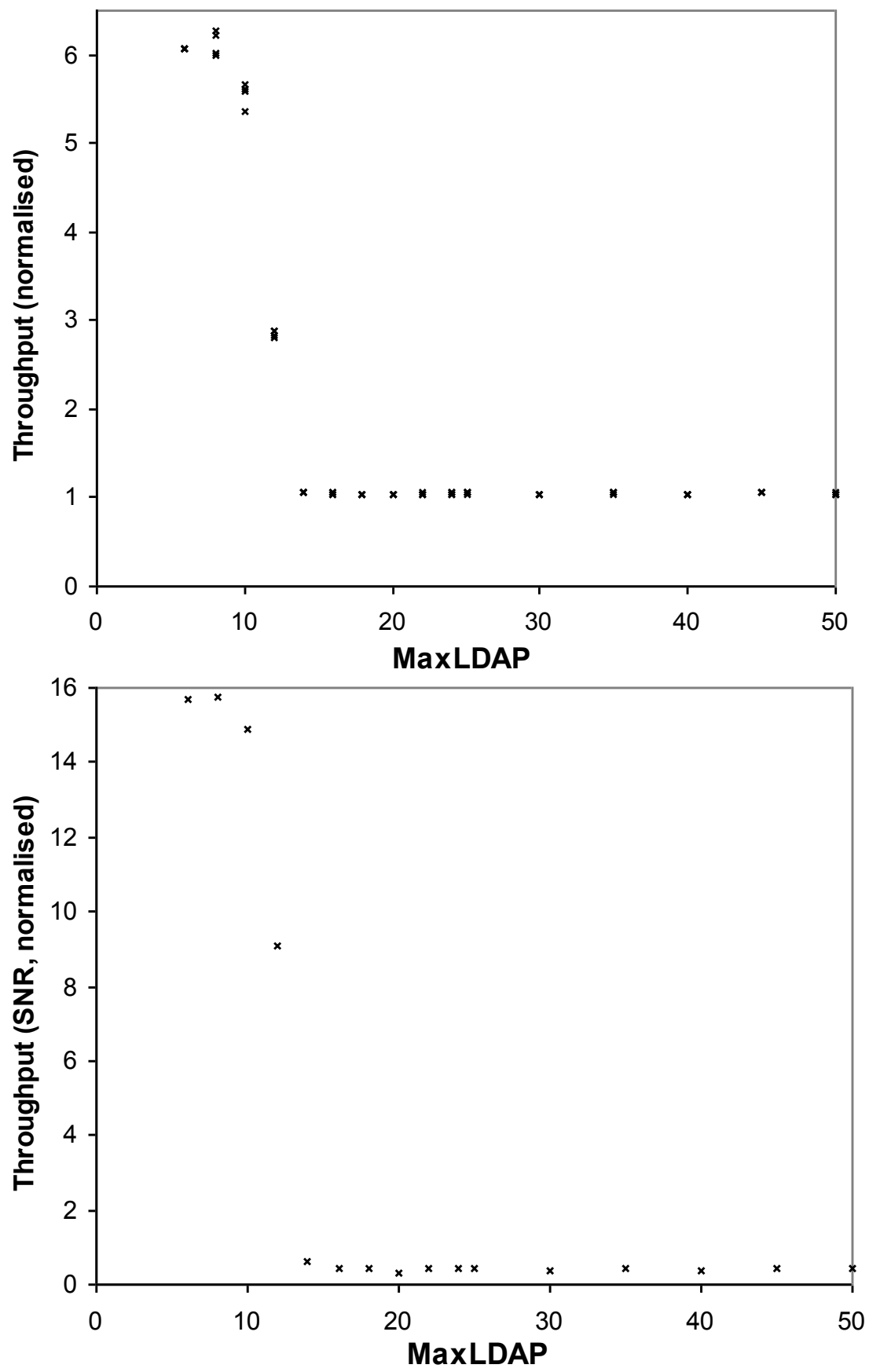

**Figure 5.29: Main effect of `MaxLDAP`, showing phase change**

## 5.3 Conclusions

This chapter presented experiments with the target systems DC-MailServer and DC-Directory. Experiments with DC-MailServer demonstrated the difficulties posed by variability in the behaviour of complex software systems, highlighting the importance of replicating trials. Experiments with DC-Directory included using Taguchi Methods to identify high-quality combinations. Figure 5.30 summarises these results, which are discussed below.

| Combination | Factor levels | | | | Observed throughput (normalised) | | | |
|---|---|---|---|---|---|---|---|---|
| | `TNE` | `MaxLDAP` | `LDAPnum` | `DispNum` | SNR | Min | Median | Max |
| Default | 10,000 | 1,001 | 2 | 5 | 0.004 | 0.9990 | 1.0000 | 1.0030 |
| 1st prediction | 32,930 | 100 | 1 | 5 | 0.461 | 1.0526 | 1.0542 | 1.0569 |
| 2nd prediction | 58,720 | 89 | 1 | 5 | 0.503 | 1.0570 | 1.0599 | 1.0617 |
| Best so far | 50,000 | 100 | 1 | 8 | 0.536 | 1.0623 | 1.0633 | 1.0656 |
| 3rd prediction | 50,000 | 100 | 4 | 6 | 0.541 | 1.0628 | 1.0648 | 1.0648 |

**Figure 5.30: Summary of results from using Taguchi Methods with DC-Directory**

In Figure 5.30, "default" refers to the default configuration used when DC-Directory is installed. The "1st prediction" and "2nd prediction" are the predicted optimal combinations according to the models from the first phase and second phase experiments respectively (see sections 5.2.4.1 and 5.2.4.2). "Best so far" refers to the best combination found during these experiments. The "3rd prediction" is the predicted optimal combination according to the model from the follow-up experiment in which all significant two-factor interactions were investigated (see section 5.2.4.4).

The best combination found gives a clear improvement in SNR over the default combination. It gives a 6% improvement in the minimum, median and maximum throughput. This is an indication of success: most software manufacturers, including Data Connection Ltd, would be happy to achieve a 6% improvement in performance without investing experts' time [99].

The predicted optimal combinations from the first and second phases are both better than the default. However, the best combination found during these experiments was one of those initially tested. It gives a median throughput 0.3% higher than that of the predicted optimal combination. This combination could have been chosen by simply ranking the configurations tested, without need of a predictive model. However, the predicted optimal combination from the model in section 5.2.4.4, labelled "3rd prediction", is of higher quality than those found previously. This high-quality combination would have been difficult to identify without a predictive model, which validates use of predictive models when searching the response surface. The difference between the 1st and 3rd predictions highlights the importance of designing experiments well (e.g. including all significant interaction effects).

The follow-up experiments in section 5.2.4.5 show that simple models, like those from first phase experiments, cannot always accurately model the main effects of factors. However, results for DC-Directory suggest that they are sufficient for predicting a combination near the optimal in the tested region of the response surface. This validates the use of models from first phase experiments in determining the centre for second phase experiments. This in turn supports the hypothesis that Taguchi Methods are useful for identifying high-quality configurations.

# 6 Discussion

## 6.1 Testing the hypothesis

This section discusses the usefulness of automatically measuring a sequence of configurations, as proposed in the hypothesis (see section 1.1). The goals discussed below are used as a basis for testing the hypothesis:

- **Find a robust configuration**. The experiments described in section 5.2.4 identified a high quality configuration of DC-Directory (in terms of robust throughput). This configuration consistently delivered 6% higher throughput than the default, under the conditions described in the case study.
- **Find characteristics of interest**. One of the experiments, described in section 5.2.4.5, discovered a phase change. Other observations were also of interest to developers, such as the effects of `TNE` on the variability in DC-Directory's behaviour (see section 5.2.2).
- **Help to construct a predictive model of the target system**. The model in Figure 5.23 was derived from experiment results. It was used to predict an "optimal" combination for DC-Directory, which was of higher quality than the other combinations tested.

## 6.2 Use of Taguchi Methods

Below is a discussion of the assumptions made in using Taguchi Methods, and their validity when designing experiments for DC-Directory:

- **The experimenter knows which factors to vary**. The software manufacturer recommended with confidence a subset of the factors on which to concentrate, based on experience of tuning DC-Directory by hand.
- **The experimenter knows which levels to test**. This is non-trivial as some factors have thousands of legal levels (e.g. `TNE` for DC-Directory): deciding on levels to test is sometimes informed guess work. The software manufacturer recommended ranges of levels for DC-Directory's factors, but further experiments are required to investigate behaviour of combinations outside this range, to validate the original choice of levels.
- **The experimenter knows which interactions to investigate**. This is sometimes intuitive: the software manufacturer has suggested several significant interactions. However, the accuracy of a predictive model is influenced by whether all significant interaction effects are investigated. Experiments that investigate a large number of interaction effects are expensive so it is important to identify insignificant interactions that can be ignored. This requires *a priori* information about the target system's behaviour, which may be supplied in some cases by the software manufacturer.

- **The more factors involved in an interaction effect, the less likely it is to be important**. This does not always hold for complex software systems:
  - A *bottleneck* can be a significant interaction effect involving multiple factors; it can correspond to a set of factors whose levels limit the target system's performance. Changing these factors' levels removes the bottleneck, but varying other factors not involved in the bottleneck will not improve performance until the bottleneck is removed.
  - The field of *hazard analysis* [77] has revealed that a *hazard* may be due to an interaction effect involving multiple factors. A "hazard" is a (potentially large) set of *conditions* that, when all are satisfied, will lead inevitably to an accident (e.g. deterioration in performance). The "conditions" could be particular levels for a set of factors.
- **Models from first phase experiments can predict a combination that is near the optimal in the tested region of the response surface**. The optimal levels predicted by the model of DC-Directory in Figure 5.23 are close to the optimal levels observed in section 5.2.4.5, which describes an investigation of the main effects. This is true for DC-Directory despite some of its main effects being impossible to model accurately using linear and quadratic terms.

Taguchi Methods can provide insight into a target system's behaviour. Results for DC-Directory were used to produce predictive models, revealed a high-quality configuration and formed the basis for designing follow-up experiments (see section 5.2.4).

It requires expert knowledge to successfully design experiments using Taguchi Methods. The region of the response surface that should be tested depends on the configuration goal and the likely conditions of use. It is therefore important to choose well the factors to vary and the range of levels for each. To produce models that accurately predict high-quality combinations requires that the experimenter investigate (at least) all significant two-factor interaction effects.

The target system's response surface must be simple for models to accurately interpolate the response of untested combinations: use of a simple model assumes that relationships between factors and the response variable are linear or quadratic. This assumption may not hold for a given target system, so predictions of the effects of adapting factors should be used as guides and be treated with caution.

Simple models of the response surface will not predict *phase changes*. The assumption that the response surface is simple therefore makes it hard to detect phase changes. However, use of Taguchi Methods can give clues:

- Presence of significant *outliers* with respect to the linear regression model (i.e. observations that do not fit the model) suggests that the simple model is insufficient for describing the response surface. This may indicate the presence of a phase change.
- It could be argued that combinations near some phase changes give high variability in performance. Phase changes correspond to sudden shifts in the way the target system behaves (e.g. a small increase in workload can cause a dramatic increase in the rate of growth of an input queue). The point in the input space at which the phase change occurs may not be fixed (e.g. the critical workload level may depend on the level of uncontrolled factors). Therefore, performance of combinations near a phase change may be more variable than those away from phase changes.

The experiments in section 5.2.2 demonstrate the necessity of replicating measurements. For some levels of `TNE`, throughput varied by over 40%.

Other experiments, for a previous version of DC-Directory, revealed significant interaction effects due to bottlenecks. Four factors were varied, relating to the number of threads that service particular queues of requests, including LDAP requests (`LDAPnum`) and search requests (`DispNum`). DC-Directory's throughput was the same for all combinations tested: none of the factors varied had a significant effect on throughput, under the conditions described in the case study and with all other factors set to their default levels. This illustrates two important points: effects of factors can depend on bottlenecks caused by other factors' levels, and DC-Directory's behaviour varies between versions.

## 6.3 Complementing on-the-fly adaptation

Running experiments before the target system goes into use has a number of limitations compared to configuring the target system on-the-fly:

- It is only feasible if the target system can be taken off-line, or for changes that can be planned and tested before the target system first goes into use.
- Some configurations are only available on-the-fly, such as those involving new versions of a component.
- Replicating faithfully a customer's environment can be difficult, particularly for distributed heterogeneous systems operating over wide area networks. Conditions may also change after the system goes into use, which can make observations from past experiments inapplicable.
- Producing a workload that is representative of the customer's likely usage patterns can be difficult. Logging a customer's usage, where possible, allows an identical set of inputs to be used during experiments, or a synthetic workload to be developed based on characteristics of the logged input. An alternative is to use domain-specific benchmarks that describe common usage patterns.

Despite these limitations, exploring the behaviour of a target system before it goes into use can be beneficial and can complement on-the-fly adaptation. Advantages of exploring behaviour before the target system goes into use, over exploring behaviour on-the-fly, include the following:

- Each trial can be run in a consistent manner, setting explicitly the state of the target system and its conditions of use. This is important for estimating robustness and for accurate comparison of target system configurations.
- Statistical techniques, such as Taguchi Methods, can be used for the design of experiments as there is explicit control over the conditions for each trial.
- Testing of specific conditions allows experimenters to:
    - validate adaptation tactics for on-the-fly adaptation;
    - find appropriate configurations for particular conditions;
    - suggest new adaptation tactics.

- Detrimental effects on performance (either temporary or permanent) are not a problem as there are no real users and the target system is restored to a known state before each trial: the target system can be restarted or even reinstalled as required.
- There are no ill effects from testing configurations that are invalid or that exhibit incorrect behaviour, assuming that errors are detected and reported during the trials. This assumes that the target system is isolated from the outside world when running experiments so errors cannot affect external systems.
- There is generally more time available than when configuring the target system on-the-fly, so the experimenter can test a wider range of configurations.

## 6.4 Experimental adaptations on-the-fly

Lehman states that "an E-type program that is used [i.e. a target system embedded in a real-world domain] must be continually adapted else it becomes progressively less satisfactory" [74]. If the target system cannot be taken off-line, this necessitates on-the-fly adaptation. A new version of ACT has been designed to support on-the-fly evolution of software systems (see section 7.3).

If an appropriate adaptation is not known and the target system cannot be taken off-line, it is necessary to make experimental adaptations while the target system is in use. However, it is hard to measure the effect of an adaptation on-the-fly because uncontrolled factors can affect the target system's behaviour: it is not possible to run experiments in a controlled and isolated environment. For example, the workload may change, components may fail and the execution platform may perform unsolicited actions such as garbage collection.

The SNR metric can be used to measure the robustness of a target system's behaviour on-the-fly. Repeated observations of behaviour give an indication of the target system's consistency and its insensitivity to variation in uncontrolled factors. However, repeating observations necessitates a delay in responding to deterioration in performance.

On-the-fly adaptation is potentially risky due to the disruption it can cause to users (e.g. deterioration in performance, loss of data or system failure). Techniques for allowing adaptations to be reversed include: use of a check-pointing mechanism to roll back to a known state [62], and adapting a duplicate component instead of the original. However, some adaptations are irreversible at run-time (e.g. involving I/O, causing space leaks, changing data that is not backed up, or affecting the encapsulated state of an external component) so any disruption they cause is permanent.

The set of available adaptations is influenced by:

- the set of adaptation mechanisms that can be used on-the-fly;
- the target system administrator's policy on the acceptable level of risk;
- any available predictions of adaptation costs (i.e. disruption caused while adapting), chance of success (i.e. chance of improvement versus chance of a detrimental effect) or implications of a detrimental effect.

"Careful" experimental adaptations made on-the-fly may reveal good configurations. Iterative improvement algorithms (see section 4.2) may prove appropriate for making a series of small adaptations to the target system's configuration, using observations of the behaviour to decide whether to accept or reverse each adaptation.

# 7 Future work

## 7.1 Further experiments

Current work focuses on the use of Taguchi Methods to explore DC-Directory's behaviour. Further experiments are discussed below, which would be run if this research were to be continued.

Taguchi Methods could be used to design experiments involving dozens of factors. Additional factors of DC-Directory could include:

- aspects of the workload (e.g. number of clients, ratio of *addressing* to *modify* requests, etc);
- concurrency policies (e.g. number of threads and thread priorities);
- distribution policies for deploying DC-Directory across multiple nodes of the Beowulf cluster.

The set of fitness metrics could be increased to improve understanding of the target system's behaviour. Current experiments collect measurements of throughput, latency[32] and resource usage. However, only throughput is used in the configuration goal. Future work is required to expand the set of fitness metrics and produce appropriate aggregating functions, based on the advice of target system experts (i.e. administrators and developers).

Additional experiments are required to further investigate the suitability of multiple linear regression for modelling DC-Directory, and to explore the use of other curve fitting techniques (e.g. splines) for modelling non-linear or discontinuous response surfaces. Data mining techniques could also be used to extract information from experiment results [50]. Feedback from experts could be used to validate models and to suggest improvements such as other factors to include and other levels to test. It would be interesting to further investigate the usefulness to developers and administrators of predictive models, and to investigate how closely models mirror current understanding of the target system.

Once the techniques have been successfully applied to DC-Directory, Taguchi Methods could be used to design experiments for DC-MailServer. The case study described in section 5.1, which did not use Taguchi Methods, revealed that coping with variability in behaviour is a big problem. Taguchi Methods could be used to cope with this variability:

- Trials could be replicated for each combination. SNR could be used to measure consistency and thus estimate each combination's insensitivity to variation in uncontrolled factors.
- The TACT process (i.e. the combination of Taguchi Methods and ACT) could be used to search for combinations that give high values of SNR.
- Taguchi's *robust design* could be used to identify target system configurations that are insensitive to variation in noise factors (see section 4.3.6). This would involve explicitly varying the target system's conditions of use, to identify and exploit the interaction effects between control factors and noise factors.

[32] DirectoryMark measures the latency for each addressing request. Measurements obtained during a single trial could be combined using SNR to reward consistently low latency.

The behaviour of other software systems could be explored to investigate further the applicability of Taguchi Methods and other search strategies. However, ACT's use is not restricted to just software systems: ACT could coordinate experiments for any process that can be run, measured and configured without human intervention.

## 7.2 Complementing other work

Use of ACT could be beneficial for other research programmes that assume appropriate target system configurations are known *a priori*. Current plans are to integrate ACT into ArchWare [4]. Other programmes that could make use of ACT include DASADA [9] and Autonomic Computing [63] (see section 2.8.3 for a description of these programmes).

Another possible avenue of further work is in the integration of ACT with *Generative Programming*: "a software engineering paradigm based on modelling software system families such that, given a particular requirements specification, a highly customised and optimised intermediate or end-product can be automatically manufactured on demand from elementary, reusable implementation components by means of configuration knowledge" [41]. The user's requirements (i.e. configuration goal) drive the assembly of the software system. A *framework* for a family of software systems has *variation points* into which components are plugged to produce a complete system. Each variation point corresponds to a factor, and each valid component is a possible level. ACT could run an experiment to help find a suitable combination: to help identify and configure appropriate components for each variation point.

## 7.3 ACT 2.0

A new version of ACT, ACT 2.0, has been designed with four main aims:

- **Extend the use of ACT 1.0 to support long-lived evolution processes**. ACT 2.0 will help make target adaptations to produce new and improved versions of the target system. In ACT 1.0, this is left to the experimenter and target system developers.
- **Provide support for on-the-fly adaptation of target systems**. ACT 2.0 will assist target system developers to continually adapt target systems throughout their lifetimes.
- **Provide a more flexible infrastructure**. Use of an event-based architecture will impose a looser coupling between ACT 2.0 components. This will make it easier to incorporate third party components, such as gauges developed under the DASADA programme.
- **Make explicit the policies, mechanisms and information in the ACT 2.0 infrastructure**. All information will be explicitly available when deciding on policies for using exposed mechanisms. This will make it easier to guide and tailor the evolution process. This contrasts with ACT 1.0, where probes and gauges used by a *run function* were all private to that function.

Observation of a target system will help drive its evolution. ACT 2.0 will support long-lived evolution processes that involve multiple experiments, changing configuration goals, and target adaptations. Experiments will identify beneficial adaptations, evaluate new versions of the target system and help focus the testing and maintenance efforts of target system developers. This work will contribute to the ArchWare project by helping to recommend architectural changes. The ArchWare ADL will be used to describe adaptations and the ArchWare environment will support adaptation of the target system [4].

ACT 2.0 will use observation and adaptation mechanisms available at run-time to evolve the target system on-the-fly. An *evolution strategy* will decide on appropriate adaptations, making use of the following mechanisms:

- ***Gauge* components**. Gauges will interpret observations of the target system and will generate events to describe the target system at a higher level.

- ***Advice* components**. These will encode expert knowledge of the target system's behaviour, which includes adaptation tactics, predictions of patterns of behaviour, and the effects on fitness metrics of factors and interactions.
- ***Model* components**. A *model* component will contain a model of the target system (e.g. a description of the architecture using an ADL). It will use the model as a basis for interpreting observations, reasoning about possible adaptations and recommending appropriate configurations.
- ***Search strategy* components**. Experiments will be run both before the target system goes into use and on-the-fly to search for beneficial adaptations. The latter could be useful as a "last resort", when degradation necessitates change but an appropriate adaptation tactic is not known (see the discussion in section 6.4).

The event-based architecture of ACT 2.0 will provide a more flexible infrastructure, imposing looser coupling between components. This will aid reuse of observation, adaptation and interpretation mechanisms.

Fundamental to ACT 2.0's design is the explicit representation of policies, mechanisms and information. Following the principles of *Compliant Systems Architecture* [83], the aim is to make explicit the mechanisms exposed by ACT 2.0, to define precisely the extent to which it is open. Decisions about how to use the mechanisms exposed by a component could be made by other components, promoting a separation of concerns between policy and mechanism. For example, an evolution strategy will decide how to use a target system's adaptation and observation mechanisms. Entities that represent information explicitly will include:

- events, which will make communicated information explicit and available to all eligible ACT 2.0 components;
- *advice* components, which will support the capture, evolution, use and reuse of expert knowledge – in ACT 1.0, advice is implicit in the choice of factors, levels and search strategy.

### 7.3.1 Event-based architecture

ACT 2.0 components will communicate over an event bus, with events encoded in XML to promote openness. The Siena content-based routing system [106] will be used for dissemination of events, which is similar to the approach described in [61]. The format of messages will conform to the *Smart Events Schema* [6] to promote interoperability with probes and gauges developed in the DASADA programme.

A *target wrapper* component will be present on each of the target system's nodes, to receive and send events. Incoming events will either be *infrastructure events* for the probes, or events that request adaptation or control of the target system. Each *target wrapper* will have an *event handler* to interpret incoming events. These events will be mapped to low-level operations on the probes or on the target system. *Target wrappers* will also generate events that represent the probes' observations and the state of ACT 2.0 components.

### 7.3.2 Evolution strategies

An *evolution strategy* component will coordinate the observation and adaptation of a target system. It will be target-independent (i.e. no inbuilt knowledge of the target system). It will use *gauge*, *advice* and *model* components as helpers to provide semantic information about the target system. These helpers, along with *search strategy* components (see chapter 4), will assist decision-making. They will be queried and/or will proactively generate events to interpret observations, recommend when and what to measure, and suggest when and what to adapt.

*Evolution strategies* will use a *component factory* to instantiate helper components appropriate to a target system and configuration goal. Target system administrators will recommend specific helper components, which will themselves instantiate other components.

For simplicity of initial implementation, *evolution strategies* will maximise the output of a single *gauge* that indicates how well the target system meets a particular configuration goal. The ability to maximise a *gauge*'s output would be powerful: the *gauge*'s implementation, and thus the configuration goal, could be arbitrarily complex. Adapting the gauge or changing which gauge is used would change the configuration goal.

A *meta-strategy* will control the choice, instantiation and configuration of an *evolution strategy*. If there is more than one *evolution strategy*, the *meta-strategy* will orchestrate their concurrent usage to ensure that conflicting instructions are resolved. For example, one evolution strategy could request frequent adaptations when attempting to improve the target system's performance, while a second evolution strategy with higher priority could request infrequent adaptations to preserve the target system's correctness.

### 7.3.3 Use of advice

*Advice* components will contain specific pieces of information about the target system. They will provide recommendations of how and when to observe and adapt the target system. An *advice* component could include state information about recent observations of the target system, or could be stateless. To use an advice component, it will not be necessary to have knowledge of the target system's architecture or an understanding of why advised actions are appropriate.

Making *advice* components explicit will aid automated generation of advice by providing an infrastructure in which information is represented and managed. ACT 2.0 will automatically generate *advice* components to encode information gleaned from past observations of the target system.

Figure 7.1 shows an example of advice for DC-MailServer[33], in the form of two *event-condition-action* rules. The $Advice_{MS}$ component will consume events concerning either throughput or the length of the IMS queue, which contains unprocessed e-mail messages. If throughput drops below a threshold of 100 messages per second, it will generate an event to recommend the activation and querying of a probe ($Probe_{imsQ}$) to measure the length of the IMS queue. If the IMS queue length exceeds a threshold of 200, it will generate an event to recommend that a new *message store* component be created.

| $Advice_{MS}$ | **Rule 1** | **Rule 2** |
|---|---|---|
| **Event** | Observed throughput | Observed IMS queue length |
| **Condition** | Throughput < 100mps & $Probe_{imsQ}$ not active | IMS queue length > 200 |
| **Action** | Activate and query $Probe_{imsQ}$ to measure IMS queue length | Create new *message store* component |

**Figure 7.1: Example rules for an advice component**

### 7.3.4 Use of models

A model is a representation that exhibits some property of the target system (see section 2.3). Models are important for successful on-the-fly evolution of a complex software system: they can form the basis for deciding on adaptations by providing a context for reasoning about observations and possible adaptations. Advice and gauges are special kinds of models: advice is a piece of information about the target system (i.e. an incomplete model), and a gauge interprets observations in the context of some property of the target system.

[33] This example *advice* component is simplistic, ignoring other explanations for a large IMS queue. There would need to be additional *advice* components to check for other causes, to decide when to decrement the number of *message store* instances, etc.

Models can include *constraints*, which can be used to guide observation and adaptation of the target system. Two types of constraint are *structural* and *behavioural*. An example of a structural constraint is that DC-Directory must have at least two *dispatcher search threads*. An example of a behavioural constraint for DC-Directory is that latency for addressing requests must be less than 100ms. Absence of constraint violations can be used to validate whether a target system configuration is legal and whether it would meet the configuration goal.

*Model* components will contain and use models to recommend how and when to observe and adapt the target system. *Model* components will:

- register to receive appropriate events, particularly those events pertinent to the target system properties being modelled;
- generate events that describe a property of the target system, and/or that recommend what to observe and adapt;
- instantiate and/or communicate with other ACT 2.0 components.

A set of *model* components will together form a (potentially distributed) model of the target system, where each *model* component contributes some details to the overall model. Each *model* component will represent *aspects* (i.e. sets of related properties) of the target system, where the representation is tailored to suit the aspects being modelled. Collaboration between *model* components is a challenging area requiring further work. A *weaver*, similar to those used in *aspect-oriented programming* [70], could perhaps combine the models to form the basis of a cohesive evolution strategy.

### 7.3.5 Challenges

There are a number of challenges involved in the design and implementation of ACT 2.0. Appendix B contains a list of general issues relating to the feasibility of observation-driven evolution and the policy decisions involved. Some design issues for ACT 2.0 are raised below:

- What will be the format of the *experiment description* (i.e. the input to ACT 2.0)? It should include target-independent information for the *evolution strategies* and *search strategies*, and target-specific information about how to interact with the target system. The extensibility of XML is well suited to representing target-specific information.
- How will events be mapped to operations on the probes and target system? A balance is required between keeping event handling code reusable and keeping the implementation of operations simple.
- How will multiple adaptations be coordinated, particularly when adapting interdependent distributed components [102]? Should adaptations be transactional with the ACID properties of *atomicity*, *consistency*, *isolation* and *durability* [62]? An open research issue is if/when these properties are required, and how they can be provided.
- What will be the interfaces for using *gauge*, *advice*, *model* and *search strategy* components? How will a database of past observations be maintained and accessed?
- What discovery mechanism will be used for identifying available ACT 2.0 components? It is envisaged that the *component factory* will maintain a list of available components, and that some components will hard-code details of other components deemed appropriate.
- How will ACT 2.0 components be configured? A *meta-object* for each component will control the configurable aspects of the component's behaviour, and respond to queries about the services the component provides.

## 7.4 Further versions of ACT

If work on ACT were to be continued, a long-term goal would be to develop future versions of ACT that drive and coordinate all agents involved in the evolution process. Fundamental to this would be the development of evolution strategies for various activities such as performance tuning, porting to new environments, and adding functionality. Observation, decision-making and adaptation would be performed either automatically or by external agents, such as target system developers implementing a new version of a component to meet a given specification. ACT would provide a framework for these tasks, to produce new versions of a target system and evaluate their behaviour in terms of the configuration goal. Realising this vision would require at least integration of systems for:

- *process management* to coordinate agents involved in the evolution process, throughout the organisation;
- *requirements engineering* to specify explicitly and formally the configuration goals;
- *resource management* to allocate resources required by the running system and additional resources required during adaptation;
- *architectural modelling* to maintain and evolve the high-level architecture of the target system;
- *configuration management* to manage changes made to the target system, in particular new versions of components and new architectural configurations;
- *deployment* to install, upgrade and activate components;
- *testing* to ensure the correctness of the target system during evolution by using techniques such as regression testing and code coverage.

## 7.5 Summary

The work described in this thesis has shown the feasibility of running automated experiments to assist target system administrators in comprehending and configuring software systems. In particular, use of Taguchi Methods for identifying robust configurations is a promising direction of research.

The applicability of Taguchi Methods for complex software systems requires further investigation. It is yet to be demonstrated whether the techniques scale for large numbers of factors, and whether the assumption of a simple model limits applicability in practice. ACT 1.0 could be used to investigate these issues by running more experiments for DC-Directory and DC-MailServer. This would require the continued cooperation of Data Connection Ltd (DCL) to validate results and evaluate their usefulness in an industrial context.

ACT 2.0, which is at the design stage, is intended to support long-lived evolution processes that involve multiple experiments, changing configuration goals and target adaptations. ACT 2.0 will coordinate observation and adaptation of the target system both before the target system goes into use and on-the-fly.

# 8 Conclusions

This thesis proposed the following hypothesis: automating the empirical measurement of a sequence of target system configurations can find robust configurations of the target system, can find characteristics of interest and can help construct a predictive model. This led to the novel use of Taguchi Methods for configuring software systems and the development of ACT (*Automated Configuration Tool*). The hypothesis was tested and validated in an industrial case study for DC-Directory.

ACT can be used to explore the behaviour of a wide variety of target systems using a variety of search strategies. It has been used in two industrial case studies with products from Data Connection Limited (DCL): DC-MailServer and DC-Directory. Experiments to-date have explored the effects on target system performance of caching policies, concurrency policies, number of processors and workloads. Quality was measured by comparing observations of the target system's behaviour against the configuration goal specified by the experimenter.

The combination of Taguchi Methods and ACT yielded the *TACT process*. This was used to design and run experiments in two phases, which produced models of the target system to identify *robust* configurations (i.e. configurations that would deliver consistently high performance). The first phase involved designing fractional factorial experiments to explore a region of the response surface specified by the experimenter. Multiple linear regression was used to produce a model, which predicted a combination near the optimal in the investigated region. The second phase tested combinations in a small region around the predicted optimal to produce a more accurate model of the response surface in that region. This second model was used to more accurately predict the optimal combination.

Use of the TACT process with DC-Directory revealed that the assumptions of Taguchi Methods were valid for DC-Directory. The combination predicted as optimal in the first phase experiment performed well, and was confirmed to be near a high-quality combination.

A configuration of DC-Directory was found that consistently delivered 6% higher throughput than the default configuration, under the conditions described in the case study. This is an indication of the technique's success: most software manufacturers, including DCL, would be happy to achieve a 6% improvement in performance without investing experts' time [99].

A secondary use of the predictive model for DC-Directory was to rank untested combinations in terms of quality, using estimates of the main effects and a selection of interaction effects. However, the model of DC-Directory did not always accurately interpolate the response variable's value for untested combinations. Instead, the model provided rules of thumb to predict the effects of varying the factors' levels.

It requires expert knowledge of the target system to successfully use Taguchi Methods: to choose which region of the response surface to investigate, to select levels for each factor, and to identify significant interaction effects to investigate. Target system developers and administrators can sometimes provide this information.

The experiments with DC-Directory have shown that the combination of Taguchi Methods and ACT, forming the *TACT process*, can be used to identify high-quality configurations. TACT is therefore a valuable addition to the techniques available for configuring software systems. It has been suggested that use of Taguchi Methods has saved other industries, such as car manufacturing, hundreds of millions of dollars by helping to produce robust products [2]. It is hoped that software engineers will adopt these methods and will obtain similar rewards, finding target system configurations that deliver consistently high performance.

In the future, the TACT process could be a common technique in software engineering. It could help to produce robust systems that are not affected by variation in uncontrolled factors. Target system developers could also use the TACT process to improve understanding of the target system's behaviour: to guide adaptations when progressing to new and improved versions of the target system.

The TACT process could become an integral part of software deployment. At the time of a target system's installation, the customer would indicate the expected workload. The target system would be tuned automatically to select a configuration appropriate for the customer's conditions. A tool like ACT 2.0 could also be deployed with the target system. It could monitor the target system on-the-fly, identify when and how the target system should adapt, and configure the target system as required.

The work presented here has demonstrated the usefulness of the TACT process. It has taken the first steps towards the vision of the future outlined above, where Taguchi Methods and ACT are used in a wide range of situations to improve the quality of software systems.

## Appendix A: Glossary

Below is a glossary of terms that relate to the design and use of ACT for exploring the behaviour of software systems, and statistical terminology relating to Design Of Experiments.

| | |
|---|---|
| ACT implementer: | The author of the **Automated Configuration Tool** (ACT); the programmer who produces the core of ACT, which provides an infrastructure for repeatedly running **trials**. |
| Adaptation: | A deliberate change to the target system's configuration or its conditions of use. |
| Adaptation, experimental: | Speculative adaptation of the target system's configuration, where the effect of the adaptation is not known in advance. |
| Adaptation function: | A function in the **target wrapper** that adapts the target system's configuration or its conditions of use. Each adaptation function sets the **level** for a particular **factor**, using an **adaptation mechanism**. |
| Adaptation, target: | An adaptation known to produce a desirable configuration. It involves adapting the target system T to progress to a new target system T' that is deemed desirable. |
| Adaptation tactic: | A rule for choosing an appropriate configuration of the target system, based on observations of the target system and its conditions of use. |
| Adaptation mechanism: | An entity offering the capability to adapt the target system's configuration or its conditions of use. It is a mechanism for setting the **level** of a **configurable aspect** or **usage aspect**. |
| Aliasing: | Where the **experimenter** cannot infer which of several **main effects** and/or **interaction effects** affected the **response**. The effects whose influence cannot be separated are said to be *aliased*. |
| Architecture: | "The fundamental organisation of a system embodied in its **components**, their relationships to each other, and to the |

environment, and the principles guiding its design and **evolution**" [1].

Architecture description language: A formal notation for describing the architecture of a software system. Typically identifies the **components** of the software system and the inter-component communication, defined by **connectors**. Abbreviated to ADL.

Automated Configuration Tool: An infrastructure to explore a target system's behaviour, without human intervention; it repeatedly configures and observes a target system under various conditions, using a **search strategy** to decide on the combinations to **test**. Abbreviated to ACT.

Black box system: A target system for which there is no knowledge of, or access to, its internal workings. Interaction with the target system is solely through the interfaces it exposes. Contrast with **white box systems**.

Combination: A target system configuration and a condition under which it is run; specifies the **level** for each **factor**.

Complex software system: A software system with **emergent properties** and/or **non-deterministic** behaviour. Complex software systems generally have many possible configurations, and it takes a long time to empirically measure each.

Component: A computational unit that forms part of the **target system**.

Condition: The **workload** that drives a target system, and the environment in which it runs.

Connector: A link between two or more **components**, across which they can interact.

Configurable aspect: An implementation detail, exposed by the target system, that can be set explicitly during an **experiment**. Referred to as a **factor** when varied during an experiment.

| | |
|---|---|
| Configuration: | Specifies a **level** for each **factor** of the target system, to describe a deployment of the target system. |
| Configuration goal: | Specifies behaviour desired of the target system in terms of a potentially conflicting set of **fitness metrics**. Observations of these metrics are combined using an aggregating function to produce a single **response** for each **trial**. |
| Configuration process: | The set of activities involved in adapting a target system's configuration to meet a **configuration goal**. It encompasses both **tuning** and **evolving** the target system, and can be done either before the system goes into use, or **on-the-fly**. |
| Customer: | The organisation (or individual) who will use the target system in a real-world situation, for whom the **configuration process** is performed. |
| Design Of Experiments: | A methodology for planning **experiments**, in which **main effects** and **interaction effects** can be inferred from the experiment results. Abbreviated to DOE. |
| Distribution: | The probabilities of an observation of the **response** having various values. |
| Effect, interaction: | An interaction effect between **factors**, say A and B (denoted AB), refers to the degree to which A's effect on the **response variable** depends on the **level** of B, and vice versa. This can cause the optimal level for a factor to vary, depending on the others' levels. An interaction effect between two factors is called a two-factor interaction effect. |
| Effect, main: | The effect on the response variable caused by adjusting the level of a single factor in isolation. |
| Emergent properties: | Behaviour of the whole system cannot be inferred from its parts. The system's behaviour is not the sum of its parts, but the product of interactions among its parts and the environment [104]. |
| Evolution: | The strategic adaptation of a target system's configuration over time, to progress to new and improved versions of the target system; it is the continuous process of identifying what to adapt and when, and configuring the target system accordingly. |
| Experiment: | The **testing** of a sequence of **combinations**, to empirically measure the behaviour of target system configurations under given conditions of use. |
| Experiment design: | The set of **combinations** to **test** and the number of **replications** for each combination. |
| Experimenter: | The user of ACT, responsible for configuring and invoking ACT for a particular target system. |
| Expert: | An expert in the details of the target system's operation. See **target system administrator** or **target system developer**. |
| Factor: | A **configurable aspect** of the target system, or a **usage aspect**. |
| Factor, uncontrolled: | A **factor** that is not set explicitly during an **experiment**, for cost or technical reasons. |
| Fitness metric: | A measure of the "goodness" of a target system's behaviour, in terms of a single characteristic (e.g. throughput). |
| Fractional factorial design: | An **experiment design** in which, given a list of **factors** and a list of **levels** for each factor, only a subset of the possible **combinations** is **tested**. |
| Full factorial design: | An **experiment design** in which, given a list of **factors** and a list of **levels** for each factor, every possible **combination** is **tested**. |
| Gauge: | An entity that gathers and interprets observations (made by **probes** or other gauges) in a context meaningful for evaluating behaviour (e.g. in terms of a **fitness metric**). |
| Input space: | A multi-dimensional space whose dimensions correspond to the **factors**. Every **combination** corresponds to a point in the space, giving a **level** for each factor. |
| Level: | A value of interest to the **experimenter**, to which a **factor** can be set. In this thesis, "level" means an uncoded level that corresponds to a factor's value. In contrast, a coded level is an index into an enumeration of values. |
| Linear graph: | A graph whose edges and vertices correspond to the columns of an **orthogonal array**. Used to facilitate the assignment of **factors**, and their **interactions**, to specific columns of an orthogonal array. |
| Loss function: | A measure of the cost incurred when the **response** deviates from the optimal. The response may be minimised, maximised or there may be a nominal value (e.g. "12" is best). |
| Meta-strategy: | A component of ACT; a special **search strategy** that dynamically binds to and uses other search strategy components. A meta-strategy can configure a search strategy and even dynamically switch between strategies. |
| Model: | A representation that exhibits some property (or properties) of a target system. |
| Non-determinism: | Where the amount of **variability** in a target system's behaviour does not meet the **replicability** demanded by the **customer**. |
| Observation: | A measurement. Values of measurements can have a scale type of *nominal*, *ordinal*, *interval* or *ratio*. These mean respectively that the values refer to categories, that the values are ordered, that the values increase in regular step sizes, or that there is a fixed zero point so that relations such as "twice the value" are meaningful. |
| Observation mechanism: | An entity offering the capability to observe an aspect of the target system or its conditions of use. |
| On-the-fly: | Refers to a running target system that is in use. |
| Orthogonal array: | A matrix representing the set of **combinations** to be **tested** in an **experiment**. Each row represents a combination and each column represents a **factor** and/or an interaction between factors. The |

matrix has the special property that every pair of columns includes every combination of **coded levels** an equal number of times.

Parameter design: Taguchi's technique for creating a **fractional factorial design** for an **experiment**.

Population: The set of all possible values of the **response variable** for **combinations** with a particular characteristic (e.g. a given **factor** at a particular **level**).

Probe: An entity that makes observations, possibly at run-time, by interacting with the target system and its environment.

Probe effect: Change in the target system's behaviour caused by the act of observing the target system.

Quality: A measure of how well a **combination** meets the **configuration goal**. Taguchi defines a high quality combination as one that imparts little loss to society from the time the target system is shipped. A **loss function** estimates the loss to society based on a single **response**: **replicated** measurements of the response can be used to estimate quality (e.g. using **signal to noise ratio**).

Recovery function: A function in the **target wrapper** that restores the target system to a stable state in the event of failure, allowing the **experiment** to continue.

Repetitions: Repeated measurements obtained during a single **trial**, without restoring the target system to a consistent state between measurements. Repetition gives a more accurate measure of a single **sample**, whereas **replication** gives measurements of multiple samples.

Replicability: The consistency of **responses** from **replicated trials**. The **distribution** of responses is an important part of replicability: the distribution desired depends on the demands of the **customer**.

| | |
|---|---|
| Replications: | Measurements of a single combination obtained from multiple **trials**, restoring the target system to a consistent state between measurements. Replication gives measurements of multiple **samples**, whereas **repetition** gives multiple measurements of a single sample. |
| Response: | A measure of a target system's behaviour. Each **trial** gives a single response. |
| Response variable: | A measure of the behaviour of interest to the **experimenter**. This could be the **combination**'s **quality** or the value for a particular **fitness metric**. The set of **responses** from replicated trials combine to give a single value of the response variable for each combination (e.g. its mean response or **signal to noise ratio**). |
| Response surface: | A surface that lies over the **input space**, using the dimensions of the input space and an additional "response variable dimension." Each point on the surface shows the value of the **response variable** for a **combination**. |
| Results database: | A component of ACT; a repository of the results obtained during an **experiment**. |
| Robust: | Consistently good **responses** with low **variability**, even when there is variation in the **uncontrolled factors**. |
| Run function: | A function in the **target wrapper** that runs and empirically measures a single **trial** of the target system. |
| Sample: | A measurement of the **response variable**'s value, drawn from a **population**. |
| Search strategy: | A component of ACT that provides a policy for making **experimental adaptations** to the target system's configuration, and to the conditions of use; generates a sequence of **combinations** to **test**. |
| Search strategy implementer: | A programmer who produces **search strategy** components for ACT. |
| Signal to noise ratio: | A metric for summarising the **robustness** of a **combination**, given a set of **responses** from **replicated trials**. Abbreviated to SNR. |
| Significant: | An **effect** is said to be statistically significant if it is accepted at a given level of confidence (e.g. 95% confidence) that change in the **response variable**'s value is due to that effect rather than an alternative explanation (e.g. random variation). |
| TACT process: | The combination of **Taguchi Methods** and ACT. |
| Taguchi Methods: | A standardised set of statistical techniques for the **Design Of Experiments**. |
| Target controller: | A component of ACT; dynamically loads the functions in the **target wrapper** and uses them to interact with the target system. |
| Target system: | The (software) system to be configured. |
| Target system administrator: | The person responsible for use of the target system; writes functions for adapting, running, observing and evaluating the behaviour of the target system. |
| Target system developer: | The person responsible for the target system implementation. |
| Target wrapper: | A component of ACT, associated with a target system; consists of a set of dynamically linked libraries (DLLs) containing **adaptation functions**, a **run function** and a **recovery function**. |
| Test: | Empirical measurement of the behaviour of a target system configuration under a given condition of use. Each test involves running a **trial**. |
| Trials: | Runs of the target system, where the behaviour of a **combination** is empirically measured during each trial, and the target system and its conditions of use are configured between trials. Each trial **tests** a combination. |
| Tuning: | Configuring a target system at a given time, to find a configuration that behaves in a desired way. |
| Usage aspect: | An aspect of the conditions of use that can be set explicitly during an **experiment**. Referred to as a **factor** when varied during an experiment. |
| Variability: | The variation in **response** from **replicated trials**. It is the inverse of **replicability**. |
| White box system: | A target system for which there is access to its internals (e.g. source code). Interaction with the target system can delve behind the interfaces that it exposes. Contrast with **black box systems**. |
| Workload: | Facets of how the target system is used. |

## Appendix B: Issues in observation-driven configuration

The questions below concern the feasibility and policy issues involved in applying the techniques described in this thesis. These questions should be answered when producing and using an automated tool for the observation-driven evolution of a target system. The terms used are defined either in the glossary in Appendix A or in the description of ACT 2.0 in section 7.3. The column on the right indicates whether the question is pertinent to running experiments before the system goes into use (B), to on-the-fly evolution (F) or to both (B/F).

- Is the target system suitable for automated observation, adaptation and control? B/F
  - What aspects of the target system can be observed? B/F
    - What in-built facilities are available for observation? B/F
    - What techniques are appropriate for inserting additional probes? B/F
    - How reliable are observations? Are probe effects significant? B/F
  - What aspects of the target system can be adapted without human intervention? B/F
    - What in-built facilities are available for adaptation? B/F
    - What additional techniques are appropriate for adaptation? B/F
    - Can adaptations be made on-the-fly? F
    - What perturbation will users of the target system experience during an adaptation? F
    - How is the consistency of the target system's configuration and state ensured? Can the tool reverse a change or recover from failure? F
  - Can the target system be controlled to start, drive, stop and restore it to a consistent state for each trial? B
  - Do the conditions of use realistically reflect the customer's likely workload and environment? B
- Is there a clear configuration goal? B/F

- What are the fitness metrics and what is the desired value (i.e. minimise, maximise or nominal value) for each? B/F
- What is the relative importance of each fitness metric? What is the aggregating function? B/F
- What is the aim? E.g. find a good configuration under a given condition, discover specific characteristics of the target system, produce a predictive model of the target system's behaviour, or continually adapt the target system to constantly meet the configuration goal. B/F
- What are the constraints on the target system's operation? F
    - Is the target system in use? F
    - What level of disruption to users is acceptable, in terms of deterioration in performance, down-time and risk of data loss? F

- What input does the experimenter give to the tool? B/F
    - Can additional input be given to the tool on-the-fly? F
    - What information does the experimenter provide about interacting with the target system? B/F
        - How is each configurable aspect described, including its legal levels, how to use the adaptation mechanism that configures it, and the estimated cost of adapting it to a specific level (in terms of resource requirements and expected perturbation)? B/F
        - How is each probe described, including what it measures, estimates of probe effects it causes and how it can be configured? B/F
        - How is the estimated cost of using a probe described (i.e. the cost of deployment, installation, activation, deactivation, uninstallation and removal)? F
        - How does the tool start, drive, stop and restore the target system? B

- How are *gauge*, *advice*, *model* and *search strategy* components described? B/F
    - How are the locations of components discovered? B/F
    - What is the interface for the tool to use these components? B/F
- What resources are available? What resources are required by the tool and by the target system (for normal usage, for testing, and during adaptation)? B/F
  - How are resources described? B/F
  - Can the description of available resources be changed on-the-fly? F
- What is the *evolution strategy* for deciding how and when to adapt the target system's configuration? B/F
  - How are observations interpreted to recognise when the configuration goal is met and to evaluate the benefits of an adaptation? B/F
  - What *search strategies* are available and how is one chosen? B/F
  - What target-specific information (i.e. *gauges*, *advice* and *models*) is available to guide configuration of the target system? B/F
    - How does the tool evaluate information to determine if and when it is of use? B/F
    - How is target-specific information used? B/F
    - How does the tool infer new knowledge from experiment results? B/F
  - How does the tool minimise the risk of self-introduced degradation (i.e. ensure that adaptations are appropriate and that adaptations are made in a timely manner)? F
  - How do the resources required and perturbation caused by adaptations affect the *evolution strategy*? F
  - How does the reversibility of adaptations affect the *evolution strategy*? F
  - Is there a meta-strategy for adapting the *evolution strategy* while it is in use? B/F

- Is a database maintained of past observations, and of adaptations made to the target system's configuration? B/F
  - Where is the database stored and how is it queried / updated? B/F
  - How is the database searched to find interesting behavioural characteristics and to predict the behaviour of combinations? B/F

## Appendix C: Example of an *experiment description*

Figure C.1 to Figure C.6 show sections of an *experiment description* file for the example target system of a mail server, described in section 5.1.

The `factors` section in Figure C.1 and Figure C.2 describes the configurable aspects of the target system and the usage aspects of its conditions of use, denoted `targFactors` and `conditionsFactors` respectively.

There are two configurable aspects, each denoted by the tag `targFactor`. The first sets the cache size to a given integer, the legal levels being 50, 100, 500 and 1000. The expected time to adapt the cache size is twenty seconds, given by the `timeToAdapt` tag. The function to perform the adaptation, given by the `adaptationFunc` tag, is located in the DLL named *libAppMailServer.so* and is called *setCacheSize*. The second configurable aspect sets the threshold at which the cache is considered full. This may be set to an integer between the values 70 and 100 where the acceptable granularity (`legalGranular`) for incrementing and decrementing the parameter is 5. Thus the legal levels of *FullThreshold* are 70, 75, 80, 85, 90, 95 and 100. The suggested granularity (`sampleGranular`) for changing *FullThreshold*, however, is 10; the experimenter is suggesting that only the values 70, 80, 90 and 100 should be used during trials. To adapt the *FullThreshold* is expected to take twenty seconds and the *adaptation function* can be found in the DLL named *libAppMailServer.so*, called *setFullThreshold*.

```
<?xml version="1.0"?>
<!DOCTYPE ACT SYSTEM "experimentDescription.dtd">
<ACT>
   <factors>
      <targFactors>
         <targFactor>
            <name>CacheSize</name>
            <levels>
               <enumeration TYPE="int">
                  <level>50</level>
                  <level>100</level>
                  <level>500</level>
                  <level>1000</level>
               </enumeration>
            </levels>
            <timeToAdapt UNITS="secs">20</timeToAdapt>
            <adaptationFunc>
               <funcLocation>
                  <dll>libAppMailServer.so</dll>
                  <func>setCacheSize</func>
               </funcLocation>
            </adaptationFunc>
         </targFactor>
         <targFactor>
            <name>FullThreshold</name>
            <levels>
               <range TYPE="int">
                  <start>70</start>
                  <end>100</end>
```

```
                <legalGranular>5</legalGranular>
                <sampleGranular>10</sampleGranular>
             </range>
          </levels>
          <timeToAdapt UNITS="secs">20</timeToAdapt>
          <adaptationFunc>
             <funcLocation>
                <dll>libAppMailServer.so</dll>
                <func>setFullThreshold</func>
             </funcLocation>
          </adaptationFunc>
       </targFactor>
    </targFactors>
```

**Figure C.1: Example *experiment description* – factors**

There is one usage aspect, shown in Figure C.2, denoted by the tag `conditionsFactors`. This aspect, named *POP3_user_instances*, adapts the workload by setting the number of concurrent POP3 client connections. It may be set to any integer level in the range 5 to 600. The experimenter suggests that this be changed in steps of 5 and expects the change to take negligible time (i.e. 0 seconds). The function to perform the adaptation can be found in the DLL named *libAppMailServer.so* and is called *setPopThreads*.

```
    <conditionsFactors>
       <conditionsFactor>
          <name>POP3_user_instances</name>
          <levels>
             <range TYPE="int">
                <start>5</start>
                <end>600</end>
                <legalGranular>1</legalGranular>
                <sampleGranular>5</sampleGranular>
             </range>
          </levels>
          <timeToAdapt UNITS="secs">0</timeToAdapt>
          <adaptationFunc>
             <funcLocation>
                <dll>libAppMailServer.so</dll>
                <func>setPopThreads</func>
             </funcLocation>
          </adaptationFunc>
       </conditionsFactor>
    </conditionsFactors>
 </factors>
```

**Figure C.2: Example *experiment description* – usage aspects**

The `fitnessMetrics` section, shown in Figure C.3, gives the names of the fitness metrics that are generated during each trial. These are called *Rcpt*, *Fetch*, *Apps*, *ASize*, *FSize*, *Fail*, *FInt* and *WIs*.

```
 <fitnessMetrics>
    <fitnessMetric>Rcpt</fitnessMetric>
    <fitnessMetric>Fetch</fitnessMetric>
    <fitnessMetric>Apps</fitnessMetric>
    <fitnessMetric>ASize</fitnessMetric>
    <fitnessMetric>FSize</fitnessMetric>
    <fitnessMetric>Fail</fitnessMetric>
    <fitnessMetric>FInt</fitnessMetric>
    <fitnessMetric>WIs</fitnessMetric>
 </fitnessMetrics>
```

**Figure C.3: Example *experiment description* – fitness metric names**

The `functions` section, shown in Figure C.4, gives the location of each of the functions in the *target wrapper*, which are used to control the target system. For example, the `runFunc` tag denotes the *run function*, which is located in the *libAppMailServer.so* library and is called *run*.

```
<functions>
   <runFunc>
      <funcLocation>
         <dll>libAppMailServer.so</dll>
         <func>run</func>
      </funcLocation>
   </runFunc>
   <recoveryFunc>
      <funcLocation>
         <dll>libAppMailServer.so</dll>
         <func>recover</func>
      </funcLocation>
   </recoveryFunc>
   <validationFunc>
      <funcLocation>
         <dll>libAppMailServer.so</dll>
         <func>validate</func>
      </funcLocation>
   </validationFunc>
   <newResultObjFunc>
      <funcLocation>
         <dll>libAppMailServer.so</dll>
         <func>newResultObj</func>
      </funcLocation>
   </newResultObjFunc>
</functions>
```

**Figure C.4: Example *experiment description* – functions**

The `resources` section, shown in Figure C.5, describes the resources that are available to ACT for the experiment. It states that there is a time limit of 100 hours imposed for running all the trials. The machines available to carry out these trials are listed as machine 30 and machines 32 to 34.

```
<resources>
   <timeAvailable UNITS="hrs">100</timeAvailable>
   <machines>
      <numberedList>30,32-34</numberedList>
   </machines>
</resources>
```

**Figure C.5: Example *experiment description* – resources**

The `miscellaneous` section, shown in Figure C.6, gives an upper bound on the length of time for each trial: a trial times out if it takes longer than 45 minutes. It also specifies that the maximum number of recovery attempts to be made when testing any combination is one. If the trial fails after one recovery attempt, a failure is recorded and testing continues for the next combination.

As there is no recommendation of a particular search strategy to use, ACT uses the default search strategy (i.e. *grid sampling*, which runs a full factorial experiment).

```
<miscellaneous>
   <timeout UNITS="mins">45</timeout>
   <maxRecovers>1</maxRecovers>
```

```
   </miscellaneous>
</ACT>
```

**Figure C.6: Example *experiment description* – miscellaneous**

## Appendix D: Example of *target wrapper* functions

Figure D.1 shows the signatures of the *target wrapper*'s functions for a back-end mail server. ACT discovers the names and locations of these functions from the *experiment description*. For the *adaptation functions*, the *experiment description* also specifies the type of the levels to which each factor can be set (restricted to one of *int*, *float* or *string*).

```
/**
 * Adaptation functions for target system's factors.
 */
void setCacheSize( int level );               /* set the target system cache size */
void setFullThreshold( int level );       /* set the full threshold for the cache */
void setFetchTimeLimit( float level );     /* set acceptable time limit for fetch */
void setQueuePolicy( const string& level ); /* set queuing policy for input queue */

/**
 * Adaptation functions for conditions of use.
 */
void setPopThreads( int level );             /* set number of POP3 user instances */
void setNetworkLoss( float level );            /* set simulated packet loss level */
void setScenario( const string& level );                        /* set workload */

/**
 * Other functions.
 */
bool validate( const Config & c );           /* validate the target system config */
void recover( const Config & c );           /* restore target to consistent state */
IResultObj * run();                         /* run target and measure performance */
IResultObj *newResultObj( const vector<Value *> & levs ); /* construct result obj */
```

**Figure D.1: Example functions in the *target wrapper***

## Appendix E: First phase experiment design

| Factor levels (coded) | | | | Factor levels (uncoded) | | | | Throughput (normalised) | | | | | |
|---|---|---|---|---|---|---|---|---|---|---|---|---|---|
| TNE | MaxLDAP | LDAPnum | DispNum | TNE | MaxLDAP | LDAPnum | DispNum | Rep 1 | Rep 2 | Rep 3 | Rep 4 | Mean | SNR |
| 1 | 1 | 1 | 1 | 10,000 | 100 | 2 | 2 | 1.04179 | 1.04286 | 1.03887 | 1.04011 | 1.04091 | 0.3482 |
| 1 | 1 | 2 | 2 | 10,000 | 1,001 | 2 | 2 | 1.03664 | 1.04058 | 1.03610 | 1.03802 | 1.03783 | 0.32251 |
| 1 | 1 | 3 | 3 | 10,000 | 2,000 | 2 | 2 | 1.04010 | 1.04090 | 1.03611 | 1.04088 | 1.03950 | 0.33642 |
| 1 | 2 | 1 | 1 | 10,000 | 100 | 1 | 5 | 1.01618 | 1.01351 | 1.00902 | 1.01522 | 1.01348 | 0.11621 |
| 1 | 2 | 2 | 2 | 10,000 | 1,001 | 1 | 5 | 0.87137 | 0.87319 | 0.87380 | 0.87376 | 0.87303 | -1.17944 |
| 1 | 2 | 3 | 3 | 10,000 | 2,000 | 1 | 5 | 0.99704 | 1.00065 | 1.00056 | 1.00127 | 0.99988 | -0.00108 |
| 1 | 3 | 1 | 1 | 10,000 | 100 | 4 | 8 | 0.98439 | 0.98155 | 0.98125 | 0.98179 | 0.98224 | -0.15563 |
| 1 | 3 | 2 | 2 | 10,000 | 1,001 | 4 | 8 | 0.82424 | 0.82573 | 0.82606 | 0.82351 | 0.82489 | -1.67214 |
| 1 | 3 | 3 | 3 | 10,000 | 2,000 | 4 | 8 | 0.93522 | 0.94136 | 0.93701 | 0.93796 | 0.93789 | -0.55705 |
| 2 | 1 | 1 | 2 | 30,000 | 100 | 1 | 2 | 1.04869 | 1.04978 | 1.05120 | 1.05066 | 1.05009 | 0.42448 |
| 2 | 1 | 2 | 3 | 30,000 | 1,001 | 1 | 2 | 1.05589 | 1.05607 | 1.05547 | 1.05524 | 1.05567 | 0.47054 |
| 2 | 1 | 3 | 1 | 30,000 | 2,000 | 1 | 2 | 1.05621 | 1.05371 | 1.05361 | 1.05452 | 1.05451 | 0.461 |
| 2 | 2 | 1 | 2 | 30,000 | 100 | 4 | 5 | 1.02139 | 1.02600 | 1.02556 | 1.02599 | 1.02474 | 0.2122 |
| 2 | 2 | 2 | 3 | 30,000 | 1,001 | 4 | 5 | 1.02035 | 1.01656 | 1.01214 | 1.01762 | 1.01667 | 0.14346 |
| 2 | 2 | 3 | 1 | 30,000 | 2,000 | 4 | 5 | 1.01656 | 1.01357 | 1.01418 | 1.01461 | 1.01473 | 0.12699 |
| 2 | 3 | 1 | 2 | 30,000 | 100 | 2 | 8 | 0.99739 | 0.99871 | 0.99468 | 0.99634 | 0.99678 | -0.02804 |
| 2 | 3 | 2 | 3 | 30,000 | 1,001 | 2 | 8 | 0.98185 | 0.98176 | 0.98165 | 0.98112 | 0.98159 | -0.16137 |
| 2 | 3 | 3 | 1 | 30,000 | 2,000 | 2 | 8 | 0.94995 | 0.94706 | 0.95160 | 0.94836 | 0.94924 | -0.4525 |
| 3 | 1 | 1 | 3 | 50,000 | 100 | 4 | 2 | 1.06305 | 1.06361 | 1.06563 | 1.06227 | 1.06364 | 0.5359 |
| 3 | 1 | 2 | 1 | 50,000 | 1,001 | 4 | 2 | 1.06139 | 1.06127 | 1.05931 | 1.06508 | 1.06176 | 0.52049 |
| 3 | 1 | 3 | 2 | 50,000 | 2,000 | 4 | 2 | 1.05289 | 1.05763 | 1.06576 | 1.01239 | 1.04717 | 0.39515 |
| 3 | 2 | 1 | 3 | 50,000 | 100 | 2 | 5 | 1.03708 | 1.03372 | 1.03558 | 1.03254 | 1.03473 | 0.29651 |
| 3 | 2 | 2 | 1 | 50,000 | 1,001 | 2 | 5 | 1.02495 | 1.02683 | 1.01093 | 1.02186 | 1.02114 | 0.18125 |
| 3 | 2 | 3 | 2 | 50,000 | 2,000 | 2 | 5 | 0.90870 | 0.91028 | 0.90458 | 0.90881 | 0.90809 | -0.83747 |
| 3 | 3 | 1 | 3 | 50,000 | 100 | 1 | 8 | 0.99866 | 1.00509 | 1.00254 | 1.00110 | 1.00185 | 0.01594 |
| 3 | 3 | 2 | 1 | 50,000 | 1,001 | 1 | 8 | 0.99073 | 0.99282 | 0.99061 | 0.98954 | 0.99093 | -0.0792 |
| 3 | 3 | 3 | 2 | 50,000 | 2,000 | 1 | 8 | 0.81118 | 0.81172 | 0.81286 | 0.80691 | 0.81067 | -1.82325 |

## Appendix F: Second phase experiment design

| | Factor levels (normalised) | | | Factor levels (uncoded) | | | Throughput results | | | | | |
|---|---|---|---|---|---|---|---|---|---|---|---|---|
| | TNE | MaxLDAP | DispNum | TNE | MaxLDAP | DispNum | Rep 1 | Rep 2 | Rep 3 | Rep 4 | Mean | SNR |
| Centre point | 0 | 0 | 0 | 32,930 | 100 | 5 | 1.05695 | 1.05256 | 1.05524 | 1.05311 | 1.05446 | 0.46061 |
| | 0 | 0 | 0 | 32,930 | 100 | 5 | 1.05485 | 1.05372 | 1.05452 | 1.05349 | 1.05414 | 0.45800 |
| | 0 | 0 | 0 | 32,930 | 100 | 5 | 1.05566 | 1.05424 | 1.05576 | 1.05377 | 1.05486 | 0.46387 |
| | 0 | 0 | 0 | 32,930 | 100 | 5 | 1.05681 | 1.05236 | 1.05167 | 1.05186 | 1.05318 | 0.44996 |
| | 0 | 0 | 0 | 32,930 | 100 | 5 | 1.05689 | 1.05043 | 1.05209 | 1.05568 | 1.05377 | 0.45486 |
| | 0 | 0 | 0 | 32,930 | 100 | 5 | 1.05541 | 1.05572 | 1.05274 | 1.05130 | 1.05379 | 0.45506 |
| Star points | -1.682 | 0 | 0 | 24,520 | 100 | 5 | 1.05126 | 1.05088 | 1.04638 | 1.05032 | 1.04971 | 0.42134 |
| | 1.682 | 0 | 0 | 41,340 | 100 | 5 | 1.05871 | 1.06006 | 1.05310 | 1.06057 | 1.05811 | 0.49051 |
| | 0 | -1.682 | 0 | 32,930 | 24 | 5 | 1.05263 | 1.04366 | 1.05009 | 1.05295 | 1.04983 | 0.42224 |
| | 0 | 1.682 | 0 | 32,930 | 176 | 5 | 1.05060 | 1.01334 | 1.05227 | 1.05818 | 1.04360 | 0.36682 |
| | 0 | 0 | -1.682 | 32,930 | 100 | 3 | 1.05782 | 1.05400 | 1.05609 | 1.05252 | 1.05511 | 0.46589 |
| | 0 | 0 | 1.682 | 32,930 | 100 | 7 | 1.05255 | 1.04737 | 1.05396 | 1.05339 | 1.05182 | 0.43873 |
| Corner points | -1 | -1 | -1 | 27,930 | 55 | 4 | 1.04988 | 1.05230 | 1.05233 | 1.05312 | 1.05191 | 0.43953 |
| | 1 | -1 | -1 | 37,930 | 55 | 4 | 1.05887 | 1.05642 | 1.05505 | 1.05749 | 1.05696 | 0.48113 |
| | -1 | 1 | -1 | 27,930 | 145 | 4 | 1.05717 | 1.05131 | 1.04751 | 1.04899 | 1.05124 | 0.43392 |
| | 1 | 1 | -1 | 37,930 | 145 | 4 | 1.04215 | 1.05588 | 1.05928 | 1.05687 | 1.05355 | 0.45253 |
| | -1 | -1 | 1 | 27,930 | 55 | 6 | 1.04830 | 1.05021 | 1.05177 | 1.05052 | 1.05020 | 0.42542 |
| | 1 | -1 | 1 | 37,930 | 55 | 6 | 1.05852 | 1.05211 | 1.05801 | 1.05625 | 1.05622 | 0.47503 |
| | -1 | 1 | 1 | 27,930 | 145 | 6 | 1.05465 | 1.05288 | 1.05365 | 1.05581 | 1.05425 | 0.45884 |
| | 1 | 1 | 1 | 37,930 | 145 | 6 | 1.05493 | 1.05221 | 1.05589 | 1.05815 | 1.05529 | 0.46742 |

The "normalised" factor levels describe a central composite design (see section 4.3.4), where level 0 refers to the centre point. Uncoded levels depend on the step size (e.g. the step size for `TNE` is 5000 for each unit of normalised factor).